\documentclass[usenatbib]{mnras}

\usepackage{amsbsy}
\usepackage{graphicx}

\newcommand{\arepo}{{\sc arepo}}
\newcommand{\gadget}{{\sc gadget-2}}
\newcommand{\mhd}{{MHD}}
\newcommand{\illustris}{{\sc illustris}}

\newcommand{\kms} {{\rm km~s}^{-1}}

\newcommand{\Mpc} {{\rm Mpc}}
\newcommand{\mo}{{\rm M}_{\sun}}
\newcommand{\Gyr}{{\rm Gyr}}
\newcommand{\K}{{\rm K}}
\newcommand{\gsim}{\lower.7ex\hbox{$\;\stackrel{\textstyle>}{\sim}\;$}}
\newcommand{\lsim}{\lower.7ex\hbox{$\;\stackrel{\textstyle<}{\sim}\;$}}
\newcommand{\camb}{{\sc camb}}
\newcommand{\music}{{\sc music}}
\newcommand{\G}{{\rm G}}
\newcommand{\muG}{{\rm \mu G}}
\newcommand{\nG}{{\rm nG}}

\newcommand{\rev}[1]{#1}

\defcitealias{Marinacci2014b}{b} 
\defcitealias{Planck2015}{Planck Collaboration XIX 2015} 
\defcitealias{Neronov2010}{Neronov \& Vovk 2010} 

\begin{document}
\title[Cosmological magnetic fields]
{The large-scale properties of simulated cosmological magnetic fields} 
\author[F.~Marinacci et al.]
{Federico Marinacci${^1}$\thanks{E-mail:
fmarinac@mit.edu}, Mark Vogelsberger${^1}$, Philip Mocz${^2}$ and 
R\"udiger~Pakmor${^3}$ \vspace*{0.2cm}\\
  $^1$Kavli Institute for Astrophysics and Space Research, 
  Massachusetts Institute of Technology, Cambridge, MA 02139, USA\\
  $^2$Harvard--Smithsonian Center for Astrophysics, 
  60 Garden Street, Cambridge, MA 02138, USA\\
  $^3$Heidelberger Institut f\"{u}r Theoretische Studien,
  Schloss-Wolfsbrunnenweg 35, 69118 Heidelberg, Germany}
\date{Accepted 2015 July 23.  Received 2015 July 15; in original form 2015 May 29.}

\pagerange{\pageref{firstpage}--\pageref{lastpage}}
\pubyear{2015}

\maketitle

\label{firstpage}

\begin{abstract}
We perform uniformly sampled large-scale cosmological
simulations including magnetic fields with the moving mesh code \arepo.  We run
two sets of \mhd\ simulations: one including adiabatic gas physics only; the
other featuring the fiducial feedback model of the \illustris\ simulation. In
the adiabatic case, the magnetic field amplification follows the $B \propto
\rho^{2/3}$ scaling derived from `flux-freezing' arguments, with the seed field
strength providing an overall normalization factor. At high baryon
overdensities the amplification is enhanced by shear flows and turbulence.
Feedback physics and the inclusion of radiative cooling change this picture
dramatically. \rev{In haloes,} gas collapses to much larger densities and 
the magnetic field is amplified strongly \rev{and to the same maximum intensity
irrespective of the initial seed field of which any memory is lost}.
At lower densities a dependence on the seed field strength and
orientation, which in principle can be used to constrain models of cosmic
magnetogenesis, is still present. Inside the most massive haloes magnetic
fields reach values of $\sim 10-100\,\,\muG$, in agreement with galaxy cluster
observations. The topology of the field is tangled and gives rise to 
rotation measure signals in reasonable agreement with the observations. However,
the rotation measure signal declines too rapidly towards larger radii as compared
to observational data.
\end{abstract}

\begin{keywords}
magnetic fields -- MHD -- methods: numerical -- cosmology: theory
\end{keywords}

\section{Introduction}\label{sec:intro}

Magnetic fields are ubiquitous in the Universe. They have been observed at all
scales, ranging from planets and minor bodies in the Solar system
\citep{Vallee1998} to galaxies \citep{Beck2013c} and clusters of galaxies
\citep{Carilli2002, Feretti2012}. Thanks to a variety of observational
techniques, present-day field intensities have been estimated for different
types of objects, and measures of the polarization of radio and infrared
radiation have allowed us to map the field orientation on galactic scales and
beyond. 

Magnetic fields also play an essential role in many astrophysical phenomena.
They are an important factor for the physics of accretion on compact objects
such as neutron stars and (supermassive) black holes \citep[][and references
therein]{Balbus2003}, where they are thought to  generate relativistic jets
propagating into the intracluster medium \citep{Blandford1977}. Propagation and
diffusion of relativistic particles (i.e.  cosmic rays) are heavily affected by
the presence of magnetic field \citep[e.g.][]{Kotera2011}. More in general, the
interaction between the field and relativistic particles provides both a
mechanism for their acceleration to relativistic speeds -- through the
so-called \citet{Fermi1949} mechanism -- and the production of synchrotron
radiation in disc galaxies
\citep{Beck2005,Beck2007b,Berkhuijsen2003,Fletcher2011}, radio-loud active
galactic nuclei \citep[AGNs;][and references therein]{Urry1995}, and radio
relics and haloes in galaxy clusters \citep{Ferrari2008}. On sub-galactic
scales magnetic fields are one of the major components of the interstellar
medium (ISM), providing a significant fraction of the pressure support needed
against gravity \citep{Boulares1990,Cox2005,Ferriere2001}. A vast category of
stellar phenomena  -- for instance radiation from pulsars \citep{Ruderman1975}
-- is also controlled by magnetic fields. Given their widespread presence and
their importance in almost any class of astrophysical objects, our theoretical
understanding of the Universe is incomplete without considering magnetic
fields. 

Although a large amount of observational data have been collected on the
existence of magnetic fields on scales up to galaxy clusters, a robust {\it B} field
detection on even larger scales -- at the level of cosmic filaments and voids
-- has proven to be difficult and only a few indirect constraints
\citep{Neronov2010}, that can be also interpreted in a different framework
\citep{Broderick2012}, are available. The problem in detecting {\it B} fields in such
low-density environments stems from the fact that the expected field strengths
are so small ($\lsim 1\,\nG$) compared to those commonly found in galaxies and
clusters ($\sim10\, \muG$) such that they are at or below the detection
limit of the current instrumentation, even though this should improve with the
next generation of radio instruments such as SKA \citep{Beck2007}. The
same is true for field strengths at (very) high redshift, for which only upper
limits can be placed \citep[see however][for evidence of {\it B} field in galaxies up
to $z\sim 2$]{Athreya1998, Bernet2008, Kronberg1992}.

This scarcity of data leaves us with very little observational guidance as to 
how magnetic fields are originally generated and subsequently amplified, a 
process that is also not well understood theoretically. Many models have been 
proposed, but these can be essentially reduced to two main scenarios. In the 
first scenario, seed magnetic fields are of cosmological origin and are 
generated by several processes during inflation, phase transitions or plasma 
phenomena in the early Universe \citep[for a recent review see][]{Widrow2012}. 
Another alternative is represented by the so-called Biermann battery mechanism 
\citep{Biermann1950}, which can operate in cosmological shocks  \citep{Ryu1998} 
or during reionization \citep{Gnedin2000}. Seed fields are then amplified 
through turbulent dynamo processes \rev{\citep[see e.g.][]{Arshakian2009,Federrath2011, 
Kulsrud1997,Schleicher2013,Sur2010}}, shear flows \citep{Dolag1999} or galactic dynamos 
\citep{Hanasz2004} as baryons collapse in dark matter haloes to assemble the 
structures that populate the Universe today.

In the second scenario, magnetic fields are produced within (proto)galaxies by
stars \citep{Pudritz1989,Schleicher2010} and then ejected by galactic winds
\citep{Volk2000, Donnert2009}. Also AGN activity \citep[e.g.][]{Daly1990, Ensslin1997,
Furlanetto2001, Beck2013b} can contribute to the generation and ejection of {\it B}
fields in the intergalactic medium (IGM). The ejected field can then be
amplified and dispersed by the processes described above. It is conceivable to
assume that, in this second scenario, seed fields are more spatially localized
near the sites where  galaxies form and are then gradually dispersed by gas
motions. If the dispersal process turns out to be not particularly effective,
the different spatial distribution of the seed field can be used to
discriminate between the two scenarios \citep{Cho2014}.

Numerical simulations are an important tool to investigate the close link
between the dynamical state of gas and the amplification of the magnetic field
to the present-day strengths as cosmological structures build up. Given the
importance and the vastness of the problem a number of approaches have been
attempted. A possible technique is to focus on an idealized \mhd\ setup.
For instance \citet[][but see also \citealt{Cho2014}]{Ryu2008} studied the
development of \mhd\ turbulence starting from different seed field
configurations, in order to assess the amplitude of the field amplification.
From this study they derive a model for injecting magnetic energy in a
non-radiative cosmological simulation to recover the final magnetic field
distribution.

A more attractive approach, the one considered in this paper, is to perform
cosmological \mhd\ simulations and study the amplification of the magnetic
field and the growth of cosmic structures simultaneously. Although the
resolution achieved is not comparable to that of idealized setups, a number of
authors have performed this type of simulations focusing on a variety of
spatial and mass scales. 

The co-evolution of {\it B} fields and the formation of galaxy clusters have been
studied by \cite{Dolag1999} in non-radiative cosmological simulations. They
found that an initial {\it B} field seed is amplified by the building up of the
cluster through gravitational collapse, shear flows and turbulence. The
actual seed adopted has a negligible impact on the final {\it B} field strength.
Those results are confirmed by subsequent studies, using a similar model but a
different seeding technique, which however may affect the {\it B} field distribution
in low-density regions \citep{Donnert2009}. The impact of non-ideal \mhd\
effects has also been explored \citep{Bonafede2011}. Other authors pointed out
the importance of radiative gas cooling \citep{Dubois2008} and anisotropic
thermal conduction \citep{Ruszkowski2011} in further boosting the amplification
of the magnetic field and changing its orientation.

Moving to smaller scales, the emergence and the evolution of a galactic-wide
magnetic field have been investigated by several works on isolated galaxies.
The goals of these calculations are to clarify whether dynamo
processes \citep[][]{Hanasz2009, Schober2012, Schober2013} or disc dynamics 
\citep{Kotarba2009, Wang2009}
could give rise to the observed magnetic field strength in present-day 
Milky Way-type objects or their progenitors, and to assess the impact of the {\it B} field
on galaxy properties such as their star formation history \citep{Wang2009,
Pakmor2013}. The role of supernova-driven galactic winds in explaining the
magnetization of the IGM was also considered \citep{Dubois2010}.  Magnetic
fields in Milky Way-type galaxies have also been studied in zoom-in
cosmological simulations, focusing on mechanisms for the generation of the seed
magnetic field \citep{Beck2013} and investigating how a primordial seed field
is amplified in the halo surrounding the central galaxy \citep[][]{Beck2012} or
within the galaxy itself \citep{Pakmor2014}.  In particular, \cite{Pakmor2014}
showed that it is possible to form a realistic disc galaxy with a well-defined
morphology and simultaneously predict the observed magnetic field strength and
orientation of late-type systems.

While there are many example of \mhd\ calculations on galactic and galaxy
cluster scales, global cosmological simulations (i.e. uniformly sampled boxes)
are relatively rare in the literature. A notable exception, which however
includes only adiabatic physics and treats other physical processes such as
particle acceleration in post-processing, is presented by \cite{Vazza2014,
Vazza2015a}, who performed high-resolution static-mesh simulations (up to
$2400^3$ grid and $300\,\,h^{-1}{\rm Mpc}$ box sizes) to study the small-scale dynamo in
a cosmological volume. Another interesting forthcoming project is the {\sc
magneticum} simulation suite (Dolag et al., in preparation), which will include {\it B}
fields with the scheme described by \cite{Dolag2009}. 

In the present study, we aim at pushing forward the modelling of 
uniformly sampled cosmological simulations by including magnetic fields -- 
through the ideal \mhd\ approximation -- in a series of calculations 
performed with the moving-mesh code \arepo\ \citep{Arepo}. Our goal is to study 
the general properties of \rev{cosmological} magnetic fields and their variation as a 
function of resolution, seed field and baryon physics. To this end we perform 
two sets of simulations at different resolution levels: 
one set only including adiabatic gas physics; the other featuring the fiducial 
model of baryon physics of the {\sc illustris} simulation 
\citep{Vogelsberger2014, Illustris}. To our knowledge, this is the first time that a 
successful physics model including the most important processes for galaxy 
formation is used in this type of simulations. The paper is organized as 
follows. In Section \ref{sec:method} we describe the numerical methodology that 
we adopted both to generate the initial conditions (ICs) and to run the simulations. 
In Section \ref{sec:results} we present the main findings of our simulations, which include 
the large-scale properties of the magnetic field (Sec.~\ref{sec:large-scale}),  
their dependence on the choice of the seed field strength (Sec~\ref{sec:seed 
field}), the properties of {\it B} field within haloes (Sec.~\ref{sec:haloes}), 
and Faraday rotation measure (RM) predictions (Sec~\ref{sec:faradayrot}). In 
Section \ref{sec:discussion} we discuss our results, while Section 
\ref{sec:conclusions} gives our conclusions.

\begin{table*}
\begin{tabular}{lcccccc}
\hline
Simulation & $N_{\rm tot}$ & $M_{\rm DM}$ & $M_{\rm gas}$ & Box size & $\epsilon$ & $\mathit{B}_{0}$ \\
           & (DM + cells) & $(10^{5}\,h^{-1} \mo)$ & $(10^{5}\,h^{-1} \mo)$ & $(h^{-1} {\rm Mpc})$ & $(h^{-1}{\rm kpc})$ & $({\rm G})$ \\
\hline
box-256-ad      & $2\times256^3$ & $4.21\times10^{4}$ & $7.86\times10{^3}$ & 100 &  5.0 & $(0,0,10^{-14})$ \\
box-512-ad      & $2\times512^3$ & $5.26\times10^{3}$ & $9.82\times10{^2}$ & 100 &  2.5 & $(0,0,10^{-14})$ \\           
box-256-ad-low  & $2\times256^3$ & $4.21\times10^{4}$ & $7.86\times10{^3}$ & 100 &  5.0 & $(0,0,10^{-16})$ \\
box-512-ad-low  & $2\times512^3$ & $5.26\times10^{3}$ & $9.82\times10{^2}$ & 100 &  2.5 & $(0,0,10^{-16})$ \\
box-256-ad-high & $2\times256^3$ & $4.21\times10^{4}$ & $7.86\times10{^3}$ & 100 &  5.0 & $(0,0,10^{-12})$ \\
box-512-ad-high & $2\times512^3$ & $5.26\times10^{3}$ & $9.82\times10{^2}$ & 100 &  2.5 & $(0,0,10^{-12})$ \\
\hline
box-256-fp      & $2\times256^3$ & $4.21\times10^{4}$ & $7.86\times10{^3}$ & 100 &  5.0 & $(0,0,10^{-14})$ \\
box-512-fp      & $2\times512^3$ & $5.26\times10^{3}$ & $9.82\times10{^2}$ & 100 &  2.5 & $(0,0,10^{-14})$ \\           
box-256-fp-low  & $2\times256^3$ & $4.21\times10^{4}$ & $7.86\times10{^3}$ & 100 &  5.0 & $(0,0,10^{-16})$ \\
box-512-fp-low  & $2\times512^3$ & $5.26\times10^{3}$ & $9.82\times10{^2}$ & 100 &  2.5 & $(0,0,10^{-16})$ \\
box-256-fp-high & $2\times256^3$ & $4.21\times10^{4}$ & $7.86\times10{^3}$ & 100 &  5.0 & $(0,0,10^{-12})$ \\
box-512-fp-high & $2\times512^3$ & $5.26\times10^{3}$ & $9.82\times10{^2}$ & 100 &  2.5 & $(0,0,10^{-12})$ \\
\hline
box-256-ad-ydir & $2\times256^3$ & $4.21\times10^{4}$ & $7.86\times10{^3}$ & 100 &  5.0 & $(0,10^{-14},0)$ \\
box-256-fp-ydir & $2\times256^3$ & $4.21\times10^{4}$ & $7.86\times10{^3}$ & 100 &  5.0 & $(0,10^{-14},0)$ \\
\hline
\end{tabular}
\caption{Properties of the performed simulations. Each box is simulated two 
times: the first time considering only adiabatic physics and the second 
including the most important physical processes for galaxy formation. Columns 
give (from left to right): simulation name, total number of dark matter 
particles plus gas cells, mass resolution in the dark matter component, mass 
resolution in the gas component, simulation box size, maximum physical 
gravitational softening (reached at $z = 1$), and  initial (comoving) seed 
magnetic field.} 
\label{tab:simulations}
\end{table*}

\section{Numerical methodology}\label{sec:method}

To study the amplification of magnetic field and its properties in a 
cosmological context, we run a series of ideal \mhd\ simulations of 
uniformly sampled cosmological boxes \rev{of size $100\,h^{-1}\Mpc$}. We repeat 
the simulations for two different resolution levels with a total particle number 
of $2\times 256^3$ and $2\times 512^3$, respectively. Each configuration is 
simulated twice: one time by just considering adiabatic physics, a second time 
by including the most important physical processes for galaxy formation. The 
main properties of the runs can be found in Table~\ref{tab:simulations}.

We adopt the set of cosmological parameters according to the re-analysis of
{\it Planck} data performed by \citet[][table 3]{Spergel2015}. This
features a $\Lambda$ cold dark matter ($\Lambda$CDM) cosmology with parameters $\Omega_{\rm m} = 0.302$,
$\Omega_{\rm b} = 0.04751$, $\Omega_{\rm \Lambda} = 0.698$, $\sigma_{8} =
0.817$, $n_{\rm s} = 0.9671$, and a Hubble parameter $H_{0} = 68~\kms\Mpc^{-1}$
(hence implying $h = 0.68$). We compute the transfer function for this cosmology with
\camb\ \citep{camb}, and generate the ICs with \music\ \citep{music}. 
The ICs are created for dark matter simulations only at a starting
redshift of $z = 127$ and baryons are introduced at the beginning of
the simulation through the procedure described in \citet{Marinacci2014a, Marinacci2014b}. 

We evolve the ICs created above with the moving-mesh cosmological code 
\arepo\ \citep{Arepo}, complemented with the extension to include ideal
\mhd\ developed by \citet{Pakmor2013} and successfully applied in `zoom-in'
cosmological simulations of (individual) disc galaxies \citep{Pakmor2014}.
We refer the reader to the original papers for an exhaustive description
of the code and for all the details about the implementation of its modules.
Here, we give just a brief overview of the most important code features.

\arepo\ solves gravitational and collision-less dynamics via a standard TreePM
method \citep[also used, by the popular \gadget\ code;][]{Springel2005b}, that
splits the gravitational force in a long-range contribution, computed by a
Fourier transform method on a mesh, and a short-range contribution, calculated
by an oct-tree algorithm \citep{Barnes1986}. For (ideal) \mhd, \arepo\
adopts finite-volume discretization on an unstructured Voronoi tessellation of
the simulation volume. \mhd\ equations are solved through a second-order
MUSCL-Hancock scheme \citep[e.g.,][]{Toro1999} coupled to the approximate HLLD
Riemann solver \citep{Miyoshi2005}. The Voronoi mesh is free to move with the
local fluid velocity field. This results in a manifestly Galilean-invariant,
quasi-Lagrangian numerical method that keeps the mass per gas cell
approximately constant. 

To ensure the $\nabla\cdot\boldsymbol{B} = 0$ constraint the \mhd\ module 
currently employs the divergence cleaning technique developed by 
\cite{Powell1999}. We are aware of the fact that this technique cannot guarantee 
$\nabla\cdot\boldsymbol{B}$ identically vanishing throughout the simulation domain
but our scheme has proven to yield acceptable divergence errors and results of 
the same quality as compared to constrained transport (CT) schemes 
\citep{Pakmor2013}. Nevertheless, we are working to include a CT scheme in 
\arepo, based on the approach described by \citet{Mocz2014}, to use it in future 
simulations and to address the potential shortcomings of the \cite{Powell1999}
method \citep[see][and references therein]{Hopkins2015}. 

In ideal \mhd\ simulations it is not possible to generate a magnetic field
starting from zero-field ICs. Therefore, the magnetic field must be seeded
appropriately. For simplicity, we seed a homogeneous magnetic field in the box
at the starting redshift along a prescribed direction. This ensures a
divergence-free initial {\it B} field, but leaves us with two parameters to be
chosen: the initial field strength and direction. Previous zoom-in cosmological 
\mhd\ simulations with \arepo\ \citep{Pakmor2014} have shown that the final
results are rather insensitive to the choice of these parameters, at least at
high overdensities. We will explore the situation at lower overdensities in
Section~\ref{sec:seed field}.

Our simulation set also includes a comprehensive model for galaxy formation 
physics \citep{Vogelsberger2013} specifically developed for the {\sc illustris} 
simulation suite \citep{Illustris,Vogelsberger2014}. The model was calibrated against a 
small set of key observables, such as the cosmic star formation history and the 
galaxy stellar mass function, and it is able to successfully reproduce most of 
the observed properties of the global galaxy population at redshift zero. A 
detailed description of all the components of the model can be found in 
\cite{Vogelsberger2013} and its integration within the \arepo\ \mhd\ module in 
\cite{Pakmor2014}. In our runs we use the same fiducial settings as in 
\citet{Vogelsberger2013}.

\rev{Finally, we want to note that in our simulations diffusive effects -- which are important to 
determine the level of turbulence at small scales and the level of {\it B} field 
amplification through dynamo mechanisms -- are controlled by the numerical 
cutoff scale. Therefore, we expect the magnetic Reynolds number to be of  the 
same order of the Reynolds number, implying a magnetic Prandtl number of $\sim 
O(1)$ \citep{Federrath2011}. The Reynolds number in halo of virial radius 
$r_{200}$ can be estimated in our runs as \citep[see][]{Vazza2014} 
\begin{equation}
{\rm Re} \approx \left(\frac{r_{200}}{\Delta x}\right)^{4/3} = \left(\frac{M_{200} N^3}{\Omega_m \rho_{\rm crit} L^3}\right)^{4/9},
\label{eq:Reynolds}
\end{equation}
where $\Delta x$ is the typical size of the resolution element inside the halo,
$M_{200}$ is the halo viral mass, $N^3$ is the total number of resolution element in the box,
$L$ is the box size, $\Omega_{\rm m}\rho_{\rm crit}$ is the total matter density in the Universe, 
and the last equality holds because gas is discretized in volume elements having
roughly the same mass (see also Sect.~\ref{sec:discussion}). We have implicitly assumed 
that the halo contains all the baryons associated with its dark matter halo. 
For $N = 512$, equation (\ref{eq:Reynolds}) yields ${\rm Re} \approx 100$ for a halo of
$\simeq 10^{13}\, \mo$ and ${\rm Re} \approx 500$ for a halo of
$\simeq 10^{15}\, \mo$, respectively. While these value are somewhat lower than
those achieved in more idealized setups \citep{Sur2010, Sur2012} or cosmological
simulations \citep{Vazza2014}, they are nevertheless in the regime in which a 
small-scale dynamo can be in operation \citep{Schekochihin2004}.
}

\section{Results}\label{sec:results}

\subsection{Large-scale properties of the magnetic field}\label{sec:large-scale} 
\begin{figure*}
\centering
\includegraphics[width=0.49\textwidth]{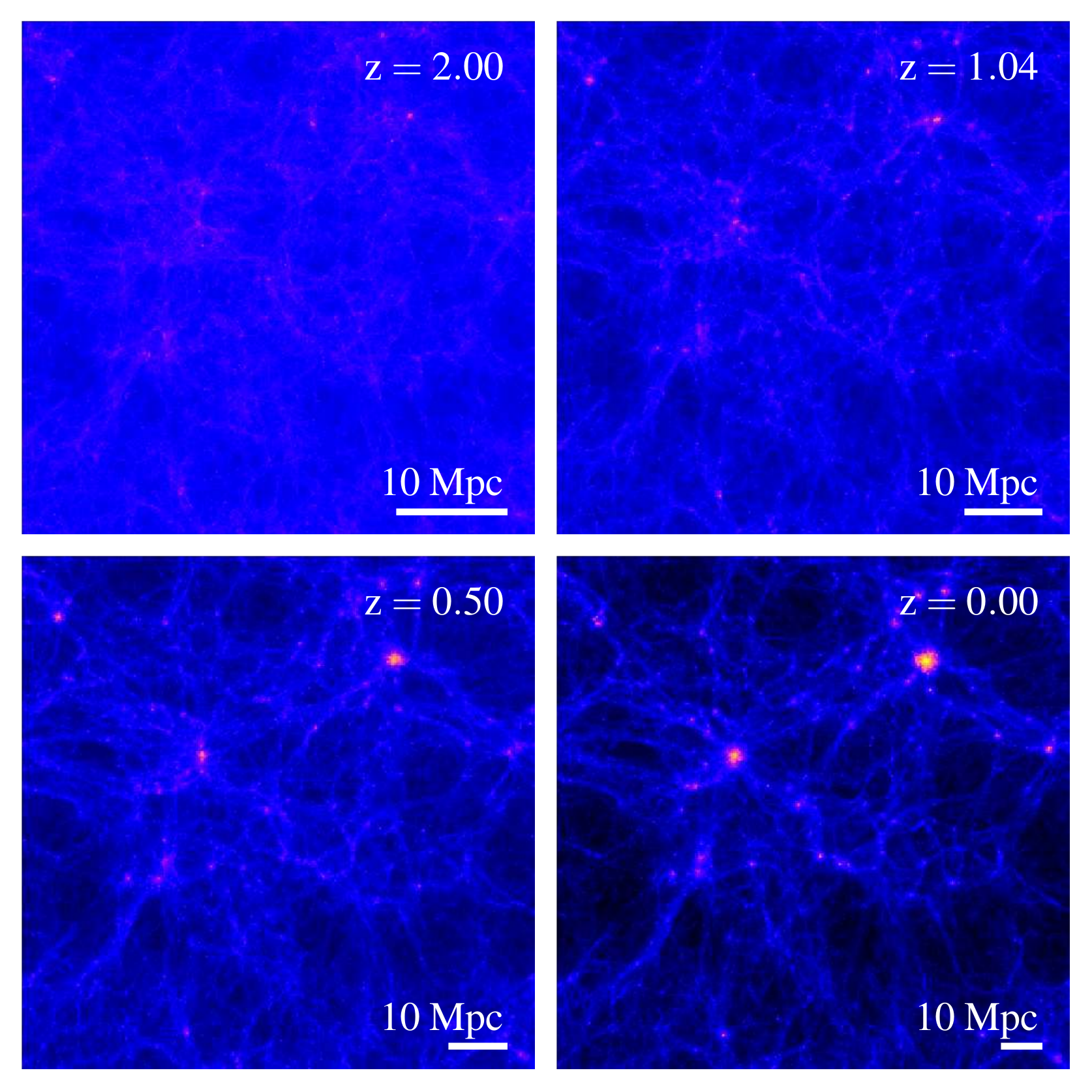}
\includegraphics[width=0.49\textwidth]{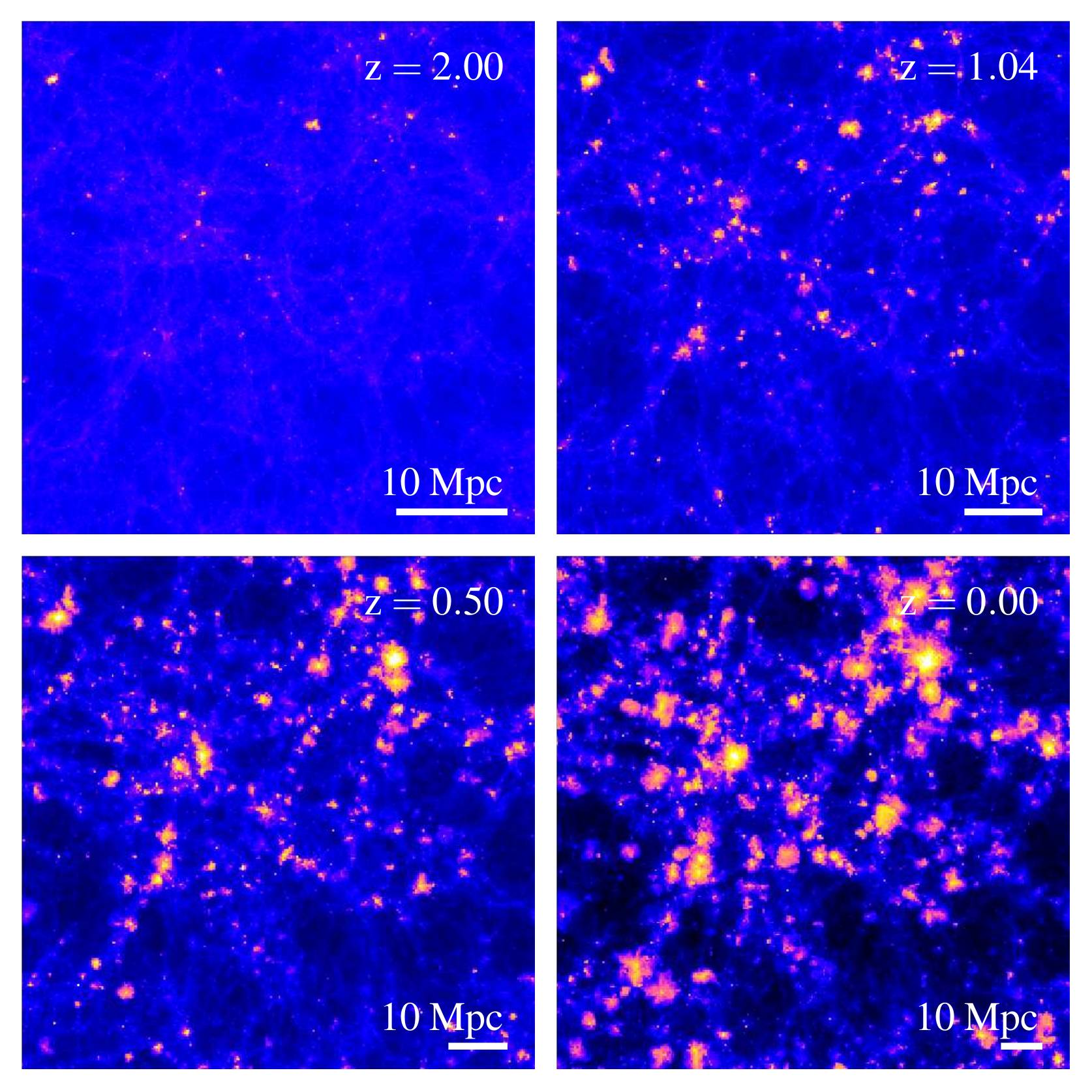}
\caption{Mass-weighted projection of the intensity of the {\it B} field at different 
redshifts, as indicated in the top-right corner of each panel, for the 
simulation box-512-ad (left) and box-512-fp (right). Each panel is $100\,\,h^{-1} {\rm 
Mpc}$ on a side (in comoving units), the full extent of the simulated 
box. The centre of the projection region corresponds to that of the simulated 
domain. The plots have been obtained by considering all the gas cells along the 
$z$-axis (the initial direction of the seed field) within $25\,\,h^{-1}{\rm 
Mpc}$ (in comoving units) from the centre, for a total thickness of 
$50\,\,h^{-1} {\rm Mpc}$. The physical scale at the corresponding redshift is 
indicated on the bottom-right corner of each panel. The colour scheme is the 
same for all the panels and maps logarithmically magnetic field intensities
in the interval $[10^{-9}, 10^{-2}]\,\muG$.}.
\label{fig:Bprojection}
\end{figure*}

We start the analysis of our simulations by
presenting the general large-scale properties of the resulting field.  In
Fig.~\ref{fig:Bprojection}, we show mass-weighted projections of the {\it B} field
intensity\footnote{Unless otherwise stated, the values of magnetic field shown
in the figures are always the physical ones.} 
at different redshifts for the
simulations box-512-ad (left) and box-512-fp (right). The intensity and
direction of the seed field are the same for the two simulations and the latter coincides
with the projection axis. The projections are obtained by considering a region
extending the full size of the computational domain (100 $h^{-1} \Mpc$ in
comoving units) perpendicularly to the projection axis for a thickness of 50
$h^{-1} \Mpc$ (again in comoving units). The centre of the projection region
coincides with that of the computational domain.

The {\it B} field evolution traces the distribution of matter in both simulations.
The largest values of the magnetic field are located at the density peaks. This
is expected, since in ideal \mhd\ the flux of the {\it B} field is conserved.
Therefore, density peaks are expected to host the largest {\it B} field, while in
voids, where the cosmological expansion is dominant, the smallest values of the
{\it B} field can be found. We recall that in the case of adiabatic (and homogeneous)
contraction/expansion flux conservation dictates that the magnetic field
evolves as
\begin{equation}
 \boldsymbol{B} = \frac{\boldsymbol{B_0}}{a^2} \propto \rho^{2/3},
 \label{eq:adexpansion}
\end{equation}
where $a = (1 + z)^{-1}$ is the cosmological scale factor, $\boldsymbol{B_0}$
is the rescaled intensity of the {\it B} field at $z = 0$, and $\rho$ is the gas density.
Equation (\ref{eq:adexpansion}) shows that the net effect of cosmic expansion is a
decrease in the physical intensity of the {\it B} field, clearly visible in the late
redshift panels of Fig.~\ref{fig:Bprojection} in underdense regions. 

\begin{figure*}
\centering
\includegraphics[width=0.24\textwidth]{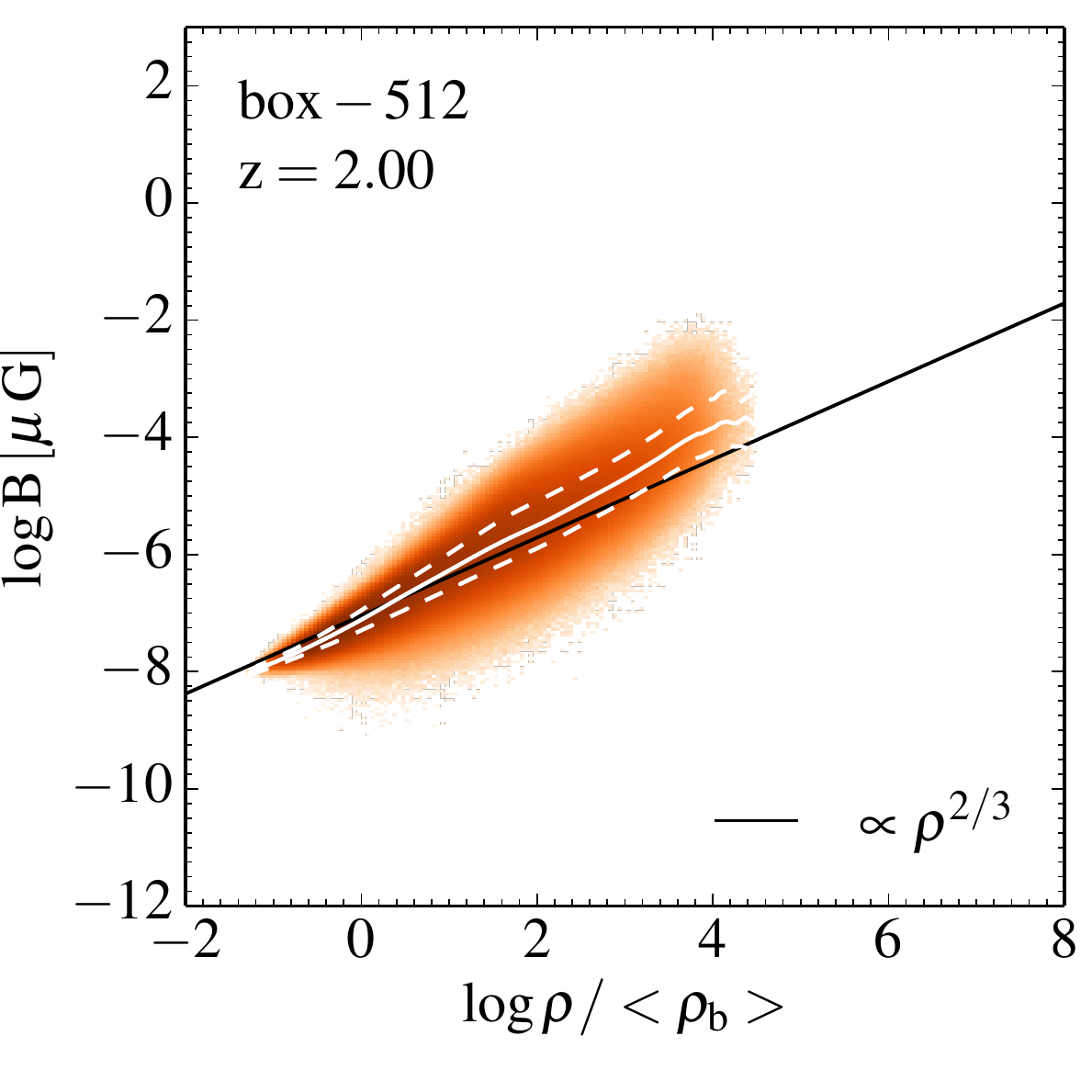}
\includegraphics[width=0.24\textwidth]{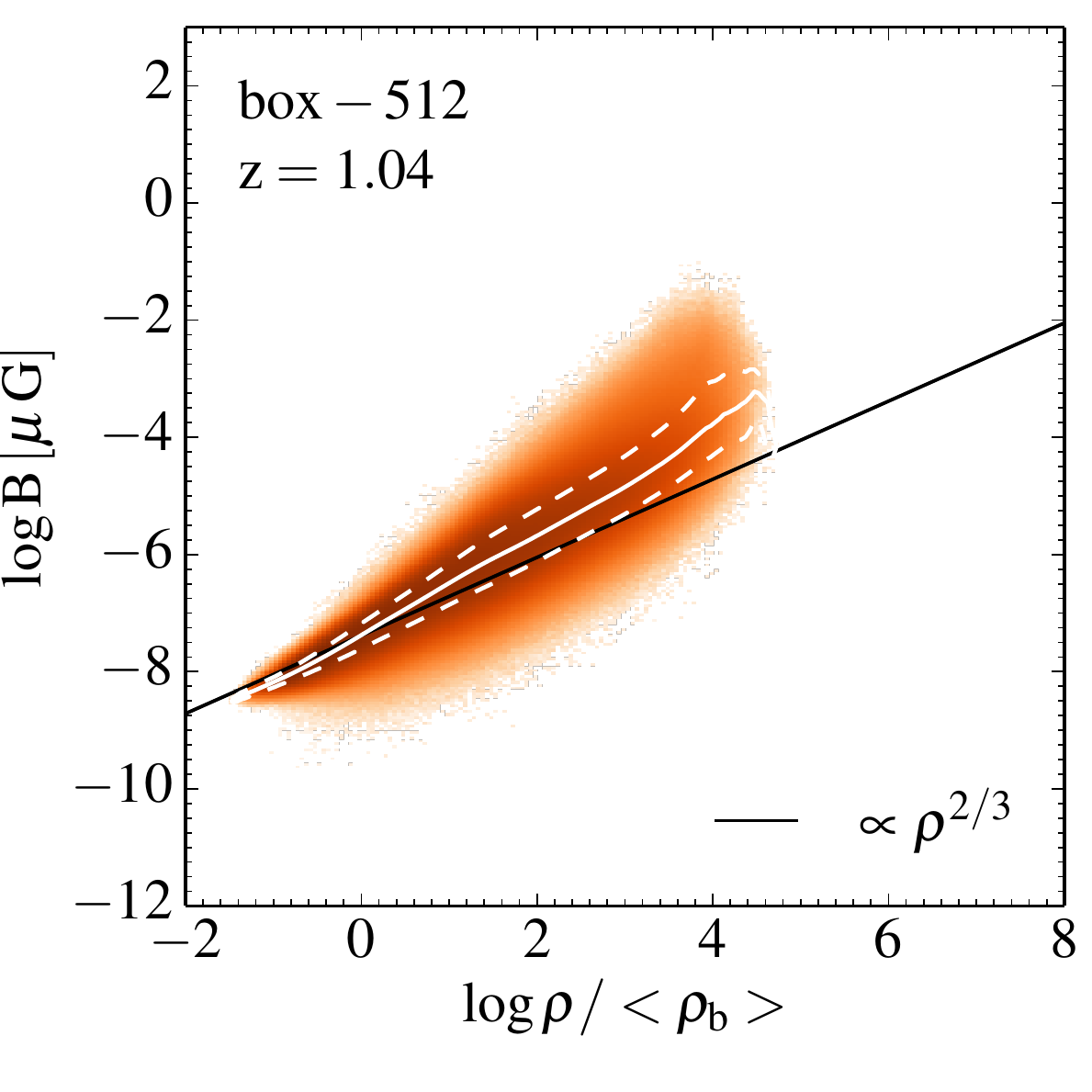}
\includegraphics[width=0.24\textwidth]{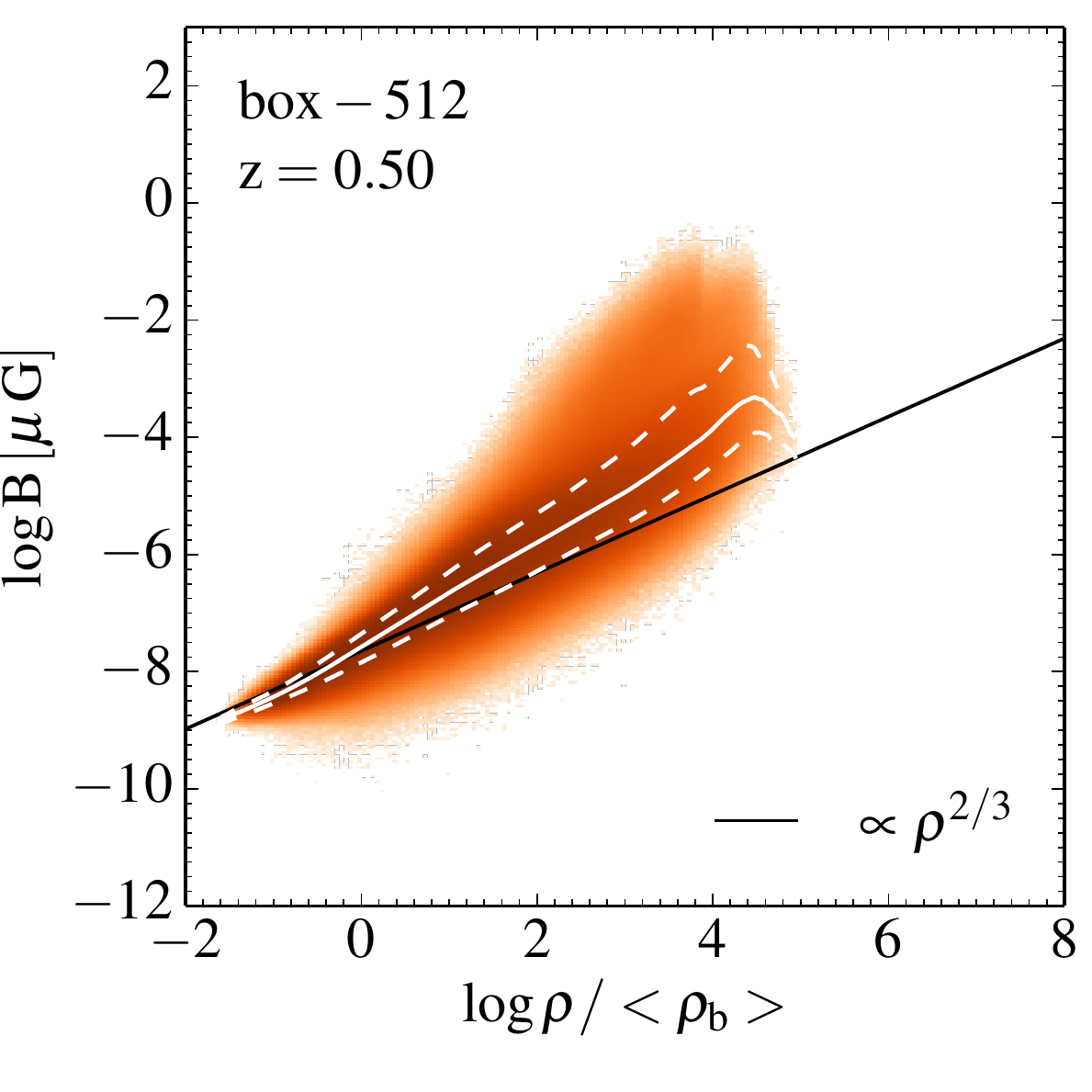}
\includegraphics[width=0.24\textwidth]{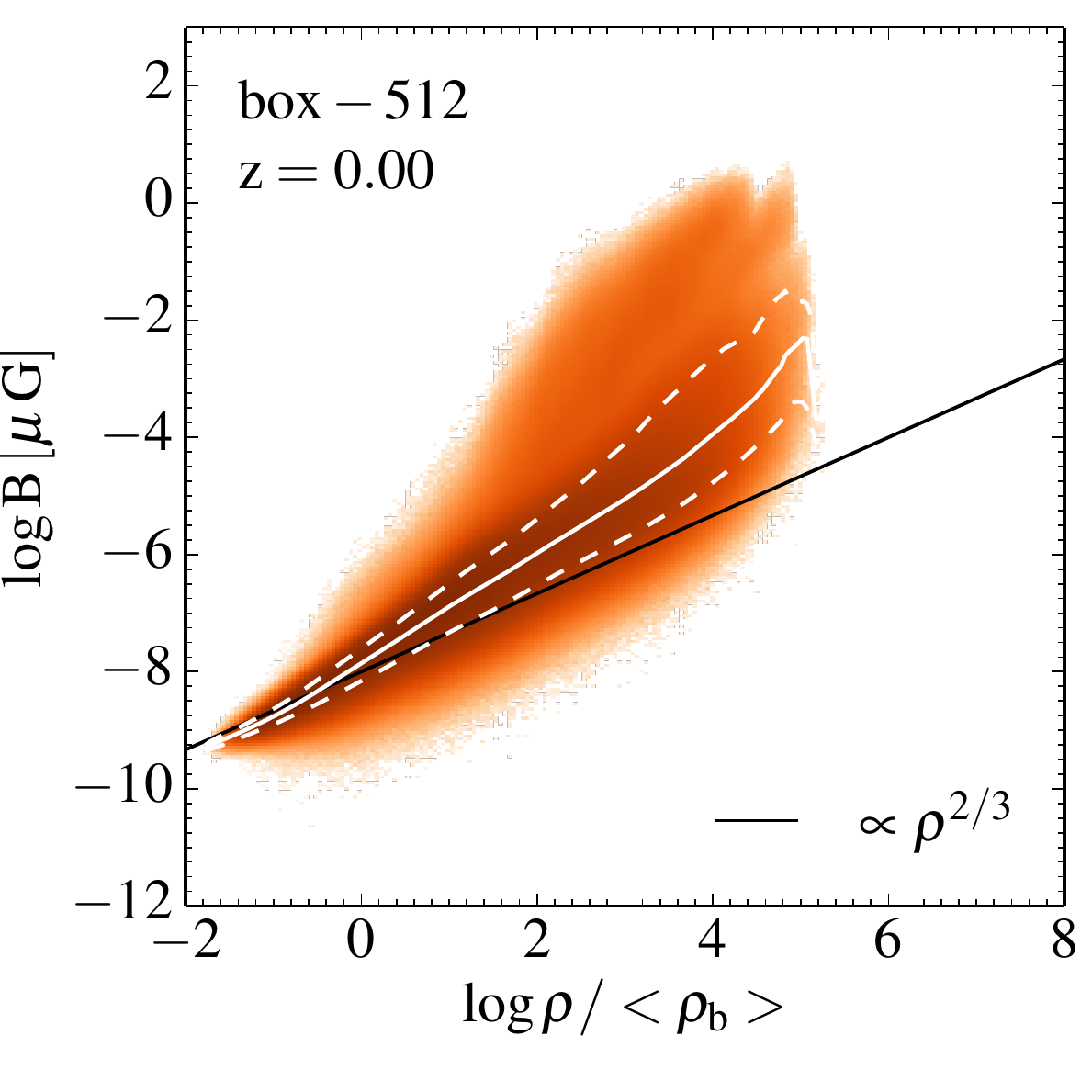}
\includegraphics[width=0.24\textwidth]{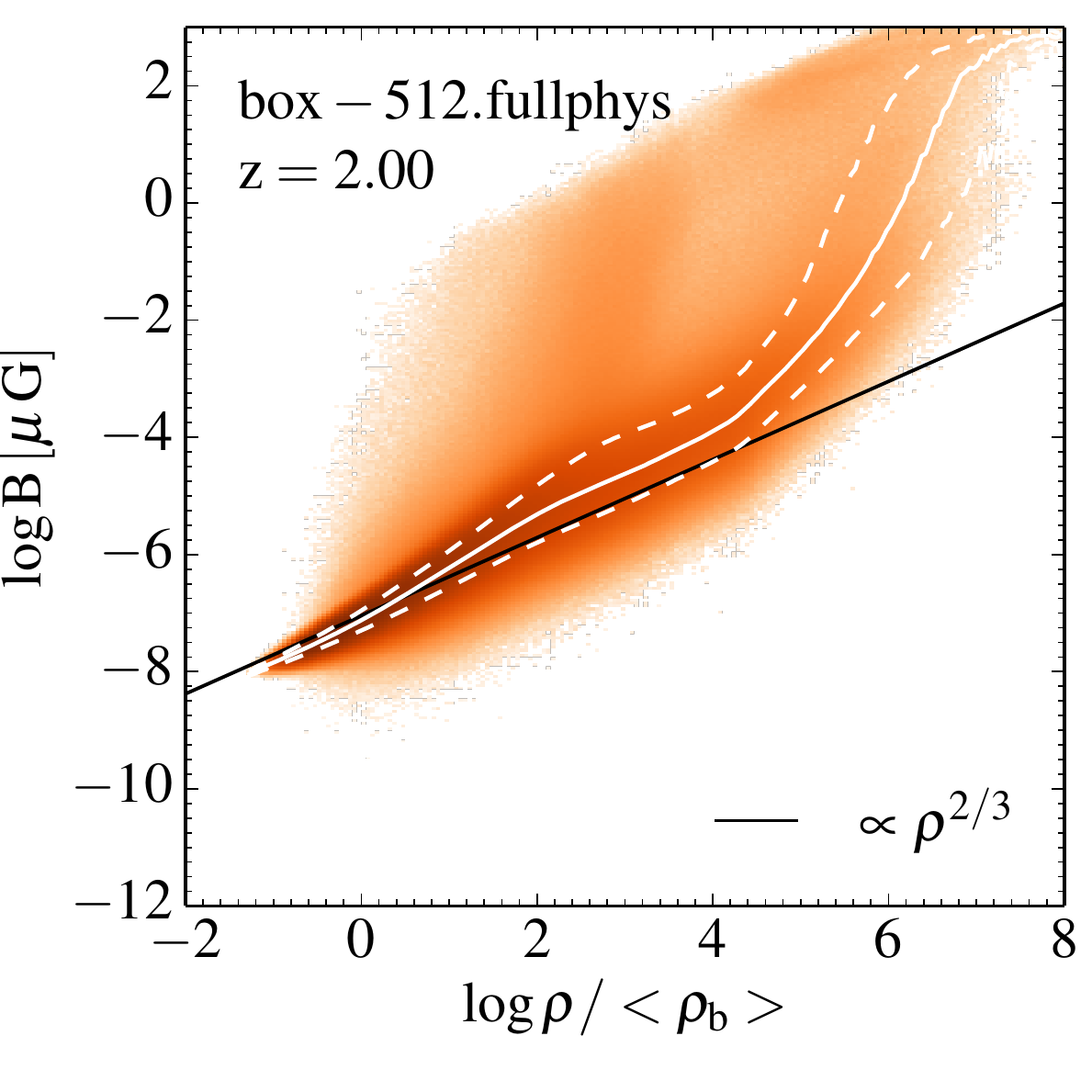}
\includegraphics[width=0.24\textwidth]{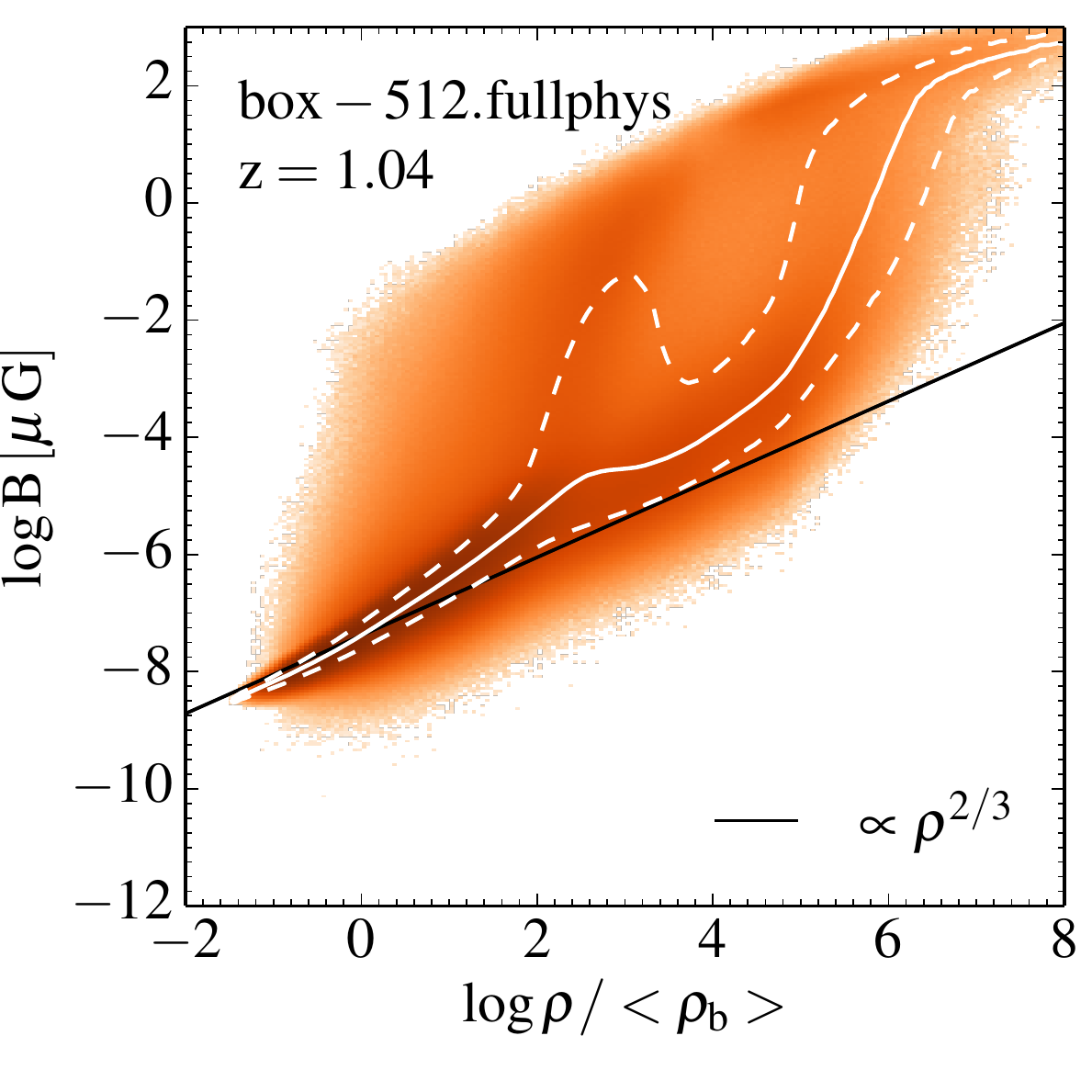}
\includegraphics[width=0.24\textwidth]{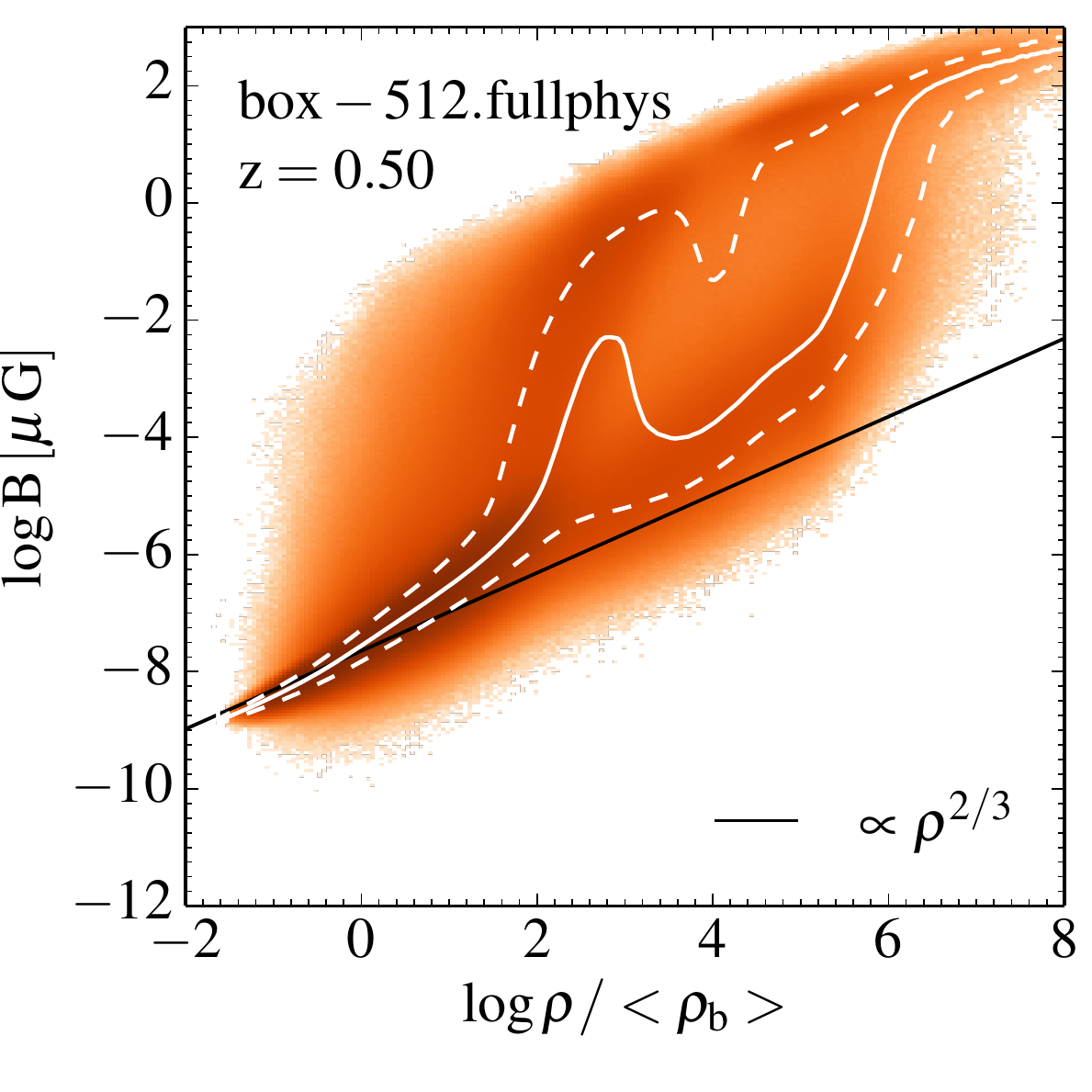}
\includegraphics[width=0.24\textwidth]{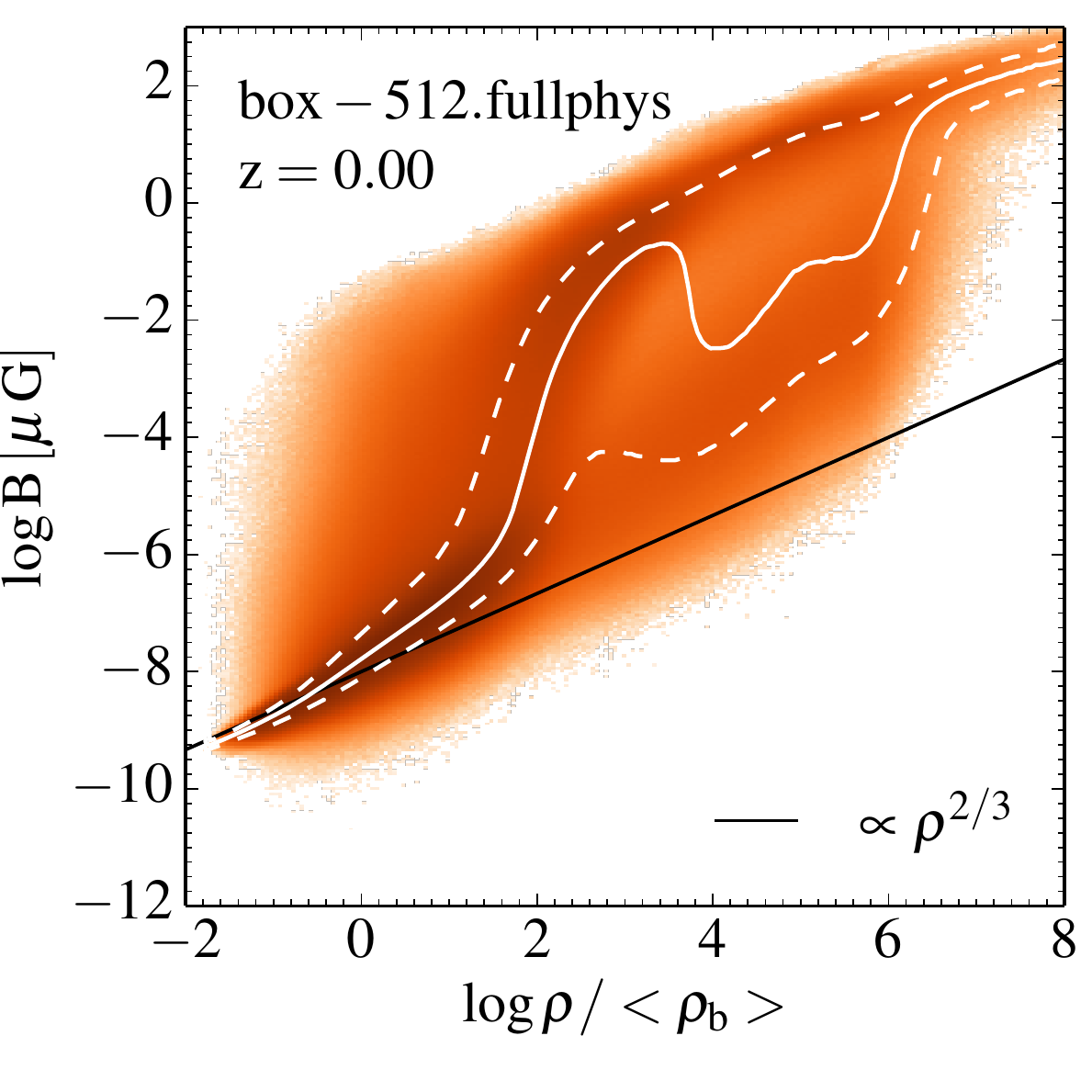}
\caption{Redshift evolution of the {\it B} field intensity versus baryon overdensity 
for the simulation box-512-ad (top) and box-512-fp (bottom). Redshift is 
indicated on the top left corner of each panel. The panels show two-dimensional 
histograms colour coded according to the mass of gas falling on to each bin 
(darker shades correspond to larger masses). White lines represent the median 
(solid) and the $16$th and $84$th percentiles (dashed) of the {\it B} field 
distribution as a function of the baryon overdensity. The black solid line is 
the expected density scaling of the {\it B} field intensity ($\propto \rho^{2/3}$) 
based on the flux conservation. For decreasing redshift is clearly visible how 
the assembling of cosmic structures (here represented by the increase of the 
dynamic range in overdensities with decreasing redshift) drives the 
amplification of the {\it B} field. At high redshift the median relation essentially 
corresponds to the scaling $\propto \rho^{2/3}$ expected from flux freezing, but 
it steepens at later time and towards higher overdensities indicating a boost in 
the {\it B} field amplification due to structure formation. In the full physics case 
the amplification is further enhanced by radiative cooling and stellar and AGN 
feedback. Also noticeable is the effect of the cosmological expansion as a 
reduction of the overall {\it B} field normalization $\propto (1+z)^{2}$ with 
decreasing redshift.} 
\label{fig:Bvsoverdensity}
\end{figure*}

The magnetic field in the full physics simulation can reach much higher values 
at the centres of massive haloes. Here the {\it B} field reaches intensities 
that are several orders of magnitude greater than in the adiabatic case (see
e.g. Fig.~\ref{fig:Bvsoverdensity}). The main cause for this is the inclusion of 
radiative cooling and feedback loops (galactic winds and AGN feedback) in the 
simulations. Radiative cooling allows gas to reach higher densities than in the 
adiabatic case, which in turn boost the amplification of the {\it B} field because of 
flux conservation. Also, the presence of gas outflows due to stellar and AGN 
feedback make the halo environment more `violent' increasing the importance of 
turbulence and shear flows, which again boost the amplification of the {\it B} 
field. The values of the {\it B} field between the two simulations tend to increase 
with time. At high redshift ($z = 2$) the projections look very similar, but as 
time goes on and the assembly of the cosmic structure progresses the differences 
become more marked. Although the morphology of low-density regions and in voids 
in particular is similar in both simulations, the {\it B} field strength is somewhat 
larger with the inclusion of baryon physics since stellar and AGN feedback can eject 
highly magnetized gas from the centres of haloes in the intergalactic space.

\begin{figure*}
\centering
\includegraphics[width=0.49\textwidth]{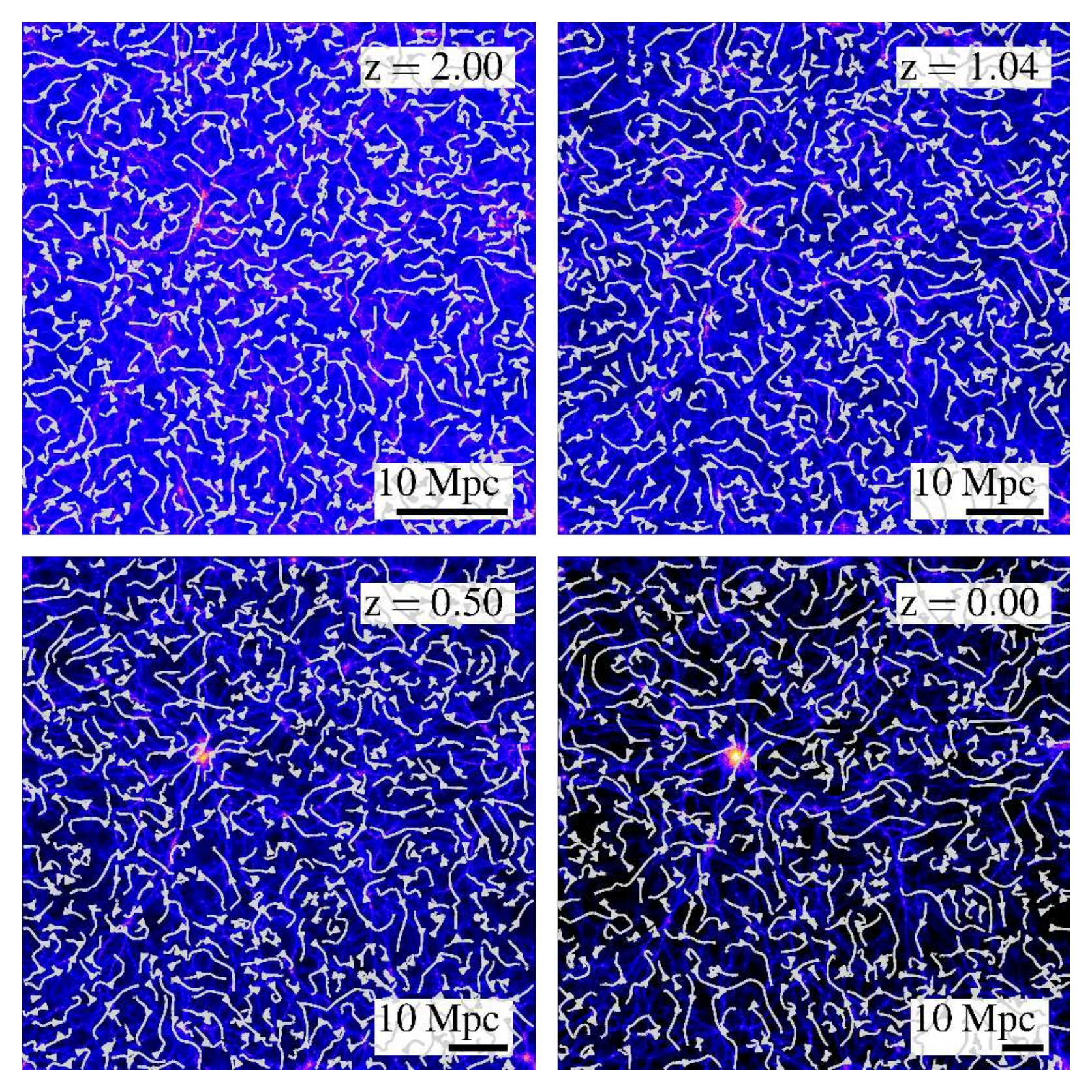}
\includegraphics[width=0.49\textwidth]{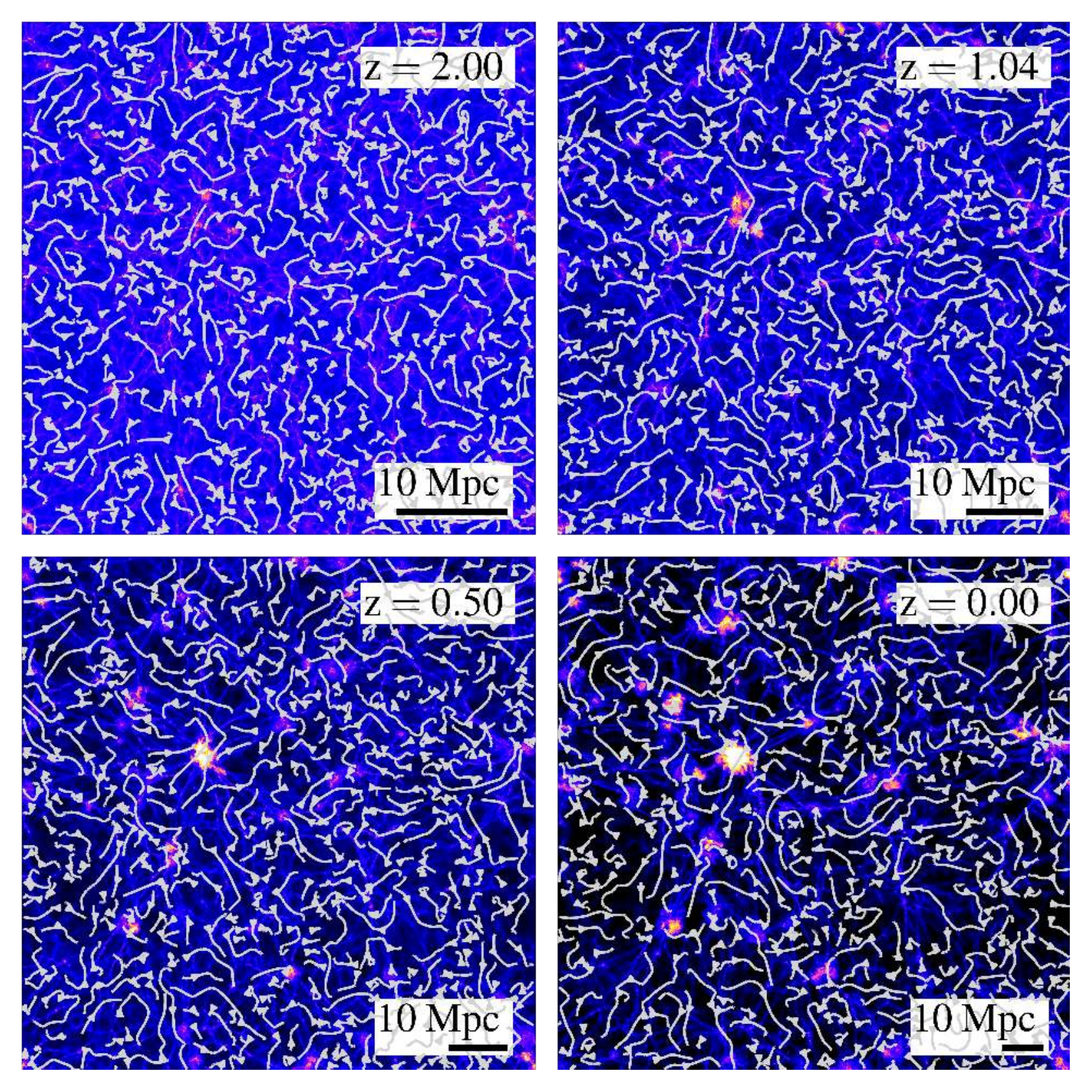}
\caption{Slice through the centre of the simulated box of the {\it B} field
intensity at different redshifts, as indicated in the top-right corner 
of each panel, for the simulation box-512-ad (left) and box-512-fp (right). 
Each panel is  $100\,\,h^{-1}{\rm Mpc}$ on a side (in comoving units), 
the full extent of the simulated box. The over plotted magnetic
field lines show the direction of the {\it B} field on to the slice plane. 
The colour scheme is the same for all the panels and maps logarithmically 
magnetic field intensities in the interval $[10^{-9}, 10^{-3}]\, \muG$.}
\label{fig:Bslice}
\end{figure*}

\begin{figure*}
\centering
\includegraphics[width=0.24\textwidth]{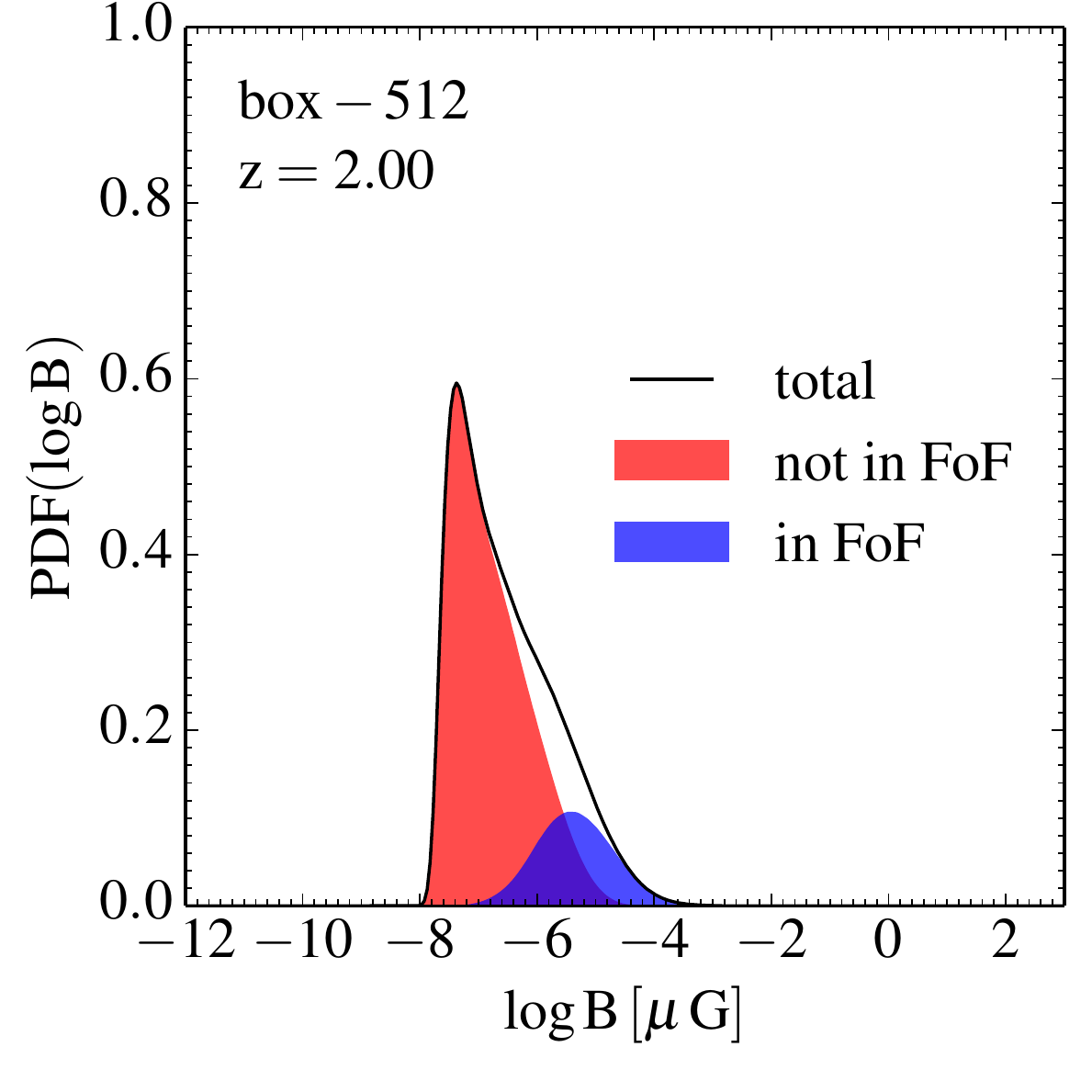}
\includegraphics[width=0.24\textwidth]{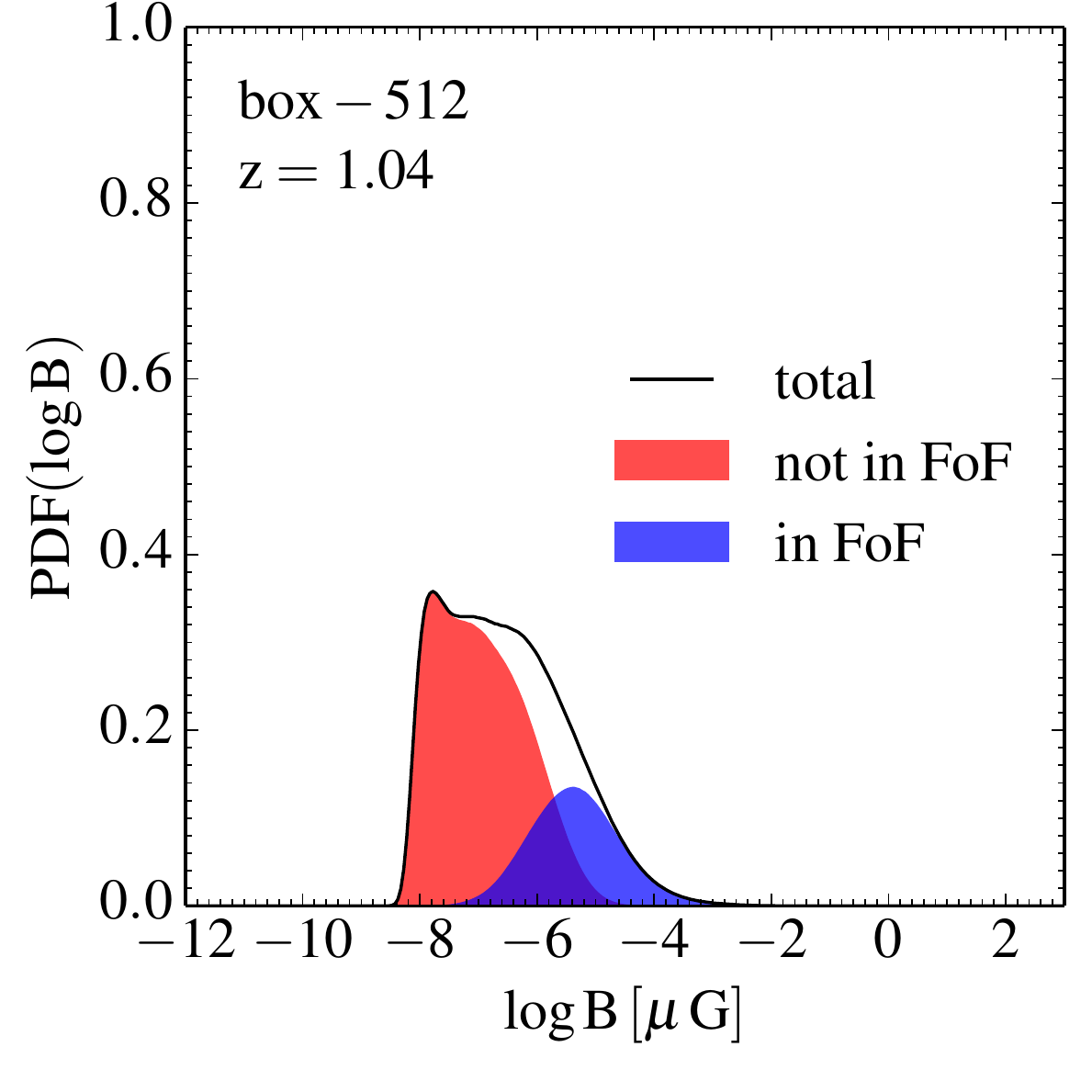}
\includegraphics[width=0.24\textwidth]{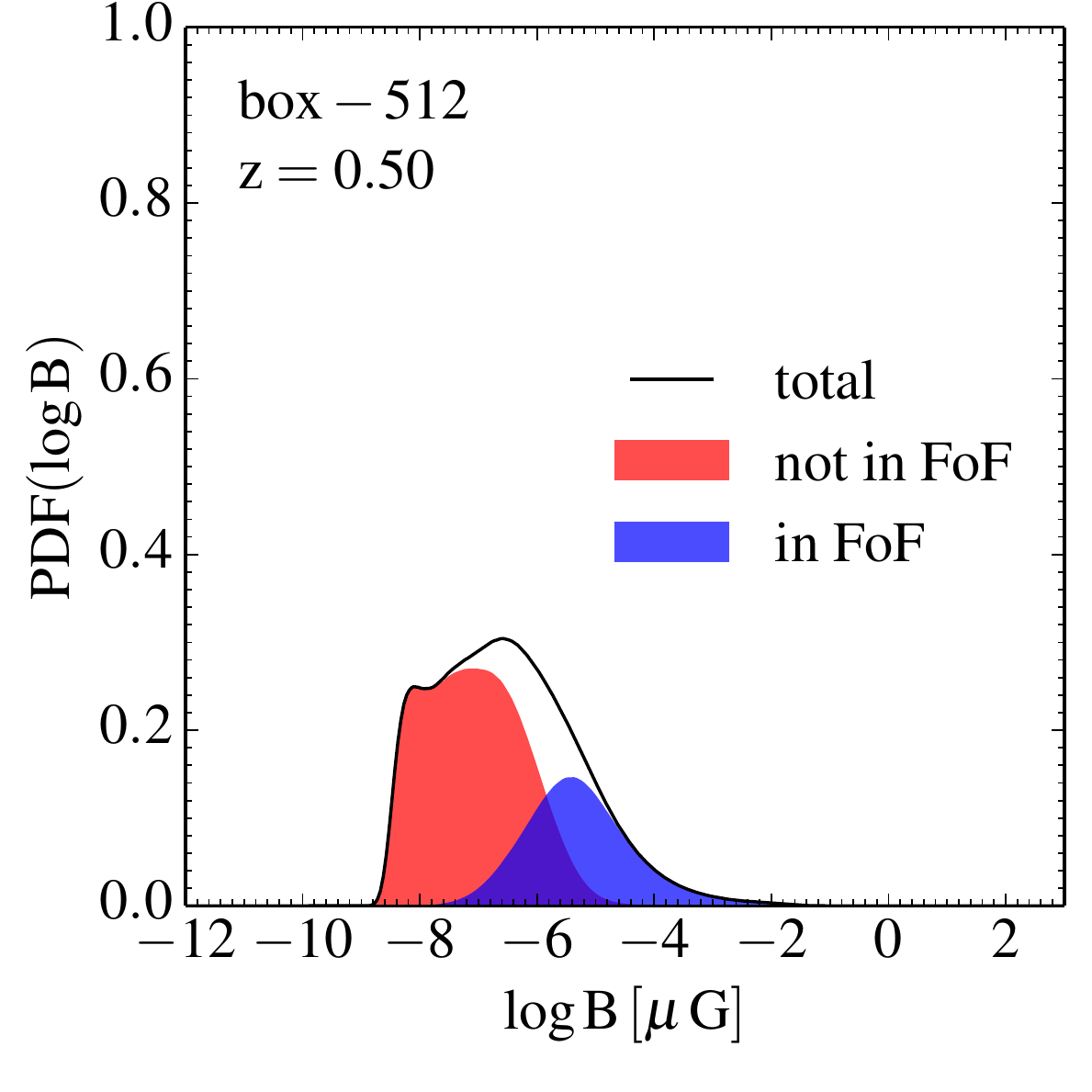}
\includegraphics[width=0.24\textwidth]{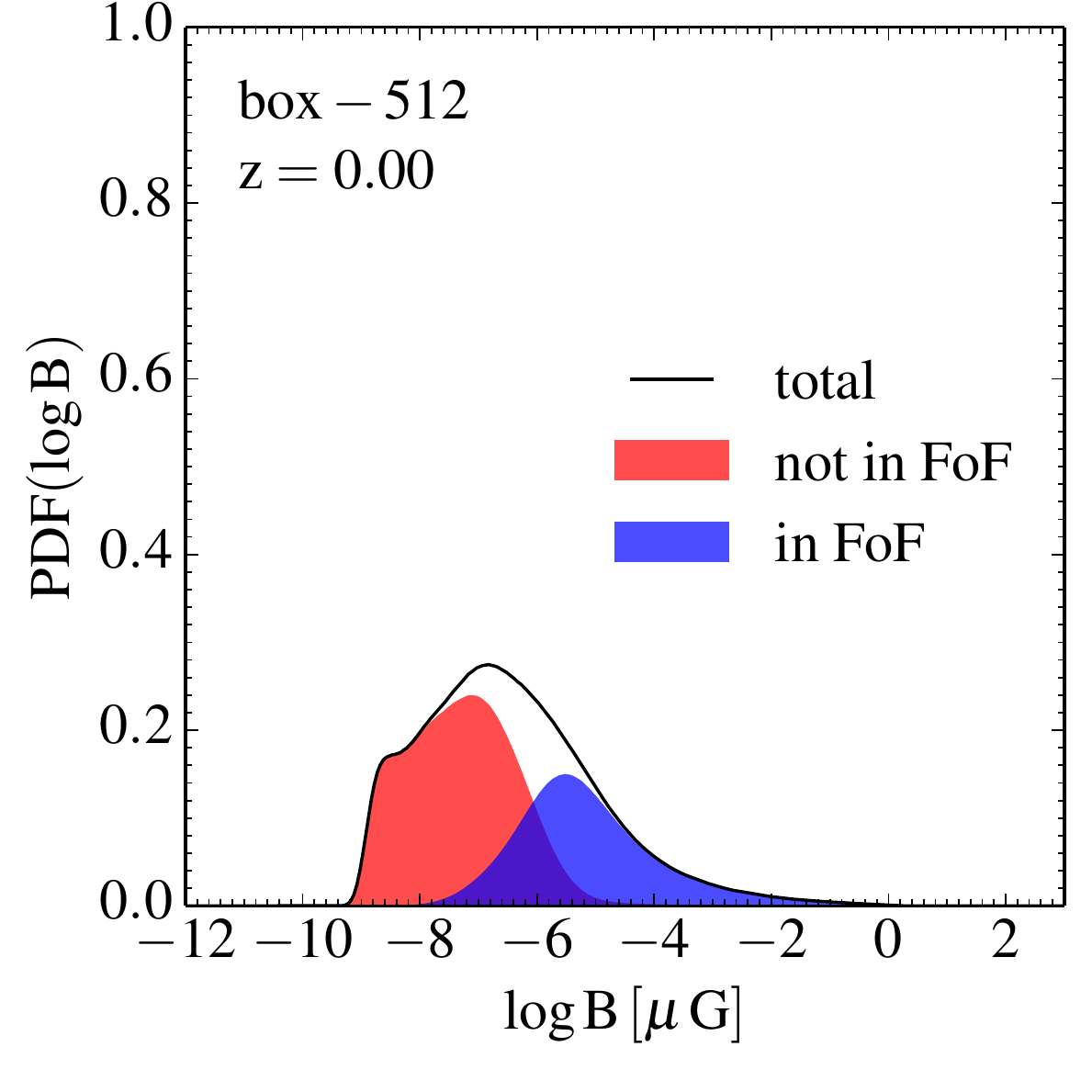}
\includegraphics[width=0.24\textwidth]{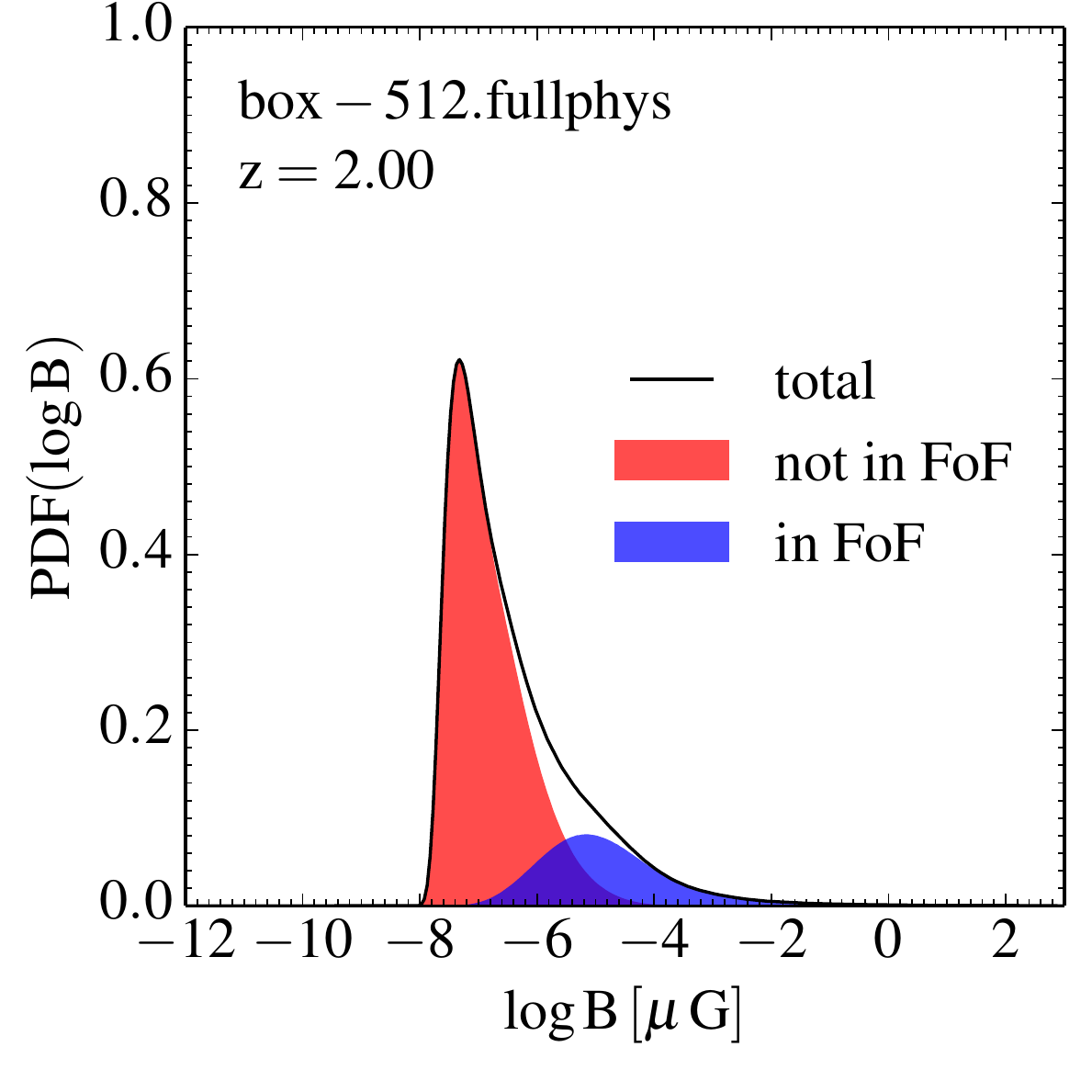}
\includegraphics[width=0.24\textwidth]{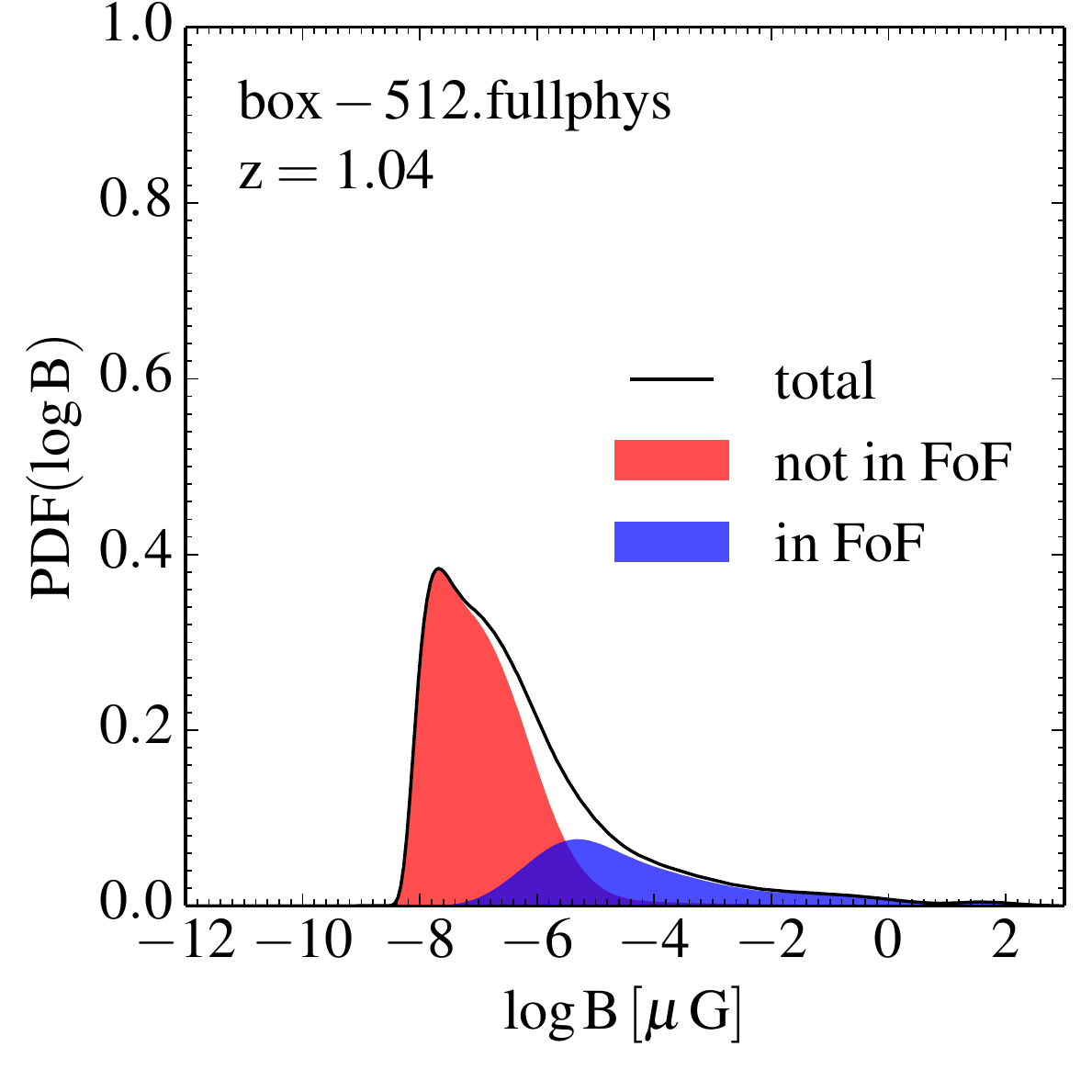}
\includegraphics[width=0.24\textwidth]{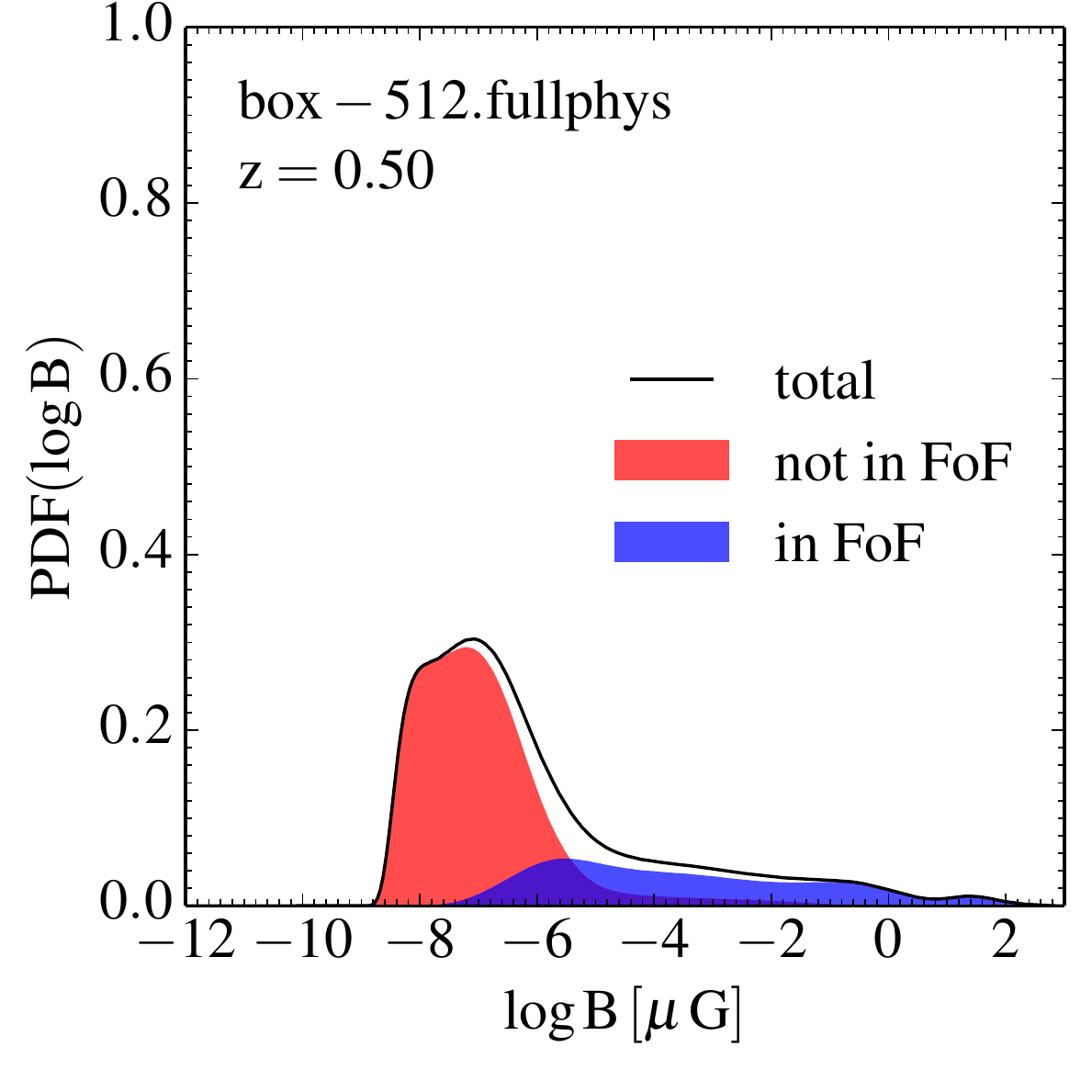}
\includegraphics[width=0.24\textwidth]{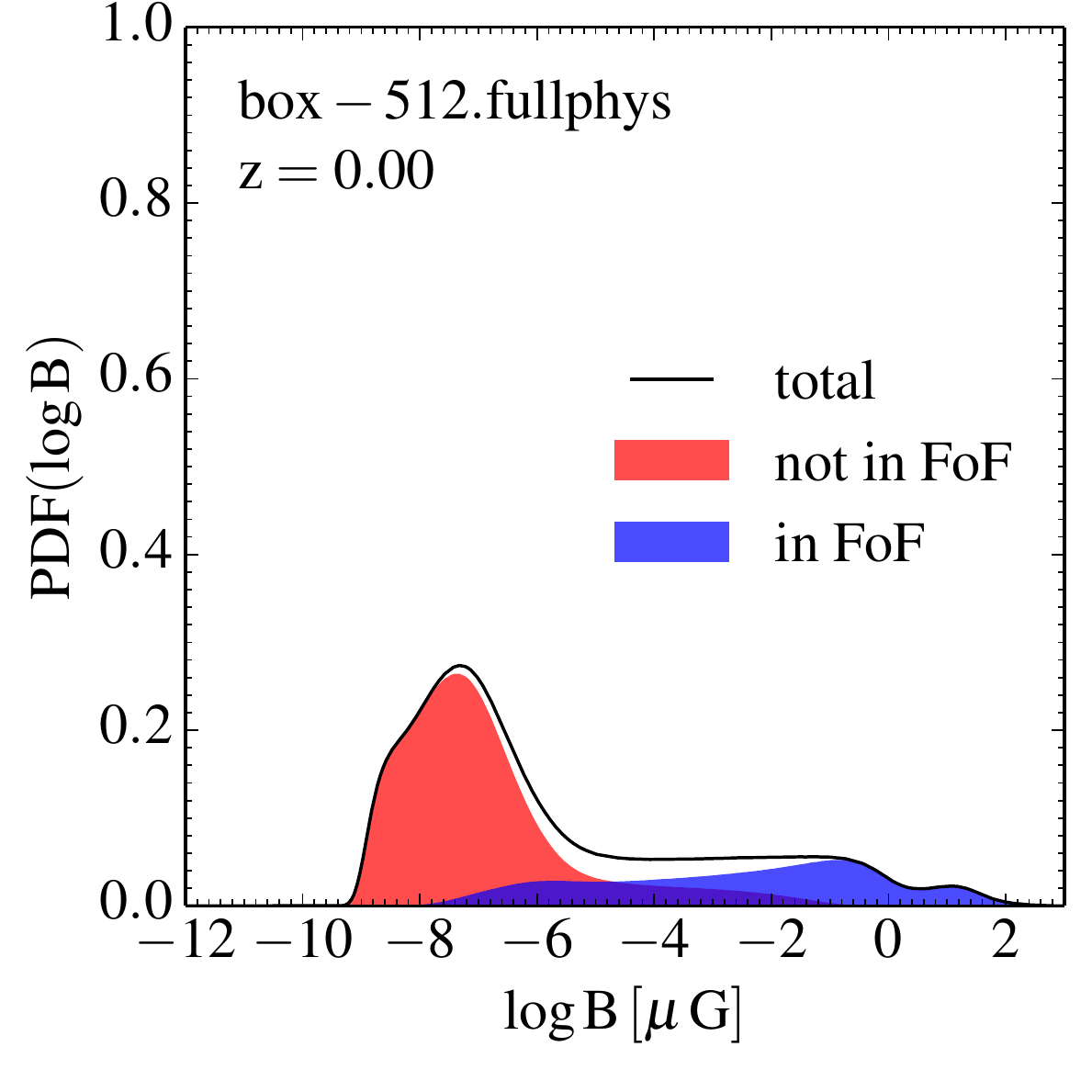}
\caption{Redshift evolution of the {\it B} field PDF for the simulation box-512-ad 
(top) and box-512-fp (bottom). Redshift is indicated on the top-left corner of 
each panel. The colour shading shows the contribution to the total PDF (black 
solid line) of gas cell in FOF-identified structures (blue) and not in FOF groups 
(red). The growth of cosmic structures can be clearly seen in the panels as the 
contribution to the {\it B} field PDF of gas cells not in FOF groups decreases in 
favour of that comprising FOF groups. The growth of cosmic structures -- 
together with baryon physics effects in the full physics run -- also drives the 
amplification of {\it B} field, visible as a more pronounced tail in the 
PDF at high {\it B} field intensity at late times. Moreover, the expansion of the 
Universe can be detected as a shift of the peak in the PDF of gas cells not 
contained in FOF groups towards lower values for decreasing redshift.} 
\label{fig:BfieldPDF}
\end{figure*}

Fig.~\ref{fig:Bslice} displays a two-dimensional slice through the centre of
the simulated domain in a direction perpendicular to the $z$-axis of the {\it B} field
intensity for the simulations box-512-ad (left) and box-512-fp (right). The
layout of the figure is the same as Fig~\ref{fig:Bprojection},
and the colour mapping has been kept the same for
both simulations. To give an idea of the orientation of the magnetic field on
to the slice plane we overplot magnetic field lines, with a small arrow
indicating the field direction.  At late times the largest {\it B} fields are found
in the largest structure that is visible in the slice. A mild magnetic field
strength enhancement is present also in the filaments that surround the main
halo, while in voids, due to the cosmological expansion, the physical {\it B} field
strength decreases at redshift zero to the value of the initial seed
field or below.

Except for the densest regions (i.e. the main halo forming in the slice) the
evolution of the magnetic field is essentially the same in the two simulations.
Even the direction of the field is very similar, as the overplotted field
lines show. This reinforces the idea that baryon physics has a fundamental role
for the amplification of the {\it B} field, but this role is limited to up to cluster
scales. On larger scales, gravitational dynamics and cosmological expansion
alone set the final field configuration.

In Fig.~\ref{fig:BfieldPDF} we present the redshift evolution of the magnetic
field probability density function (PDF) for the reference simulations
box-512-ad (top row) and box-512-fp (bottom row).  For each {\it B} field PDF the
relative contribution of gas cells contained (blue shading) and not contained
(red shading) in friends-of-friends (FOF) groups is shown. This division roughly separates scales
where baryonic effects are expected to have a strong impact on the {\it B} field
amplification from those where gravitational dynamics and cosmic expansion
dominate, instead.

From Fig.~\ref{fig:BfieldPDF} it is readily apparent that the formation and
growth of cosmic structures as a function of time lead to a decrease of the
peak of the PDF at low {\it B} field values. Almost all the contribution to this peak
comes from gas cells not contained in FOF groups. As time passes and more and
more cells are incorporated into gravitationally collapsed objects the height
of this feature decreases, partially compensated by an increase of the
contribution to the total PDF from particles included in FOF groups. However,
the peak at low values of {\it B} field never completely disappears, even at redshift
zero. The width of the {\it B} field PDF of gas cells not included in any structure
broadens with time. Here, two effects are at work: the cosmological
expansion, that tends to lower the {\it B} field and detectable as a movement of the
left edge of the PDF with decreasing redshift, and the amplification of the {\it B}
field in intermediate-density regions such as filaments, that tend to move gas
cells towards higher {\it B} field values. This behaviour is the same in both
simulations.  Indeed the shape of the non-FOF part of the {\it B} field PDF and its
evolution are remarkably similar in both runs.

The situation changes significantly for the {\it B} field PDF of gas cells contained
in FOF groups. The evolution of this part of the PDF is markedly different for
the two runs. There is a general trend of an increase of the relative
contribution of this part to the total PDF as a function of time for both
simulations, which can be readily explained by the growth of cosmic structures
driving the magnetic field amplification. Another common feature is the
formation of a peak at $\approx 10^{-6}\,\muG$. In the adiabatic case the
amplitude of this peak slowly but steadily increases with time, and the
contribution to the total PDF is symmetric with respect to the peak
location. The maximum magnetic field intensity, however, rarely exceeds
$\approx 10^{-2}\,\muG$. In the full physics simulation the formation of the
peak at $\approx 10^{-6}\,\muG$ is still visible, but its amplitude is lower
and the contribution to the total PDF more skewed towards higher values of {\it B}
field intensity at all redshifts. At redshift zero this feature disappears and
a bump forms at $\approx 1\,\muG$. The maximum value of the {\it B} field intensity
reached at the end of the simulation is $\approx10^2\,\muG$, about four orders
of magnitude larger than in the adiabatic case. This demonstrates that the full
spectrum of baryon physics is necessary to amplify primordial magnetic fields
to the values ($\sim 10-100\, \muG$) that are observed in galaxies and galaxy
clusters at low redshifts \citep[e.g.][and references therein]{Carilli2002,
Basu2013, Beck1996, Beck2013c, Feretti2012}.

In Fig.~\ref{fig:Bvsoverdensity} we present two-dimensional histograms of {\it B}
field intensity versus baryon overdensity ($\rho/\langle\rho_b\rangle$ where the latter
term is the mean baryon density) at different redshifts for the reference runs
box-512-ad (top row) and box-512-fp (bottom row). The colour shading in the
histogram represents the gas mass falling on to each bin. We also overplot the
trends of the $16$th, median and $84$th percentiles (white lines) plus the
scaling $\propto \rho^{2/3}$ (see also eq.~[\ref{eq:adexpansion}]) expected 
from magnetic flux conservation (black line). 

In the adiabatic case, a well-defined relation exists between the {\it B} field
strength and baryon overdensity. This relation is indicative of the fact that
magnetic field amplification and mass assembly in cosmic structures are tightly
linked. The degree of scatter increases with overdensity, as shown by the
diverging behaviour of the $16$th and $84$th percentile with respect to the
median relation, suggesting that at high overdensities some other mechanism
(i.e. turbulence and shear flows) other than gravitational collapse is at
work to boost the magnetic field strength.  This idea is further supported by
the steepening of the median relation with respect to the scaling expected by
magnetic flux conservation ($\propto \rho^{2/3}$) as a function of time. The
latter describes well the intensity of the {\it B} field at low overdensities, but
underpredicts its median strength in the most dense regions.

These general trends can also be observed in the full physics simulation.
However, when the histograms are compared, the difference in dynamic range in
both overdensity and magnetic field intensity with respect to the adiabatic
case is immediately noticeable. Due to the inclusion of radiative cooling, the
baryon overdensity can reach values as high as $10^8$, almost three orders of
magnitude larger than in the previous case. The effect on the magnetic field
strength is also dramatic: at $z = 2$ for overdensities $\gsim 10^3$ bins at {\it B}
field strength $\gsim 1\,\muG$ are already populated. Eventually, a second
branch in the relation at high {\it B} field forms, which is shifted from the
prediction of magnetic flux conservation by $\sim$ five to six orders of magnitude. The
larger overdensity range caused by gas cooling is not the only reason for such
a strong enhancement of the {\it B} field intensity with respect to the previous
simulation, as the upper branch of the histogram starts at overdensities of
$\sim 10^3$, well within the reach of the adiabatic run. Therefore, the
combination of radiative cooling and larger shear flows and turbulence
caused by galactic outflows (which are powered by stellar and AGN feedback) is
necessary to explain the high degree of amplification of the magnetic field in
the full physics case.

\begin{figure}
\centering
\includegraphics[width=0.45\textwidth, viewport= 0 0 350 300, clip=true]{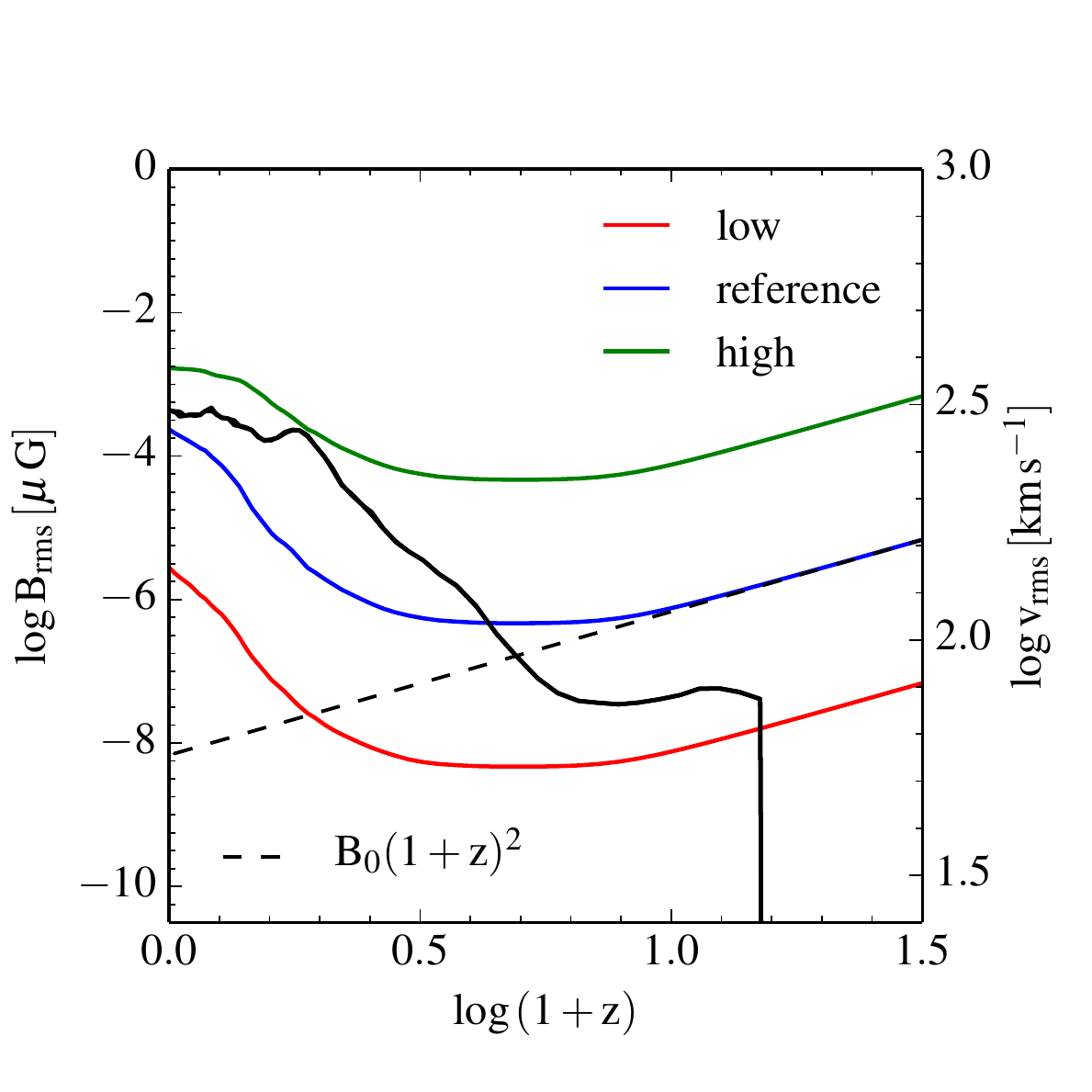}
\includegraphics[width=0.45\textwidth,viewport= 0 0 350 300, clip=true]{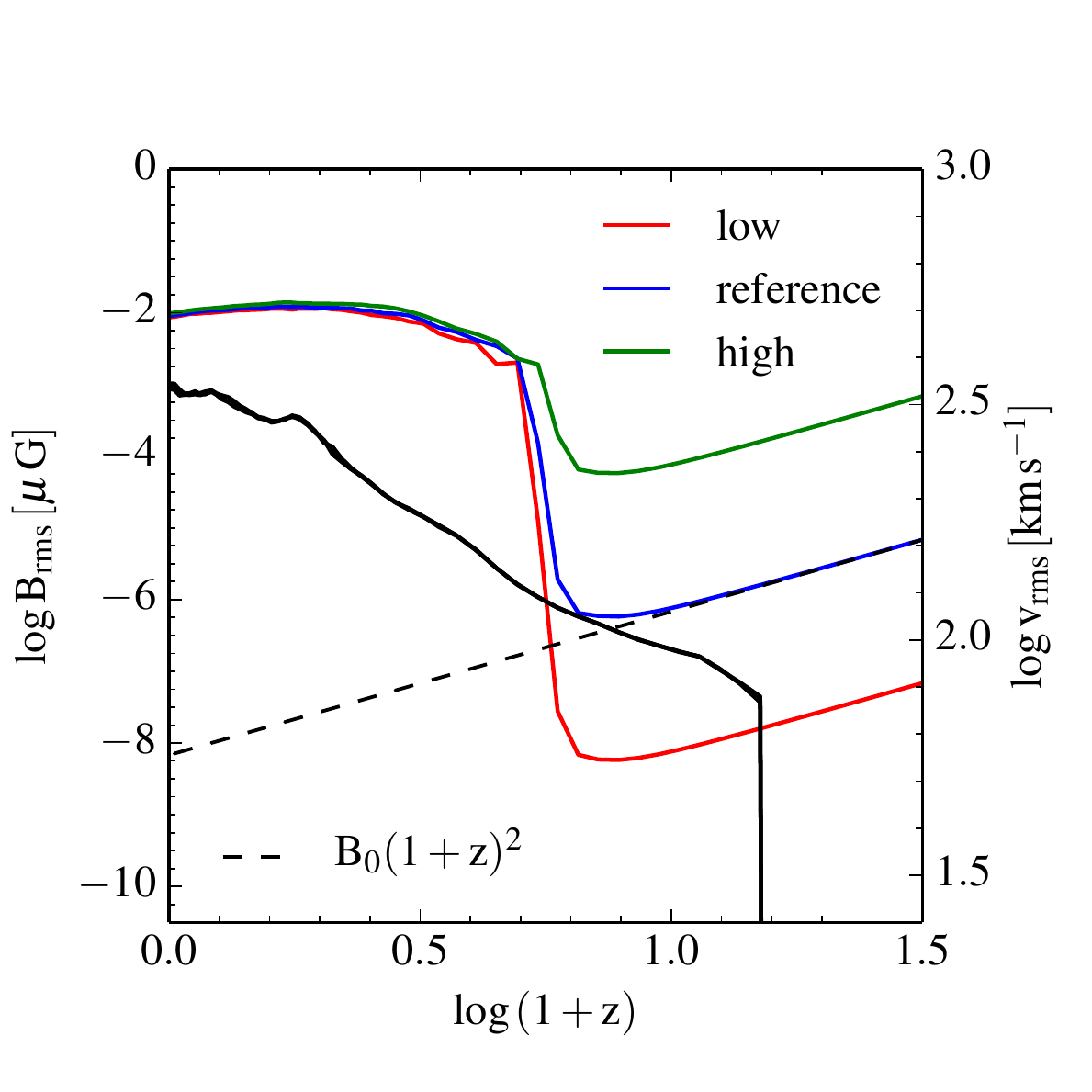}
\caption{Evolution of \rev{the rms velocity (black solid line) and} the 
volume-weighted {\it B} field rms \rev{(coloured lines)} as a function of redshift in 
the box-512-ad (top) and box-512-fp (bottom) simulations. Both calculations are 
repeated with different initial seed field strengths, as indicated in the 
legend. The black dashed lines show the {\it B} field evolution due to the Universe 
expansion under the assumption of (global) magnetic flux conservation.} 
\label{fig:Bzevolution}
\end{figure}

\subsection{Dependence on the strength of the seed field}\label{sec:seed field}
The 
seed field in our simulations can arbitrarily be chosen and there are different seeding 
strategies that have been adopted in numerical simulations, also depending on the 
scales of the simulated problem \citep[see e.g.][]{Dolag1999, Donnert2009, 
Dubois2010, Hanasz2009, Kotarba2009}. The only stringent constraint is that 
the seeding strategy must yield a divergence-free initial field. In our runs we 
adopted the simplest approach: a uniform field throughout the simulation box. 
This approach leaves us the freedom to select the intensity of the initial seed 
field and its direction as well. The present section is devoted to study how 
the properties presented in Sect.~\ref{sec:large-scale} vary as a function of 
the initial field intensity. We find that the orientation of the seed field has 
a smaller impact on the final simulation results, and we defer the discussion 
of this aspect to Sect.~\ref{sec:discussion}.

In Fig.~\ref{fig:Bzevolution}, we show the evolution of the root mean square 
volume-weighted {\it B} field -- a measure of the total magnetic energy -- in the 
whole simulation box as a function of redshift for the reference (i.e. 
$2\times512^3$) adiabatic simulation (top panel) and the its full physics 
counterpart (bottom panel). To investigate the effects of the seed field 
strength, we repeat each calculation three times with different intensities 
of the initial seed field. In our fiducial setting (blue lines) the initial seed 
field strength is $||\boldsymbol{B}_{0}|| = 10^{-14}\,\,{\rm G}$. We 
then explore two additional configurations with $||\boldsymbol{B}_{0}|| = 
10^{-16}\,\,{\rm G}$ (red lines) and $||\boldsymbol{B}_{0}|| = 10^{-12}\,\,{\rm G}$ (green 
lines), for a total variation in the initial magnetic field intensity of four 
orders of magnitude. The black dashed line in each panel is a guiding line that 
shows the evolution of the {\it B} field intensity (in the case of our fiducial set 
up) only due to the expansion of the Universe under the assumption of (global) 
magnetic flux conservation.

\begin{figure*}
\centering
\includegraphics[width=0.3\textwidth]{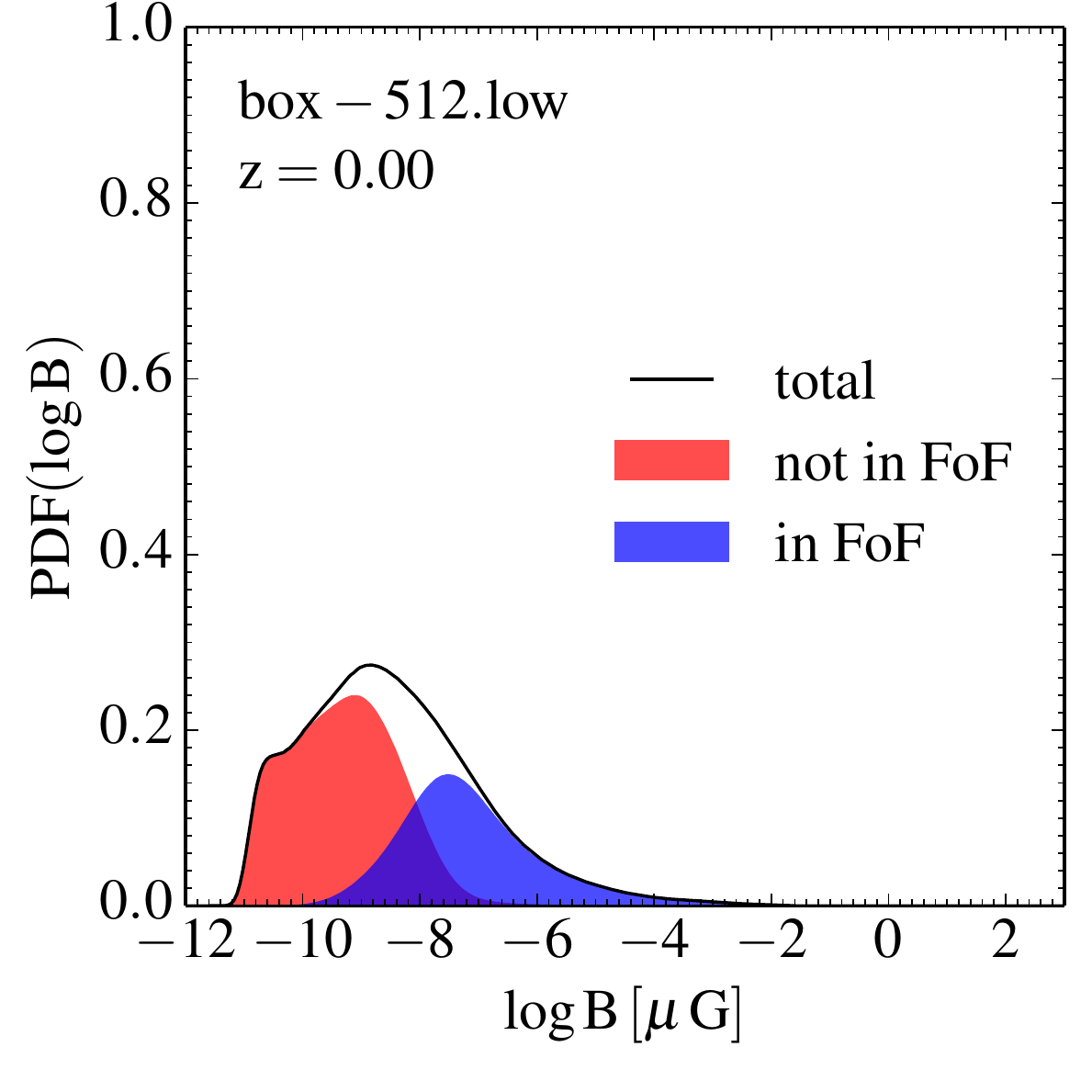}
\includegraphics[width=0.3\textwidth]{fig6b.pdf}
\includegraphics[width=0.3\textwidth]{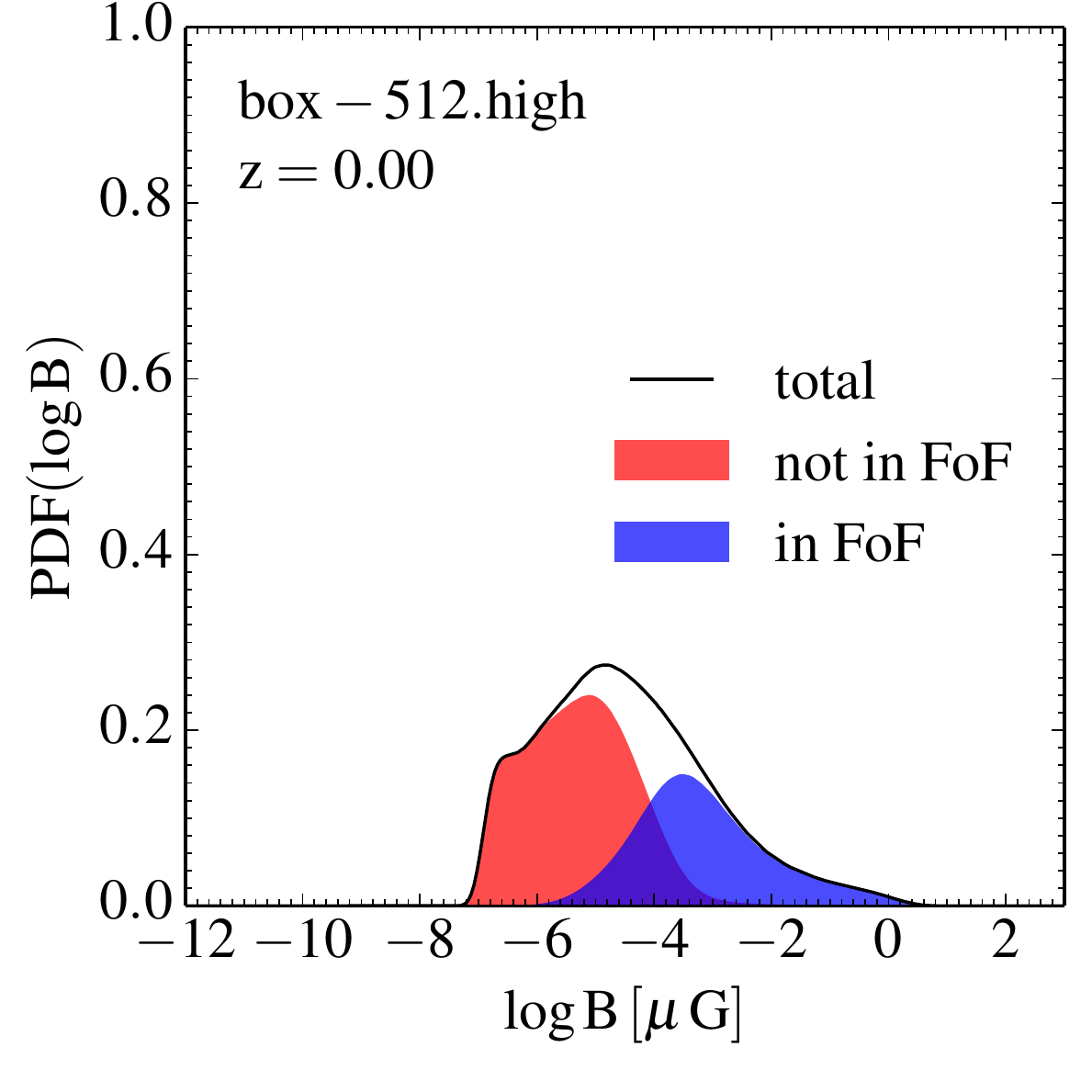}
\includegraphics[width=0.3\textwidth]{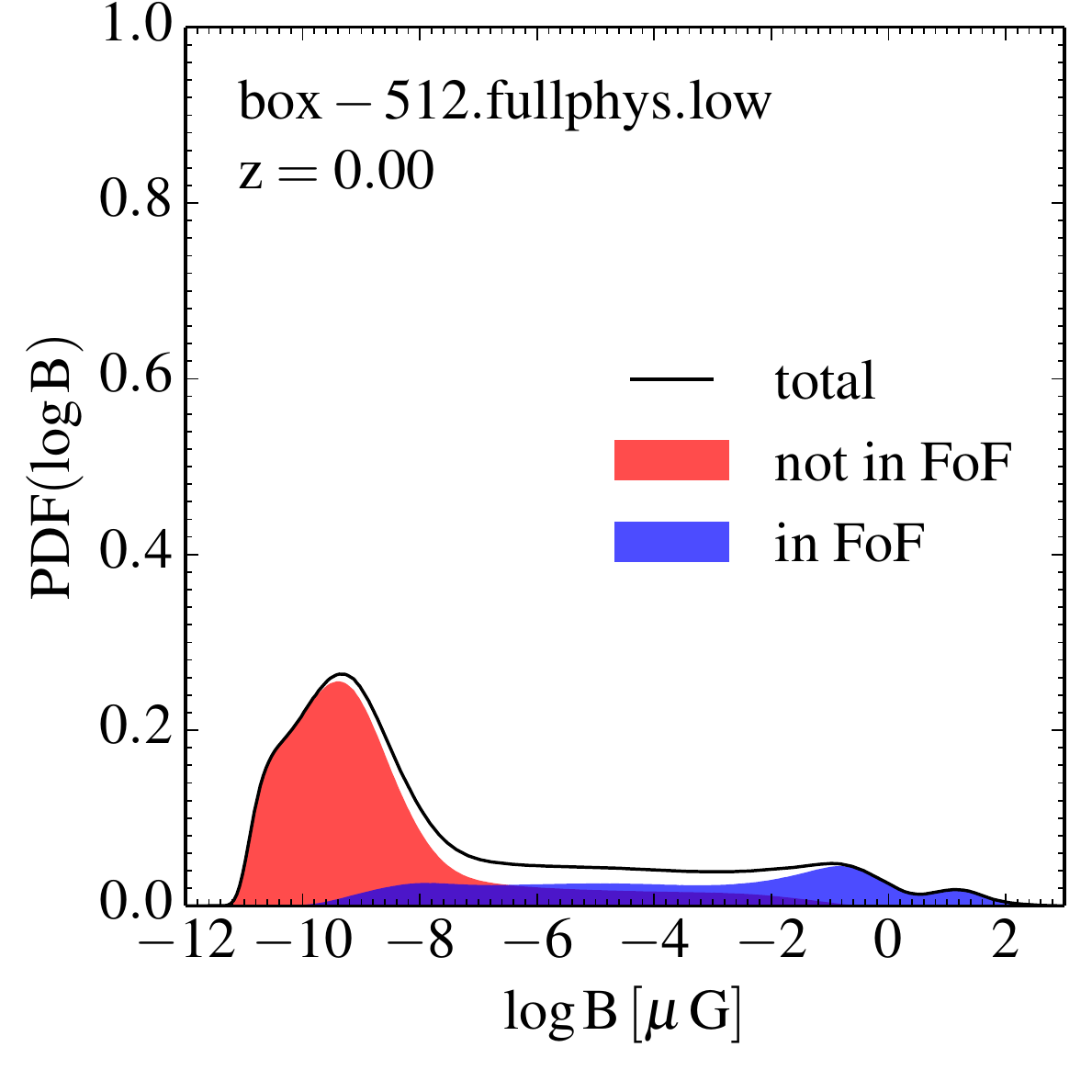}
\includegraphics[width=0.3\textwidth]{fig6e.pdf}
\includegraphics[width=0.3\textwidth]{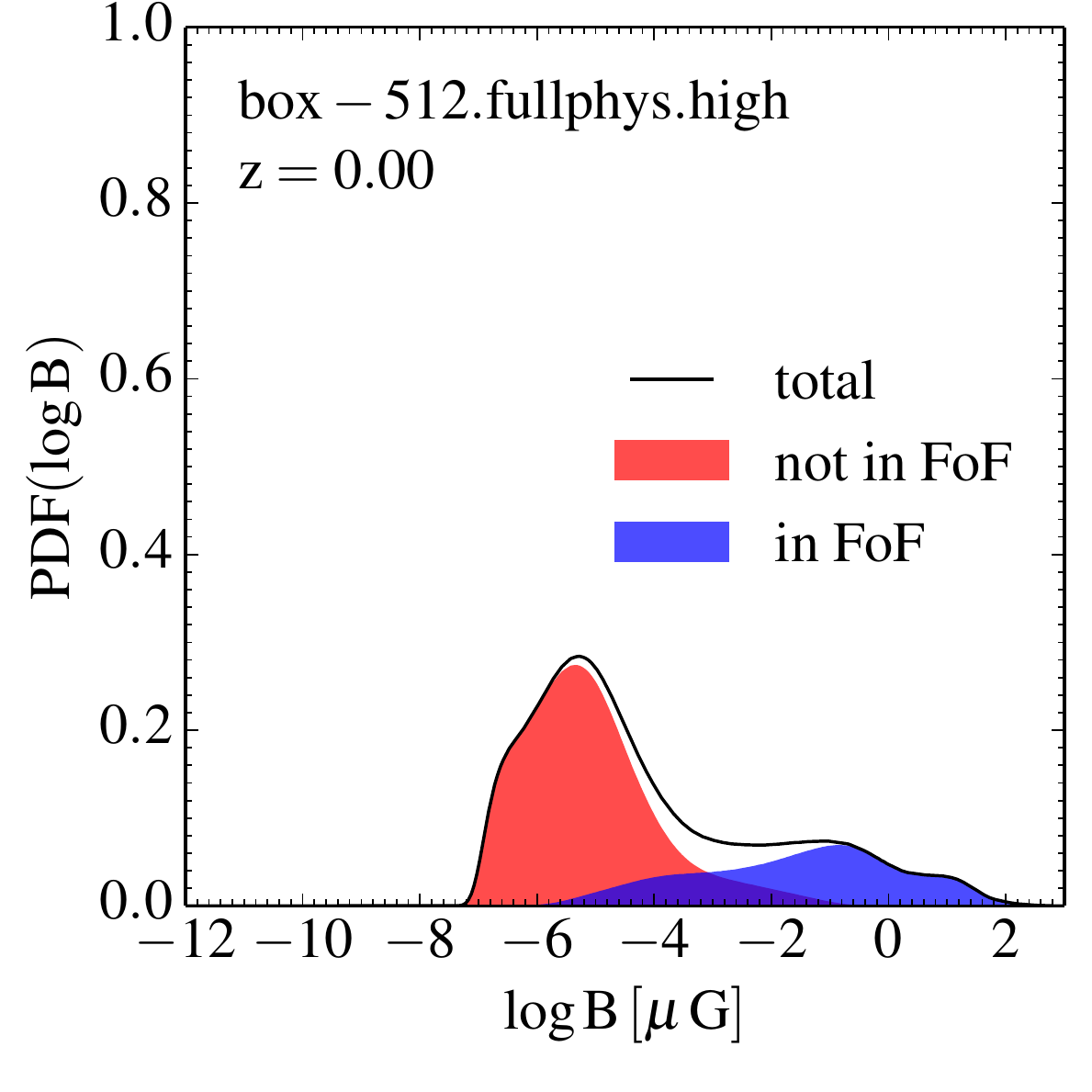}
\caption{{\it B} field PDF at redshift zero for the simulations box-512-ad (top row) and 
box-512-fp (bottom row) with varying initial {\it B} seed field strength. The initial field 
strength is $10^{-16}$, $10^{-14}$ and  $10^{-12} {\rm G}$  
from left to right. The panels show the relative contribution to the total PDF 
(black line) of gas cells not contained in any of the FOF groups (red shading)
and included in FOF structures (blue shading). 
The {\it B} field PDF is essentially unaffected by a change in the initial value of 
the seed field, except for a global rescaling of the {\it B} field strength visible as 
a horizontal shift of the plots in each row.} 
\label{fig:BfieldPDFseed}
\end{figure*}

\begin{figure*}
\centering
\includegraphics[width=0.32\textwidth]{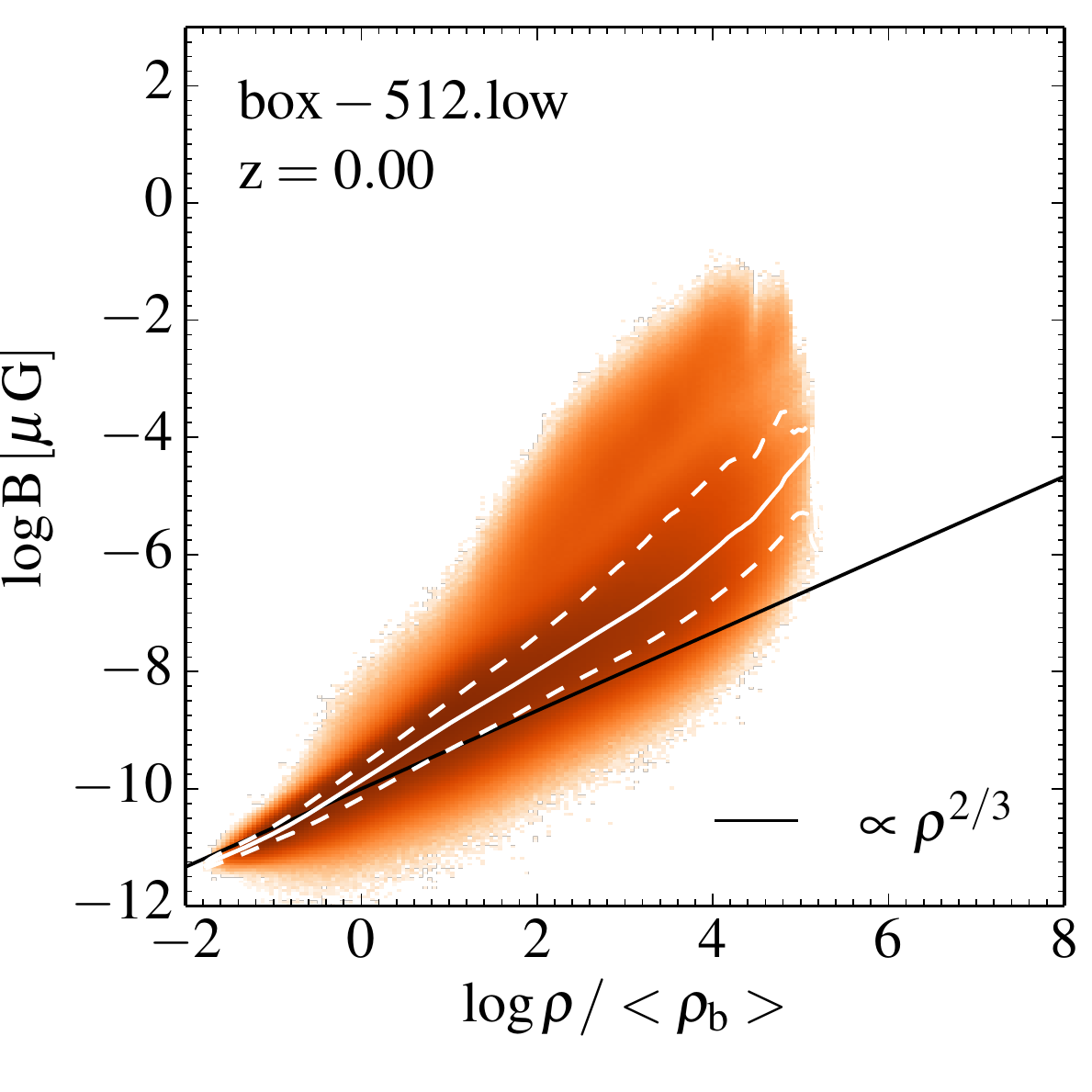}
\includegraphics[width=0.32\textwidth]{fig7b.pdf}
\includegraphics[width=0.32\textwidth]{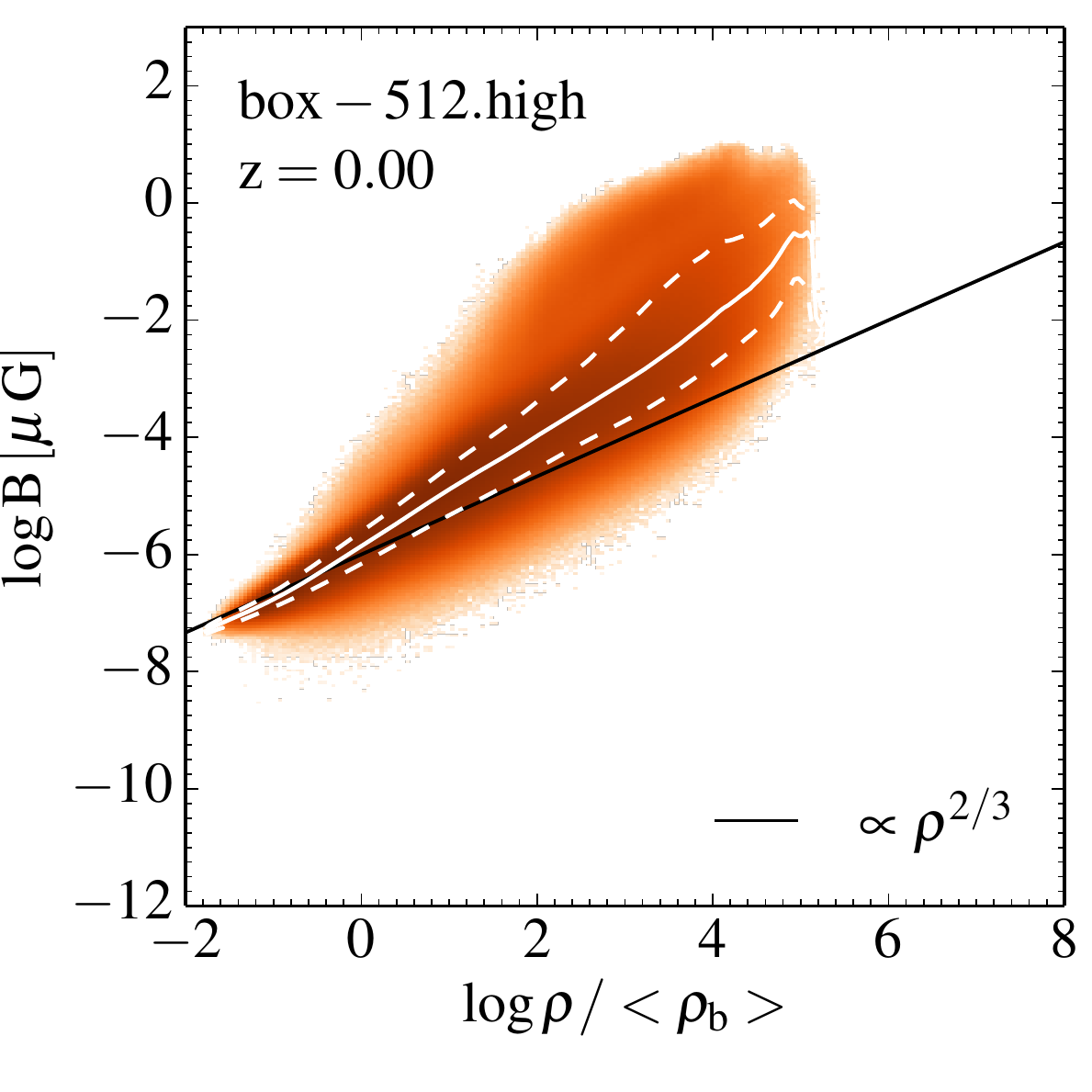}
\includegraphics[width=0.32\textwidth]{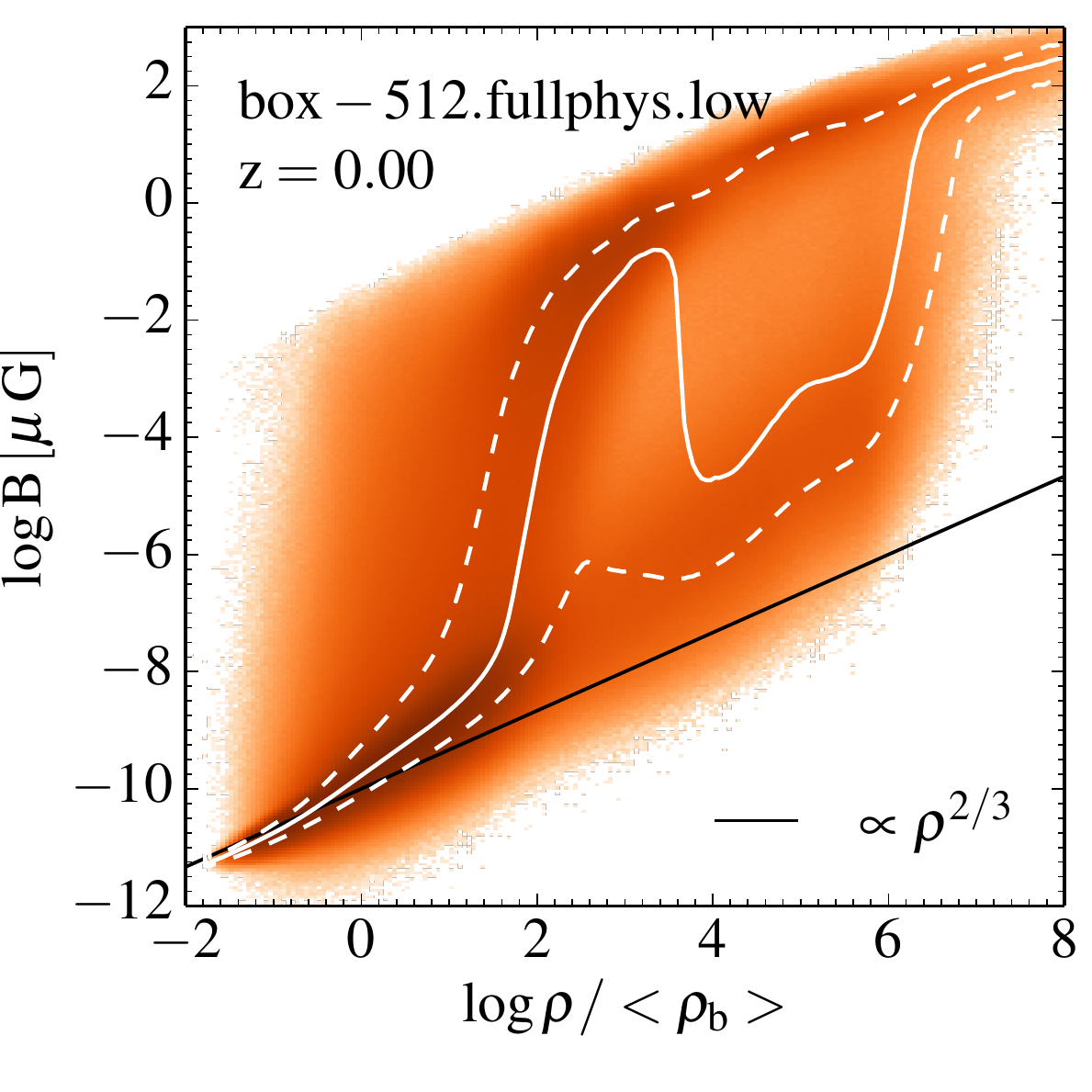}
\includegraphics[width=0.32\textwidth]{fig7e.pdf}
\includegraphics[width=0.32\textwidth]{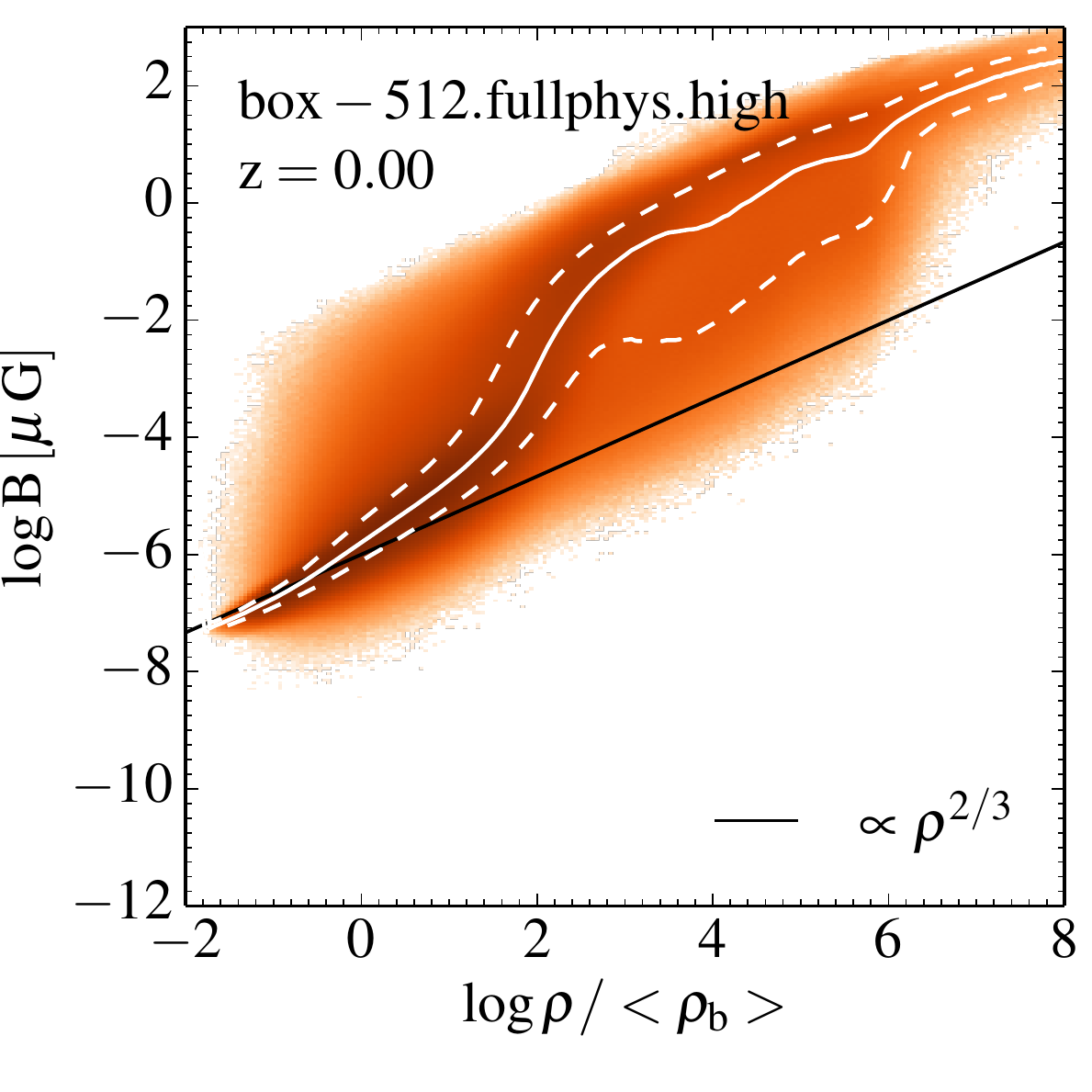}
\caption{{\it B} field versus baryon overdensity at redshift zero for the simulations 
box-512-ad (top row) and box-512-fp (bottom row) as a function of the initial {\it B} 
seed field strength. The initial field strength is $10^{-16}$, $10^{-14}$ 
and  $10^{-12} {\rm G}$ from left to right. The meaning of the 
colour scale and of the lines in each panel is the same of 
Fig.~\ref{fig:Bvsoverdensity}. For the adiabatic run there is essentially no 
change in the relation except a global rescaling of the {\it B} field strength due to 
the variation of intensity of seed field visible as a global vertical shift of 
the plot in the three panels. The same scaling is present also in the full physics run
for low magnetic field intensities. However, the region of the plot at higher {\it B}
field values is unaffected by the choice of the seed field, signalling that in those
regions saturation is reached.
} 
\label{fig:Bvsoverdensityseed}
\end{figure*}

The {\it B} field evolution at high redshift is very similar in both the adiabatic
and the full physics case, and is largely dominated by the expansion of the
Universe, with the different intensities of the seed field accounting for an
overall normalization factor. At about $z \approx 3$ structure formation leads
to deviations from this simple trend, and a clear upturn of the field intensity
can be observed. Turbulence and shear flows induced by structure formation
are the primary mechanisms by which the field is amplified then. The late time
evolution, however, is markedly different. The major difference is of course
the final level of amplification reached in the two cases, with the full
physics simulations reaching a final value of the {\it B} field strength that is at
least $\sim10$ times larger than in the adiabatic case. The other striking
difference is that in the full physics runs the final amplitude of the {\it B} field
($\sim 10^{-2}\, \muG$) is reached \textit{regardless} of the initial seed
field intensity (i.e. the amplification process \textit{saturates}
\footnote{\rev{In this work with saturation we indicate this independence of the final
field strength from the initial value of the seed
field intensity, or that at a fixed overdensity the magnetic field is amplified
to a maximum value that depends only on the overdensity but not on 
the seed field strength (see also Figs~\ref{fig:BfieldPDFseed} and \ref{fig:Bvsoverdensityseed}).}} after
$z\approx 2$ ), while in the adiabatic simulation this is not the case and the
evolution curves are merely a rescaled version (with the initial seed field
giving again an overall normalization factor) of one another. The only effect
of changing the initial seed field intensity in the full physics run is a
\rev{minimal} delay (for decreasing seed fields) of the time at which the exponential
amplification from the (global) flux-conserving evolutionary phase to the final
{\it B} field values saturates \citep[see also][]{Pakmor2014}. \rev{A fit to the exponential
amplification phase results into an e-folding time 
for $B_{\rm rms}$ of $\simeq 0.1$, $0.14$, $0.23\,\Gyr$ from the lowest to the
highest seed field strength.}

\rev{The black solid lines in both panels display the redshift evolution of the 
gas rms velocity $v_{\rm rms}$ for all cells in FOF groups. The bulk velocity of 
the FOF groups has been subtracted from gas particles before computing $v_{\rm 
rms}$, so that this quantity can be thought as a rough measure of gas turbulent 
motions. For each panel we plot the evolutionary tracks of $v_{\rm rms}$ for 
different seed fields, which however are practically indistinguishable from one 
another. This agrees with the findings of Sect.~\ref{sec:haloes} in which we show 
that magnetic fields are not dynamically important even within haloes where 
their strength is the largest. $v_{\rm rms}$ increases with decreasing 
redshift in both adiabatic and full-physics simulations. The $v_{\rm rms}$ 
values reached in the full-physics runs are larger than in their adiabatic 
counterparts at all redshifts. This is consistent with the picture discussed in 
the previous section in which feedback-induced galactic outflows are an 
important source of shear and turbulent gas motions that in turn promote 
magnetic field amplification.}

In Fig.~\ref{fig:BfieldPDFseed}, we show how the redshift zero {\it B} field PDF is
affected by the choice of the initial field strength for the two reference
simulations box-512-ad (top row) and box-512-fp (bottom row). In each row
the seed field initial strength increases from left to right. The meaning of
the shaded areas is the same as in Fig.~\ref{fig:BfieldPDF}.

In the adiabatic run, it is readily apparent that the shape of the {\it B} field PDF
and those of the contributions of gas cells in structures and outside them is
essentially unchanged for the three different choices of the seed field.  Only
the PDF position along the $x$-axis varies, and in particular it shifts by two
orders of magnitude in each panel. This is the exact value by which the seed
field is varied in each configuration. Thus, we conclude again that the initial
intensity of the seed field in the adiabatic run just sets the overall
normalization of the final magnetic field. From this follows that the {\it B} field
in the adiabatic runs is on global scales dynamically unimportant -- at least
for the intensity range explored here. As far as the intensity of the {\it B} field
is concerned, the amplification factor due to gas and gravitational dynamics is
not affected by the initial field strength: the final maximum intensity values
are different but their \textit{ratio} with the seed field strength is the
same.

The behaviour of the {\it B} field PDF in the full physics run is slightly more
complex. Here, in fact, there is a difference between low-value and high-value
{\it B} field parts. The low {\it B} field part, mostly comprised by gas cells not part of
any FOF group, is almost identical to its adiabatic counterpart and exhibits
the same shift along the $x$-axis as a function of the seed field strength. The
high {\it B} field part has a different behaviour. First of all, the largest
values that the {\it B} field can reach are almost independent of the initial seed
field strength, meaning that the amplification process has reached saturation.
Note that this result agrees with what found by \citet{Pakmor2014}, who
explored an even larger range of seed fields in `zoom-in' \mhd\ simulations
of disc galaxy formation with \arepo. Since the largest {\it B} values are fixed as a
function of the seed field, the contribution of particles in FOF structures to
the total {\it B} field PDF has a different shape in three panels, with a shorter
tail towards high {\it B} values for increasing seed field strengths.  Summarizing,
also in the full physics case the lower intensities of the {\it B} field are set by
the initial seed strength. However, due to the inclusion of baryon physics, the
amplification of magnetic field within cosmic structures -- in particular at
halo centres -- reaches saturation and the memory of the initial seed field is
lost completely.

Fig.~\ref{fig:Bvsoverdensityseed} confirms the general trends discussed above.
The figure shows two-dimensional histograms of of {\it B} field intensity versus
baryon overdensity at redshift zero for the reference simulations box-512-ad
(top row) and box-512-fp (bottom row) as a function of the seed field
intensity, growing from left to right as in Fig.~\ref{fig:BfieldPDFseed}. The
histograms and the over-plotted lines are constructed in the same way as in
Fig.~\ref{fig:Bvsoverdensity}.

In the adiabatic run, it can be seen that the net effect of a change in the
seed field strength is to shift vertically the histogram and the associated
relations. The shift is given by a factor of $10^2$ for each panel from left to
right, which is exactly the variation of the seed field in the three runs. In
the simulation featuring the largest seed field at high overdensities the value
of the {\it B} field does not fully grow up to the expected value, suggesting that
the amplification process has reached saturation, at least in the most dense
regions. Except for that, no other effect is present. In particular, the
overdensity range does not vary, which again is consistent with the fact that,
for these choices of the magnetic field strength, the magnetic field is
dynamically unimportant. 

\begin{figure}
\centering
\includegraphics[width=0.48\textwidth]{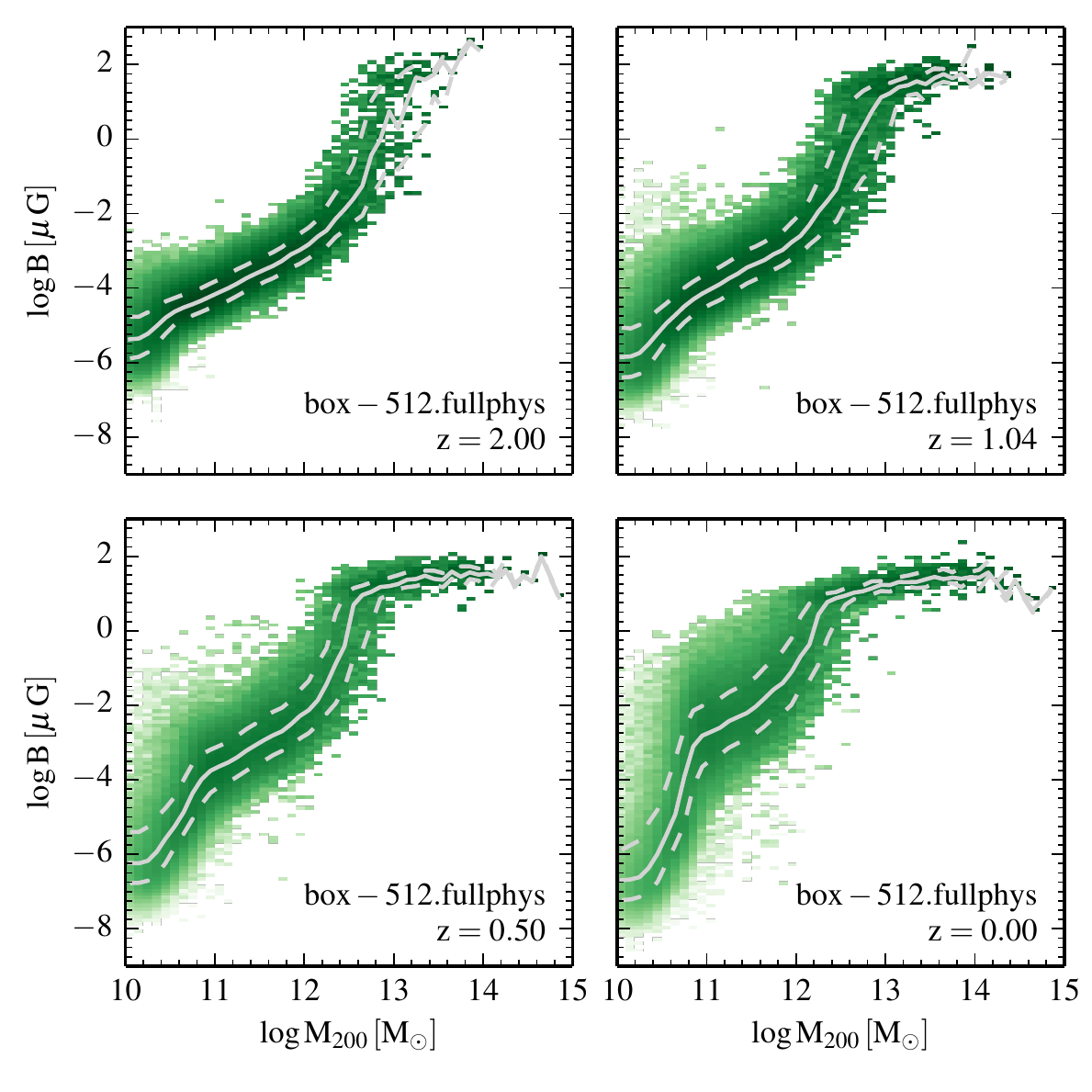}
\caption{Redshift evolution of the {\it B} field intensity versus virial mass 
($M_{200}$) of FOF groups for the simulation box-512-fp. Redshift is indicated 
on the bottom-right corner of each panel.  The panels show two-dimensional 
histograms colour coded according to the mass of gas falling on to each bin 
(darker shades correspond to larger masses). Light grey lines represent the 
median (solid) and the $16$th and $84$th percentiles (dashed) of the {\it B} field 
distribution as a function of the virial mass. Magnetic field steadily increases
with halo mass at all redshifts, forming a well-defined sequence in all the panels. 
The same trend is present in the corresponding adiabatic simulation (not shown).
However, the inclusion of baryon physics makes the relation considerably steeper 
and allows the {\it B} field to reach much larger (up to four orders of magnitude) values.} 
\label{fig:Bvsvirialmassfp}
\end{figure}

\begin{figure}
\centering
\includegraphics[width=0.48\textwidth]{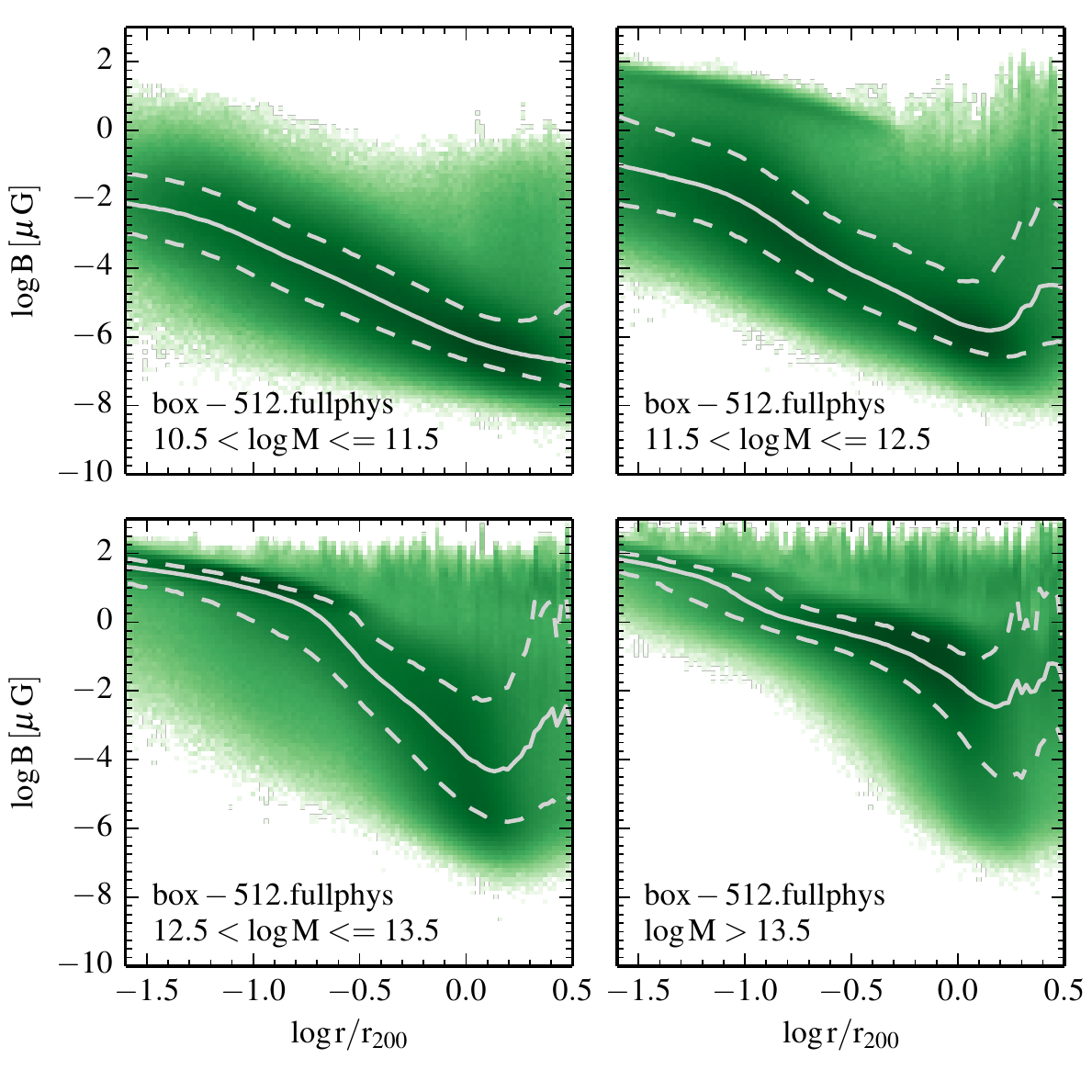}
\caption{Magnetic field profiles as a function of distance expressed in terms of 
the virial radius for the FOF groups of the simulation box-512-fp at redshift 
zero. The figure has been created by stacking the FOF groups contained in the
simulation in four virial mass bins as indicated in the bottom left corner of each
panel. The panels show  two-dimensional histograms colour coded according to the
mass of gas falling on to each bin (darker shades correspond to larger masses). 
Grey lines represent the median (solid) and the $16$th and $84$th percentiles 
(dashed) of the resulting profiles. 
} 
\label{fig:Bstackedfullphys}
\end{figure}

In the full physics run the vertical shift is also present, but it only affects
the lower values of the {\it B} field. The upper values are consistently remaining at
the same location, a further indication that saturation has been reached and
that the \rev{maximum} value of the {\it B} field lost memory of the initial strength
from which it started. Not surprisingly, the dynamic range in {\it B} field
intensities shrinks for increasing values of the seed field. Eventually, if the
value of the seed field keeps increasing, the upper and lower parts of the
diagram will have the tendency to align. If we extrapolate the results from the
right-hand panel, the alignment will occur for a critical seed field of $\sim
10^{-8}\,\,\G$. It is plausible, however, that either this high intensity of the
seed field corresponds to a larger value of the {\it B} field at saturation or that
the {\it B} field becomes dynamically important, altering dramatically structure
formation. On the other hand, for a too small value of the seed field it is
conceivable that the dynamic range spanned by the magnetic field will not
increase indefinitely, but numerical effects such as diffusion on the grid
scale \citep{Cho2009,Jones2011,Ryu2008, Schekochihin2004,Vazza2014} may prevent
the final field to grow up to the values reached in the current simulations.

\subsection{Magnetic field in haloes}\label{sec:haloes}

From the previous analysis it is clear that the amplification of the magnetic
field is particularly effective within the assembling structures and in the
presence of radiative gas cooling and of baryon processes. Therefore, in this
section we will be mostly concerned with the box-512-fp simulation that
explicitly includes baryon physics (but see Figs~\ref{fig:Bstackedpressure} and
\ref{fig:Bvstemperature}). In the adiabatic case, most of the conclusions reached
here will hold as well, although the strength of the resulting {\it B} fields is much
lower, especially in the central regions of haloes. For clarity, with halo
scales we indicate distances from the centres of haloes of the order of the
virial radius $r_{200}$ determined as the radius enclosing $200\,\rho_{\rm crit}$,
the latter being the critical density of the Universe. We will first discuss
the general {\it B} field properties of the halo population
(Sect.~\ref{sec:generalprop}) and then analyse a few individual examples
(Sect.~\ref{sec:haloexamples}).

\subsubsection{B field in the halo population} \label{sec:generalprop}

In Fig.~\ref{fig:Bvsvirialmassfp}, we present two-dimensional histograms of {\it B} field 
intensity versus halo virial masses -- determined as the total 
mass enclosed by halo virial radius $r_{200}$ -- at different redshifts for the 
reference run box-512-fp. The colour shading in the histogram represent the 
gas mass within the virial radius of each identified halo in the simulation box 
falling on to each bin. We also plot the trends of the $16$th, median and 
$84$th percentiles (grey lines). 

\begin{figure*}
\centering
\vspace{-0.2cm}
\includegraphics[width=0.95\textwidth]{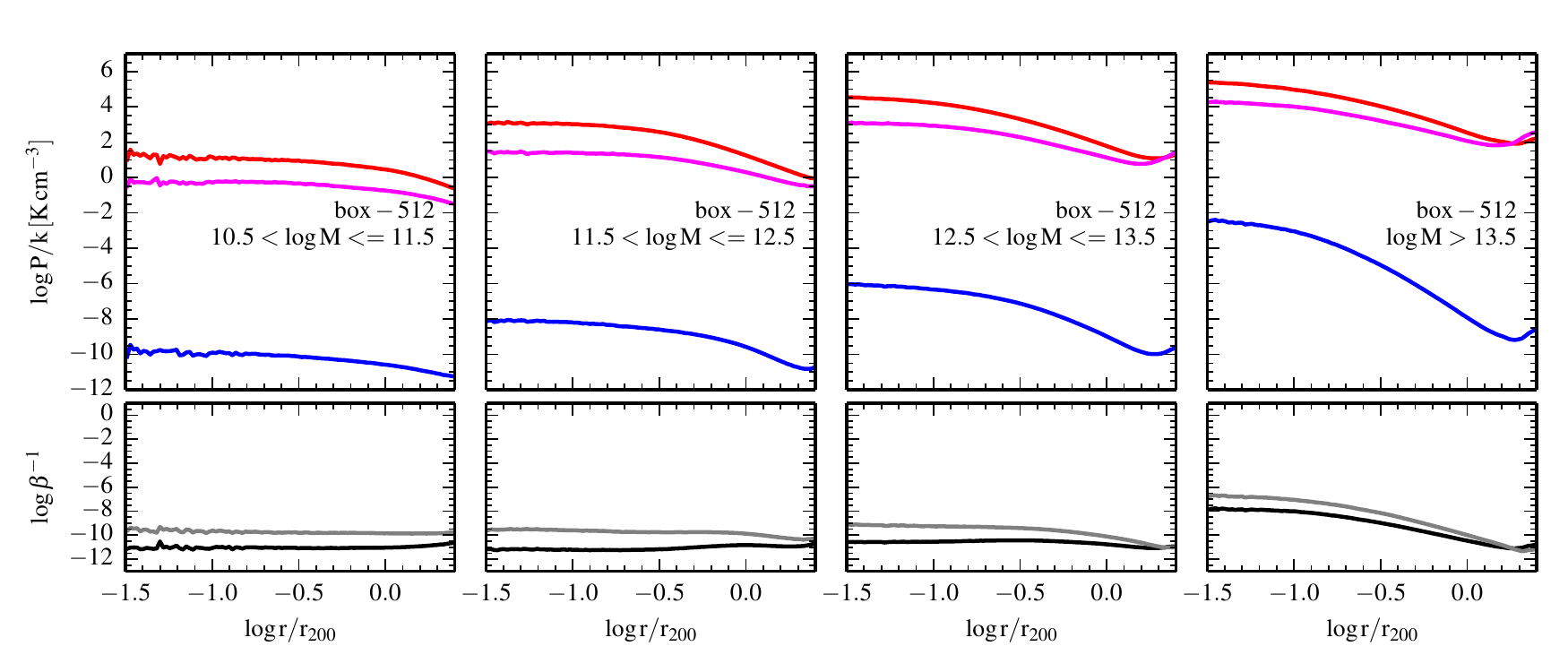}
\vspace{-0.3cm}
\includegraphics[width=0.95\textwidth]{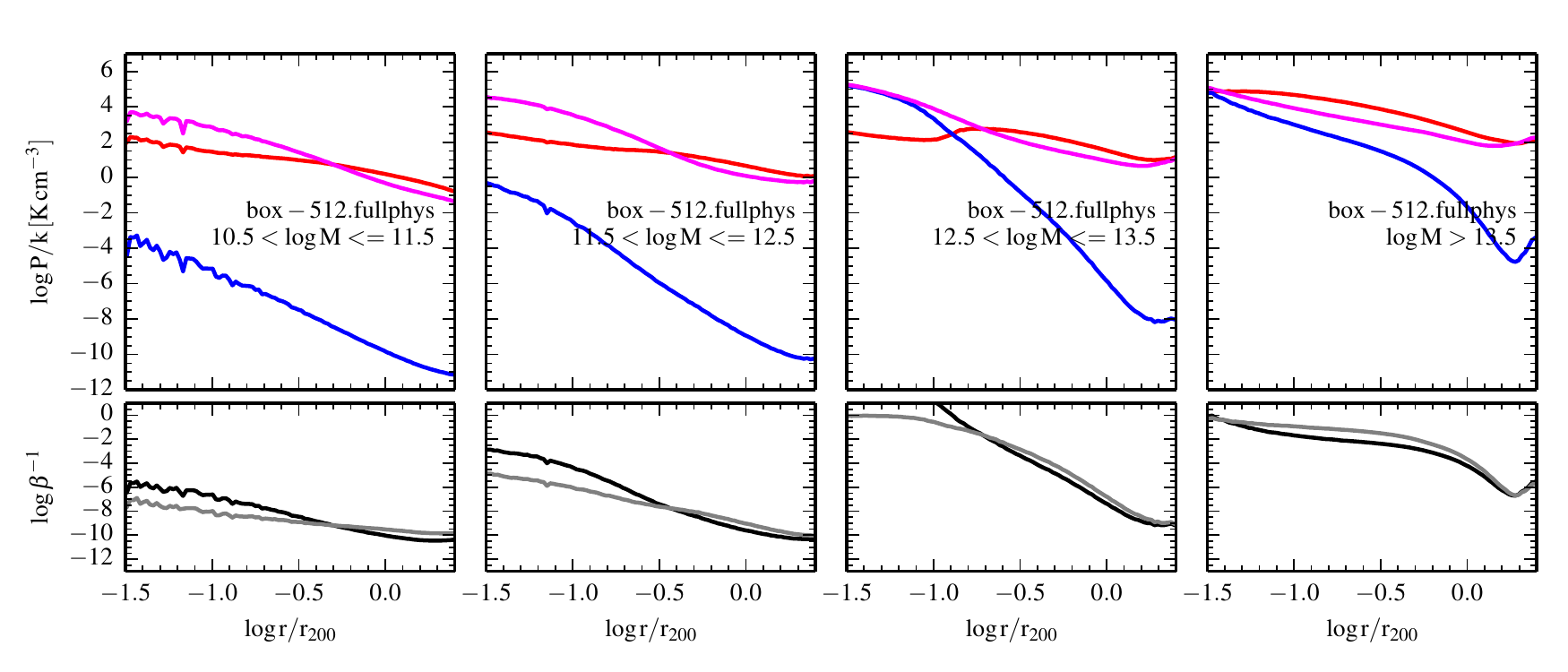}
\caption{\rev{Volume-weighted thermal pressure (red lines), magnetic 
pressure (blue lines), and kinetic energy density (magenta lines) median 
profiles} as a function of distance expressed in terms of the virial radius for 
the FOF groups of the simulations box-512-ad (top row) and box-512-fp (bottom 
row) at redshift zero. The figure has been created by stacking the FOF groups 
contained in the simulation in four virial mass bins as indicated in each panel. 
\rev{Lines in the bottom panels represent the ratio between 
magnetic and thermal pressures (black) and between magnetic pressure and kinetic
energy density (grey)}. Magnetic fields are largely subdominant and 
their contribution to the total pressure of the halo gas is negligible. However, 
in the box-512-fp simulation, in the mass bin $12.5 < \log\,M/\mo \leq 
13.5$ the magnetic field can reach values well above the gas thermal energy but this is 
limited only to the inner regions $r \lsim 0.1\,r_{200}$. \rev{In the full physics
simulations kinetic energy tends to dominate the profiles in the inner regions, 
except for the most massive mass bin.}} 
\label{fig:Bstackedpressure}
\end{figure*}

From the figure it is immediately visible that a well-defined correlation
between halo virial masses and intensity of the magnetic field is present at
all redshifts, with more massive haloes hosting larger average magnetic fields.
This is expected on the basis of the hierarchical bottom-up nature of structure
formation in a $\Lambda$CDM universe. As small structures merge to form more
massive ones, turbulence and shear flows are excited. This in turn promotes
the amplification of the magnetic field contained in the merging structures,
that was already boosted from its seed value by gravitational gas compression,
largely due to effective cooling in the halo central regions, as well as by
galactic outflows generated by stellar and black hole (for the larger
structures) feedback. Also, additional processes (such as ram-pressure
stripping), could be at work to generate turbulence in massive haloes.
Interestingly, the relationship between the halo virial mass and the
median intensity of the hosted {\it B} field steepens with time. Moreover, at the
high-mass end ($M_{200} \gsim 10^{13}\,\mo$) the magnetic field
intensity reaches a \rev{maximum} value of $\sim 10-30\,\,\muG$. At the low-mass
end, instead, there is a hint of decrease of the average field intensity as a
function of time, which can be explained by the cosmological expansion (see
eq.~[\ref{eq:adexpansion}]).

In Fig.~\ref{fig:Bstackedfullphys}, we show stacked magnetic field intensity profiles 
as a function of radius (normalized to the virial radius) at redshift zero for the 
reference simulation box-512-fp. The stacking of the FOF groups in the simulation
has been done for four different virial mass bins, indicated in the bottom-left 
corner of each panel. The colour shading encodes the gas mass falling into each bin
and the grey lines the $16$th, median and $84$th percentile of the resulting profiles.

The overall trend, for all mass bins, is that of a declining {\it B} field as a 
function of radius. For low-mass haloes (top-left panel) the decrease has a 
roughly constant slope. In the centres of these haloes the median magnetic field 
intensity is $\sim 10^{-2}\,\,\muG$, as compared to a (physical) seed field 
strength of $10^{-8}\,\,\muG$. In the external regions the median {\it B} field 
is about a factor of $10^5$ smaller. For higher mass haloes, the profiles are 
always declining but in the central regions larger median values of the {\it B} field 
intensity, which saturate at $\sim 10^{2}\,\,\muG$ for the most massive bins, 
can be reached. In the outskirts, the amplification of the {\it B} field depends on 
the mass bin considered and it is more efficient the larger the halo mass. This 
again agrees with the expectations of  hierarchical structure growth in which 
more massive haloes are more likely to have experienced more frequent merger 
events. The higher merger frequency triggers gas processes such as turbulent 
motions responsible for magnetic field amplification, leading to larger {\it B} field
values. Finally, for the most massive objects also gas stripped from substructures may 
contribute to increase the {\it B} field in the halo.

\rev{In Fig.~\ref{fig:Bstackedpressure}, we present volume-weighed magnetic 
pressure (blue lines), thermal pressure (red lines), and kinetic energy 
density (magenta lines) median profiles, together with their ratios 
(black lines for magnetic to thermal and grey lines for magnetic to kinetic)}, 
at $z = 0$ for the reference simulations box-512-ad (top row) and box-512-fp 
(bottom row). The profiles are shown as a function of the distance from the halo 
centre normalized to the halo virial radius, in analogy with 
Fig.~\ref{fig:Bstackedfullphys}, and in the same halo mass bins.

\rev{If we compare the magnetic pressure to the thermal pressure and 
kinetic energy density profiles}, we can immediately
see that the magnetic fields are subdominant; i.e. their importance for the
dynamics of the system is negligible. The only exception are the two most
massive bins of the full physics run, in which the magnetic pressure is
comparable to or exceeds the  gas thermal pressure \rev{and kinetic energy} in the inner regions ($r
\lsim 0.1\,r_{200}$). The thermal pressure profiles of both simulations in
each mass bin are similar, indicating that the support needed against
gravity within haloes is mostly provided by the ``standard'' gas pressure. Only
the $12.5 < \log\,M/\mo \leq 13.5$ mass bin in the full physics run shows
a pressure dip in the central regions, likely due to the efficient radiative
cooling of the gas, which can explain why the magnetic contribution to the
total gas pressure is so large. 

On the other hand, magnetic pressure profiles are very different between the two 
runs. In the adiabatic case, their shape closely follows that of the thermal gas 
pressure, as can be seen from the profile \rev{of their ratio}, which is 
approximately constant, at a level of $10^{-11}$, with radius and among the mass 
bins, with the exception of the most massive one where a larger deviation from a 
flat profile can be seen. \rev{This} tiny value is an additional confirmation 
of the unimportant role played in the dynamics of the haloes, but it is 
interesting to note that the absolute value of the magnetic pressure (and thus 
of the magnetic field intensity) increases for the most massive systems, 
indicating that the amplification process is more effective in high mass 
structures, \rev{which also feature a larger value of the gas kinetic energy 
density that can be used to power the amplification of the {\it B} field}. In the full 
physics simulation, magnetic pressure profiles show a much steeper decline with 
respect to their adiabatic counterpart. This is due to the fact that larger 
values of magnetic field intensities can be reached in the halo centres where 
cooling and other baryon processes are at work \rev{-- substantially increasing 
the gas kinetic energy density, which is indeed the dominant component in the 
halo inner regions for all the mass bins but the most massive --} while the 
level of amplification outside haloes ($r \gsim r_{\rm vir}$) is lower and 
similar to that of the adiabatic simulations. As in the adiabatic case, larger 
values of magnetic pressure are found for larger halo masses.

\begin{figure}
\centering
\includegraphics[width=0.37\textwidth]{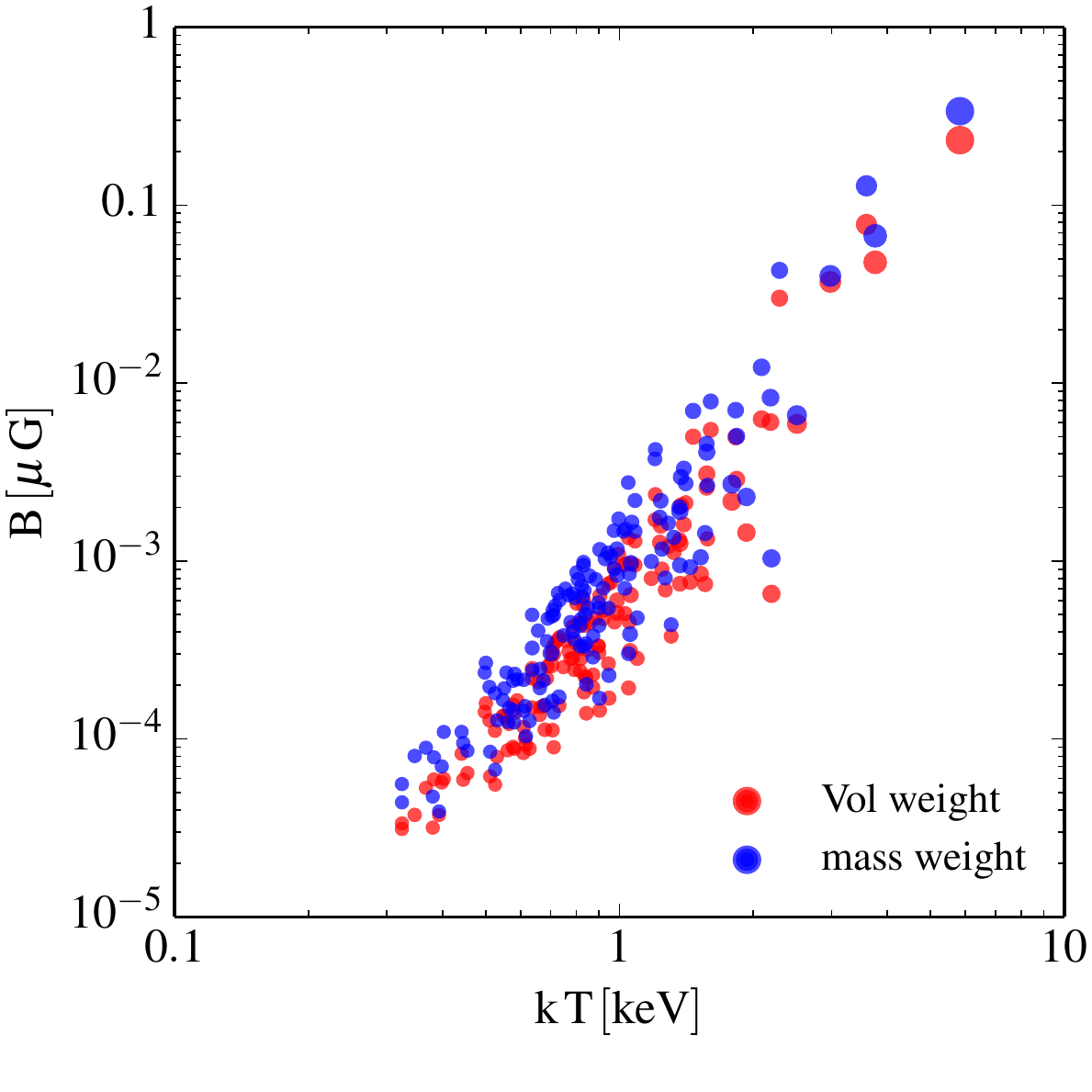}
\includegraphics[width=0.37\textwidth]{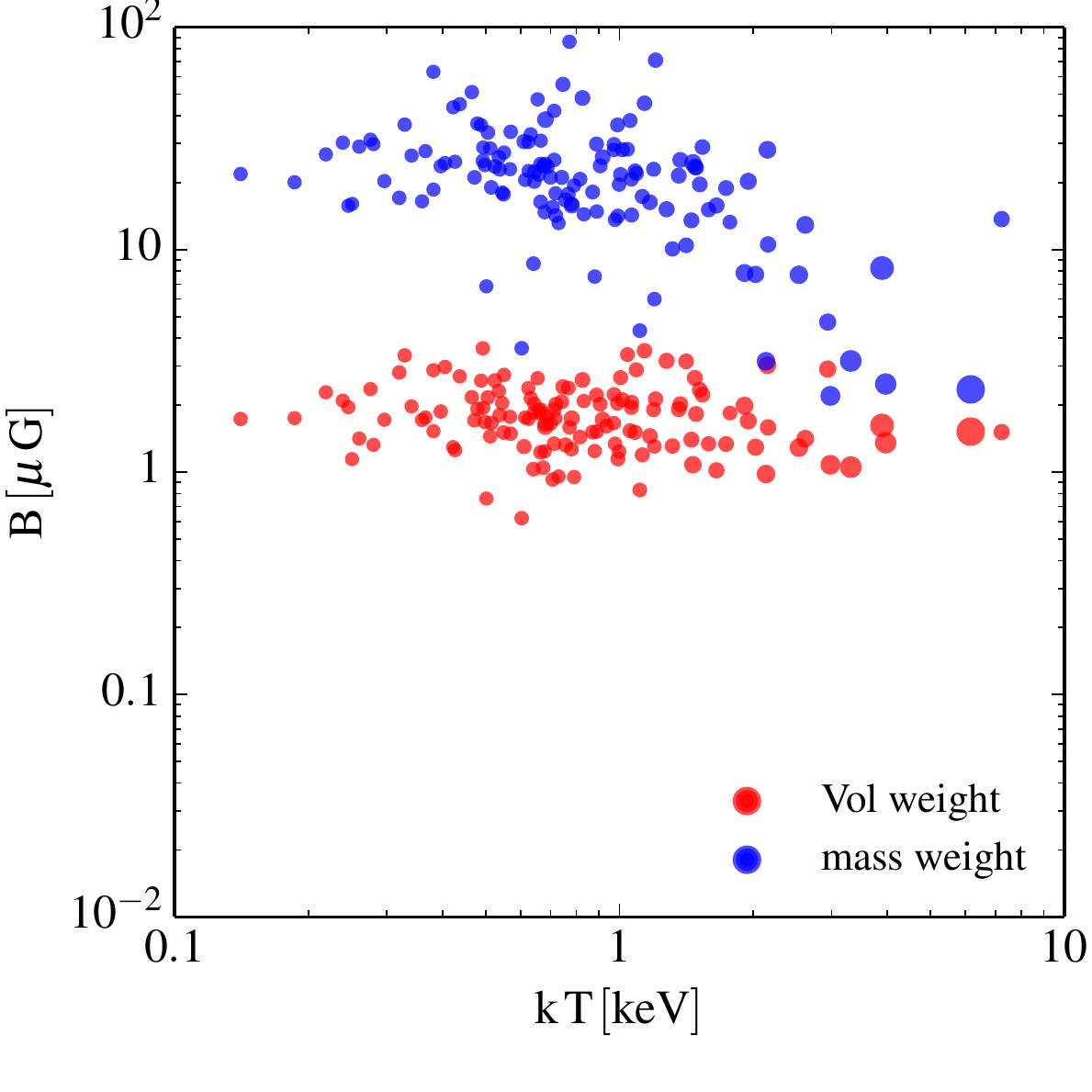}
\caption{Average volume-weighted (red symbols) and mass-weighted (blue symbols)
magnetic field as a function of the mass-weighted gas temperature for the 
box-512-ad (top panel) and box-512-fp (bottom panel) simulations at redshift zero. 
The averaging procedure for both temperature and {\it B} field has been performed in a 
region within $35\%$ of the virial radius of each halo. Only haloes more massive 
than $10^{13.5}\,\mo$ have been considered in this plot. Symbols
sizes are scaled according to halo virial masses.} 
\label{fig:Bvstemperature}
\end{figure}

In Fig.~\ref{fig:Bvstemperature}, we show the redshift zero average {\it B} field as a 
function of the (mass-weighted) average temperature for haloes more massive than 
$10^{13.5}\,\mo$ for the simulations box-512-ad (top panel) and 
box-512-fp (bottom panel). The average values have been computed in a spherical 
region centred on the potential minimum of each halo and of radius equal to 
$0.35\, r_{200}$. To compute the magnetic field value we adopt both a volume 
(red symbols) and a mass (blue symbols) weighting procedures, the latter to 
allow a comparison with a similar figure presented in \cite{Donnert2009} who 
studied the amplification of {\it B} field in cosmological simulations of galaxy 
cluster formation. Their simulations did not include radiative cooling but 
focused on a more elaborate way of seeding the primordial magnetic field via 
ejection of magnetized material by galactic outflows. 

\begin{figure*}
\centering
\includegraphics[width=0.245\textwidth]{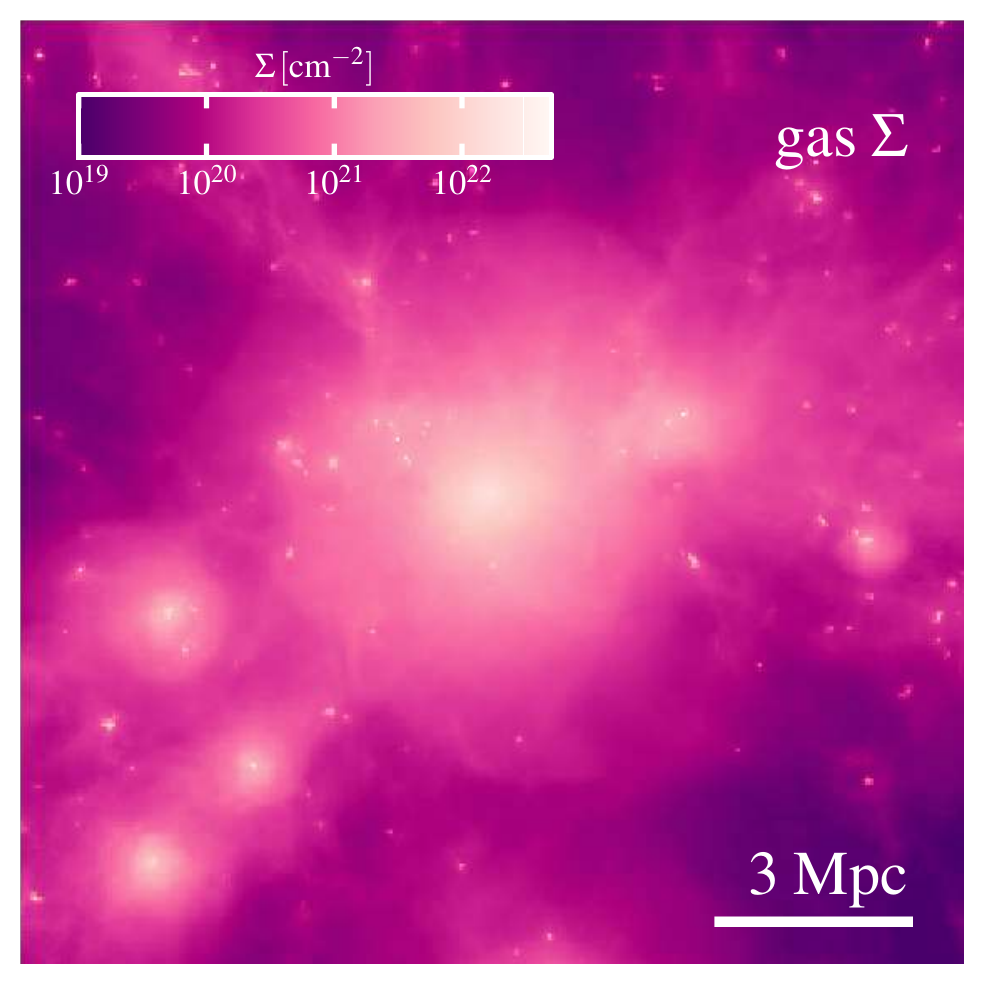}
\includegraphics[width=0.245\textwidth]{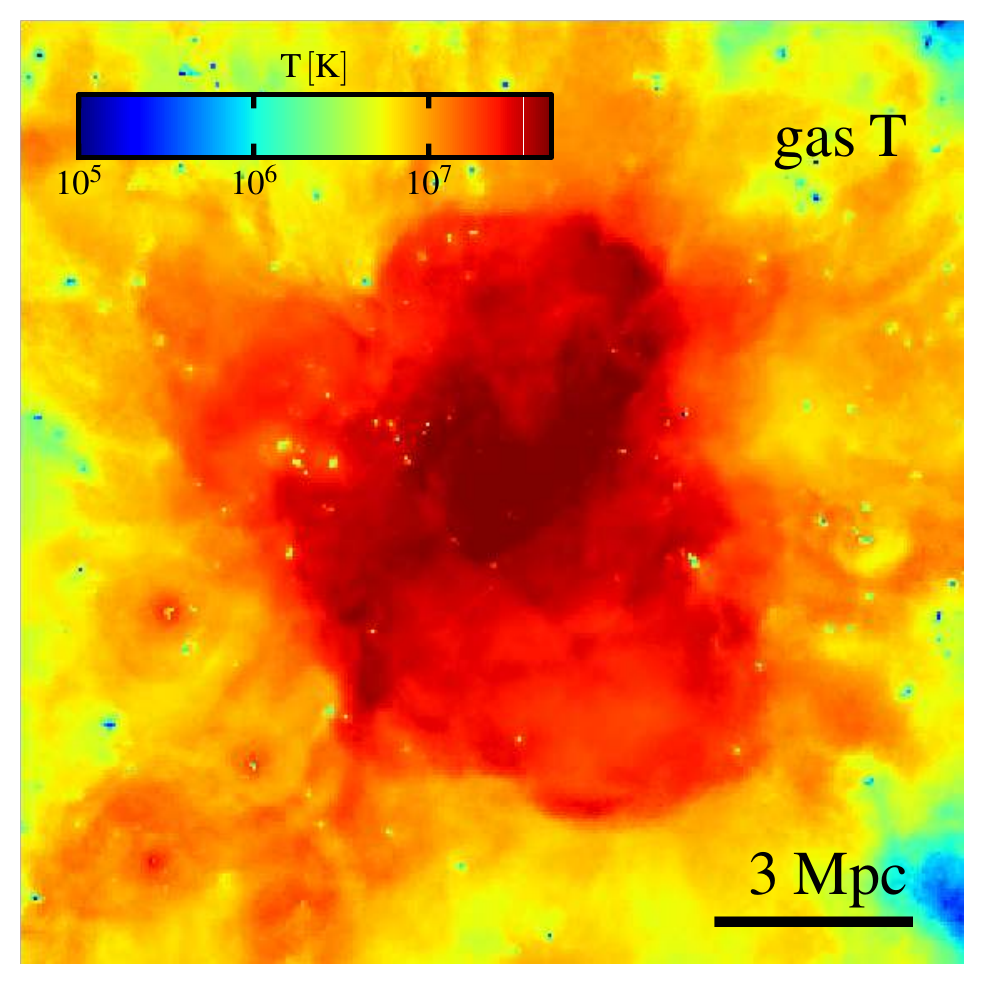}
\includegraphics[width=0.245\textwidth]{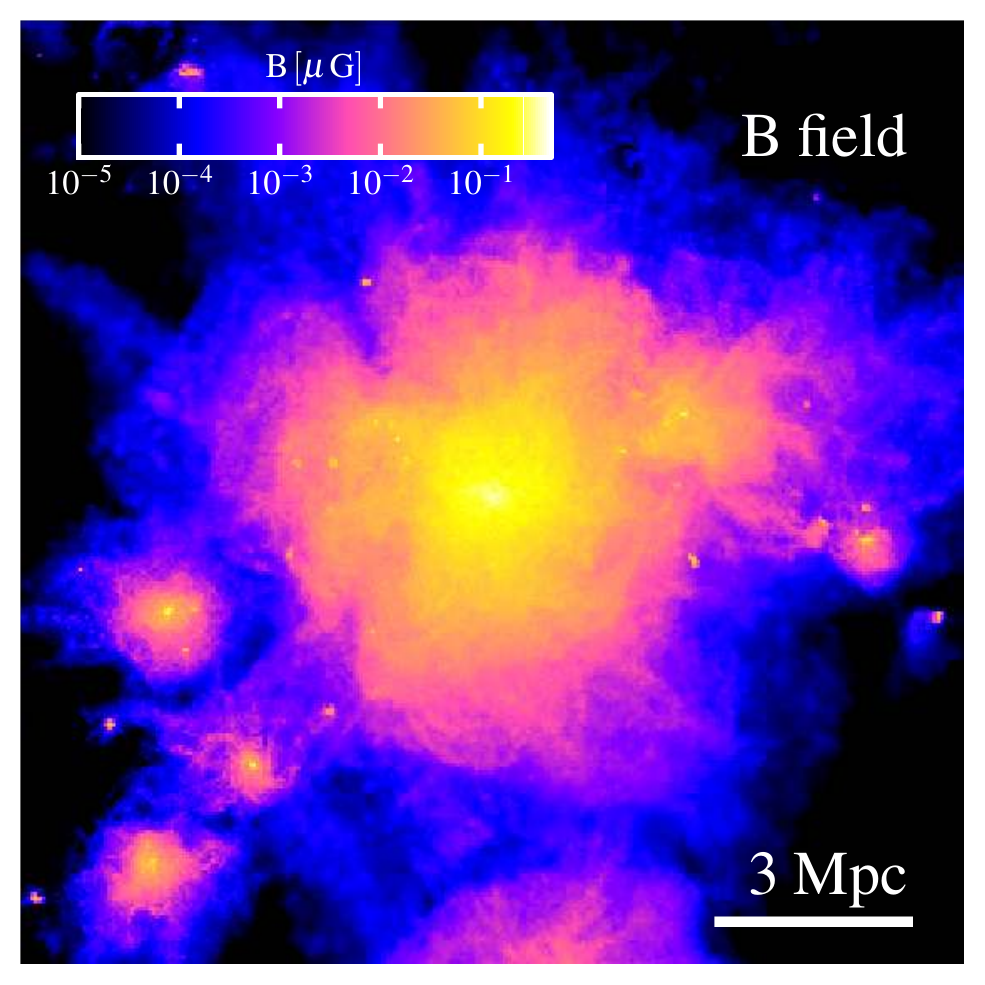}
\includegraphics[width=0.245\textwidth]{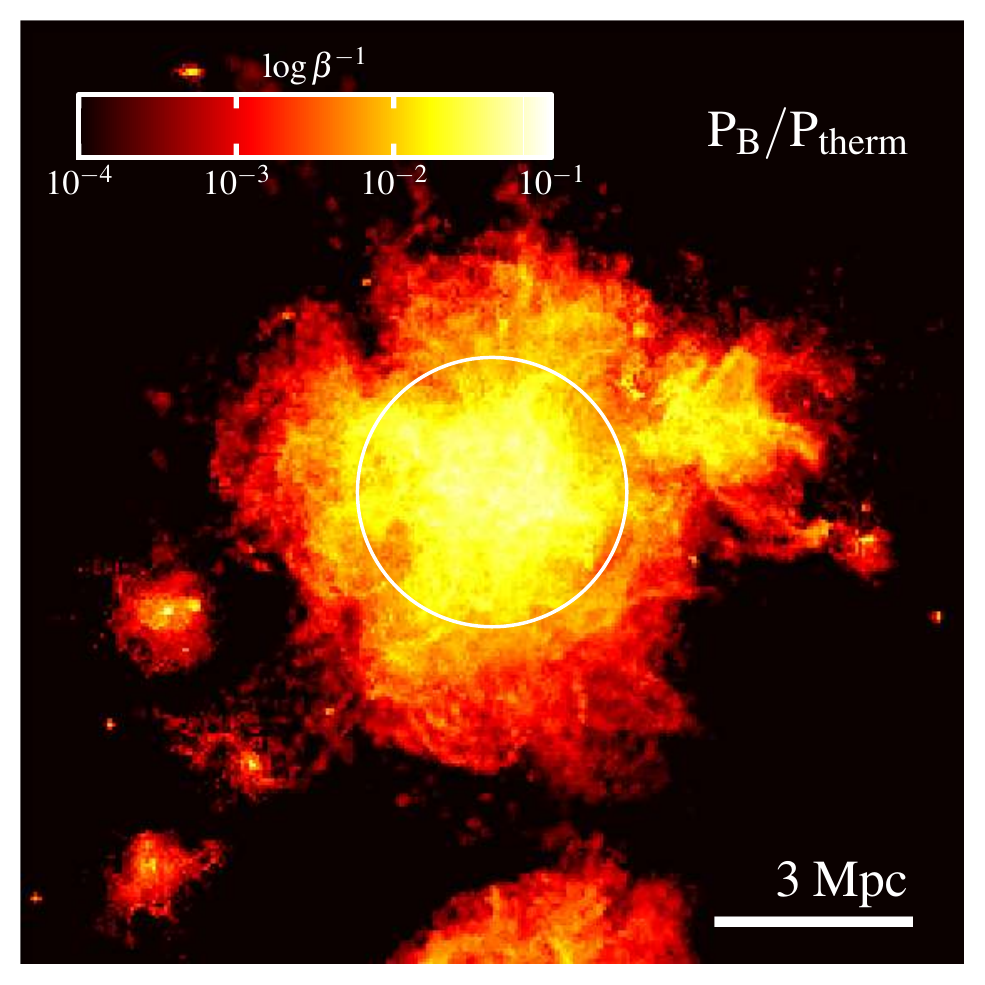}
\includegraphics[width=0.245\textwidth]{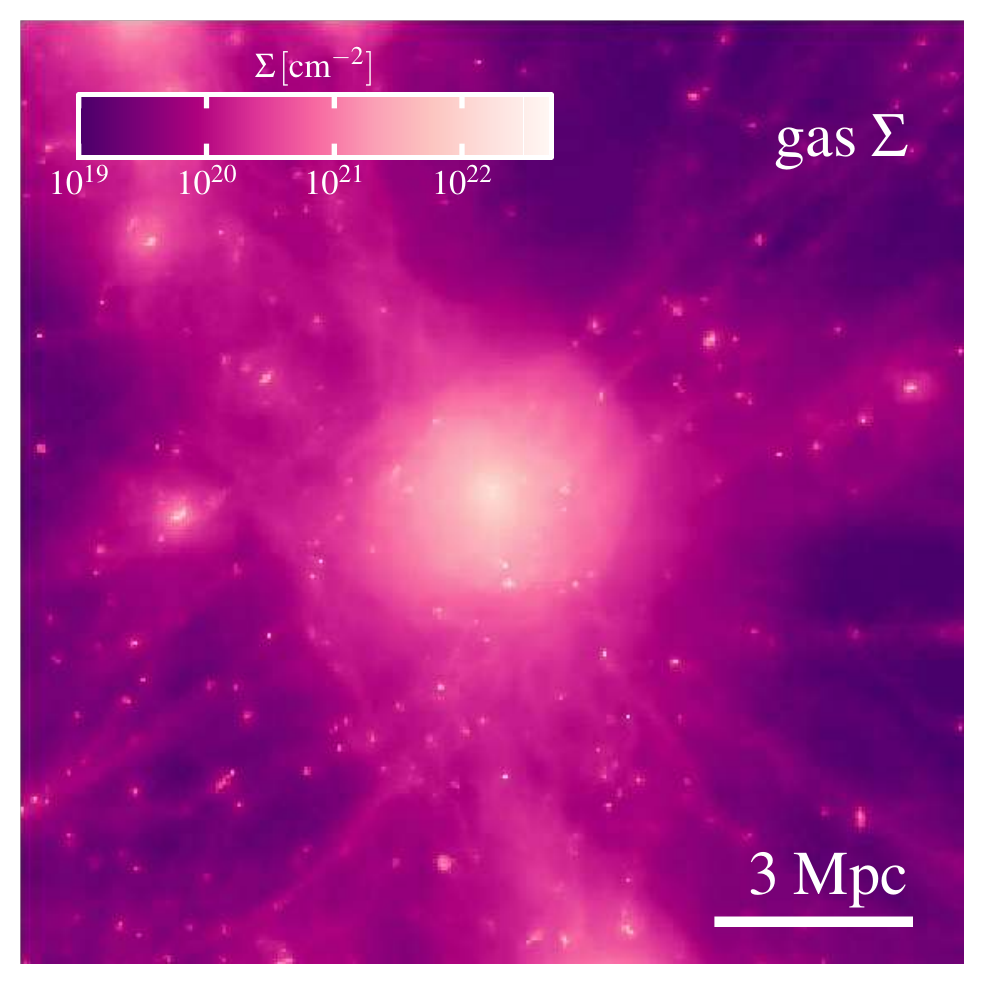}
\includegraphics[width=0.245\textwidth]{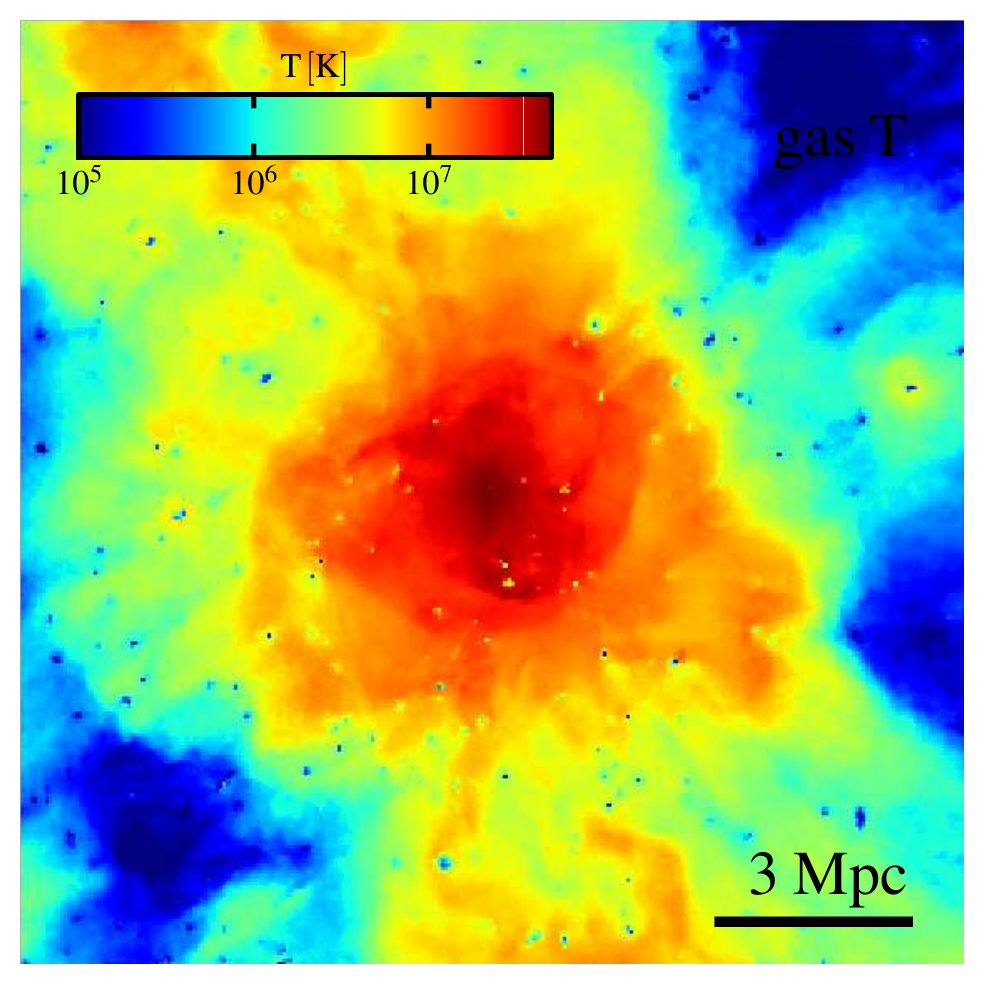}
\includegraphics[width=0.245\textwidth]{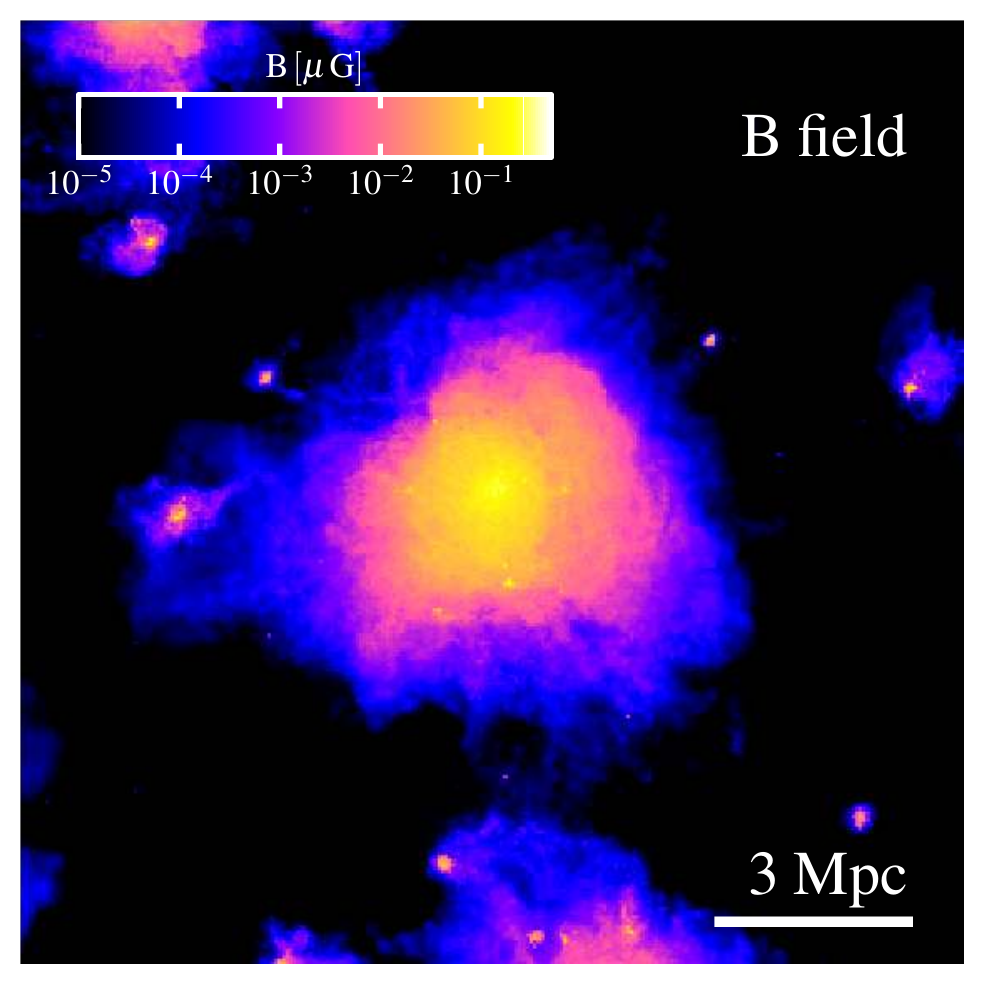}
\includegraphics[width=0.245\textwidth]{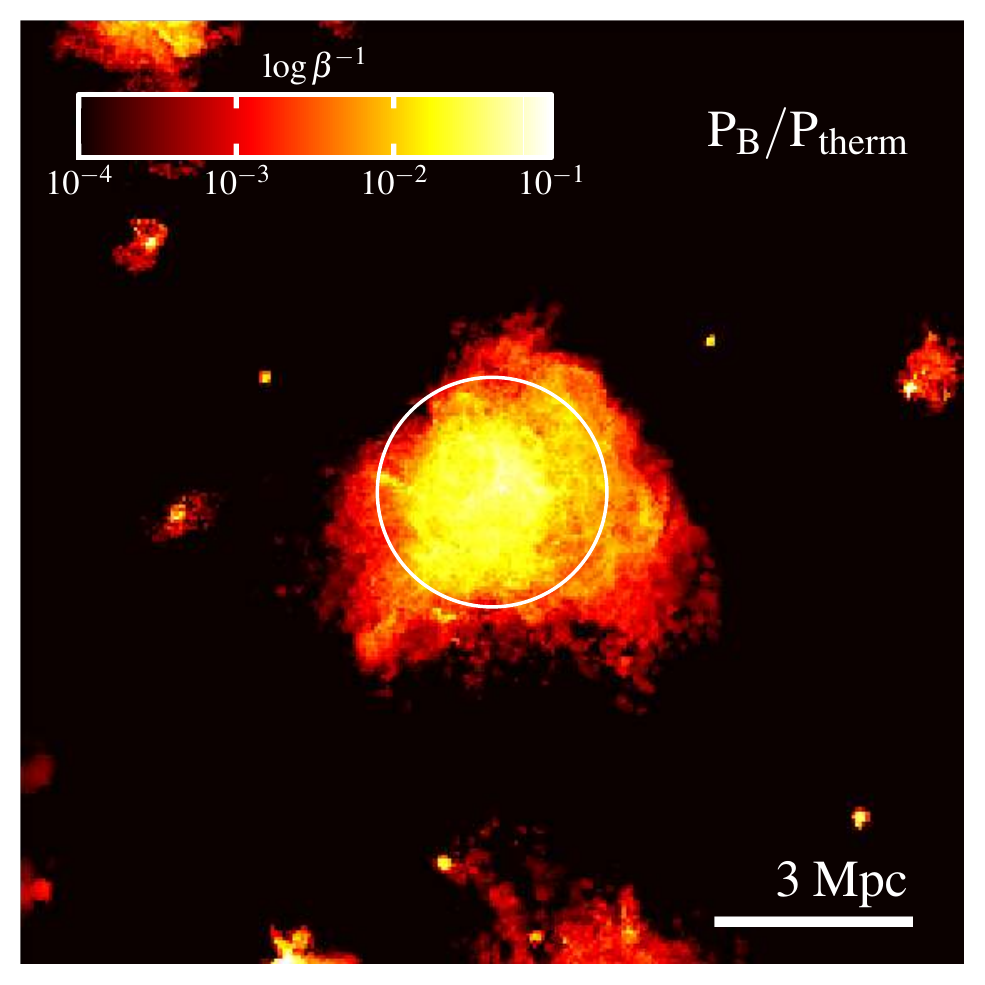}
\includegraphics[width=0.245\textwidth]{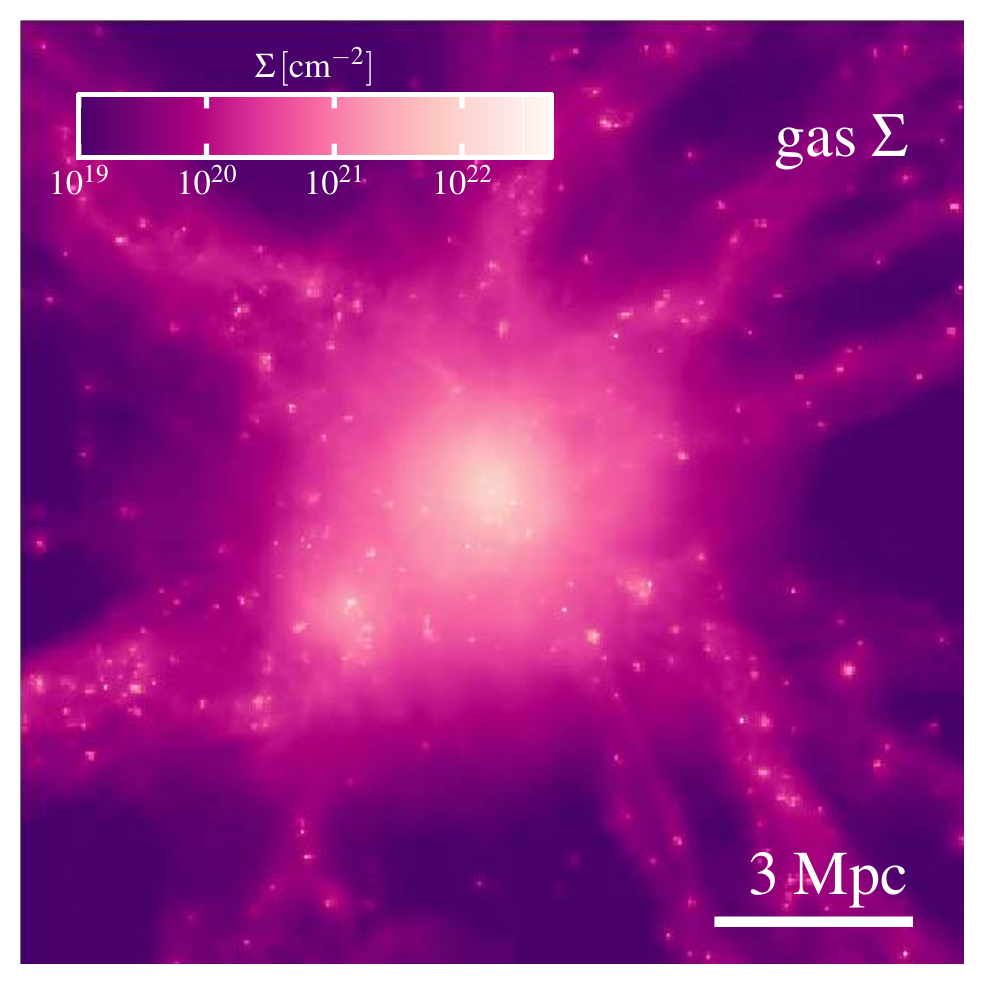}
\includegraphics[width=0.245\textwidth]{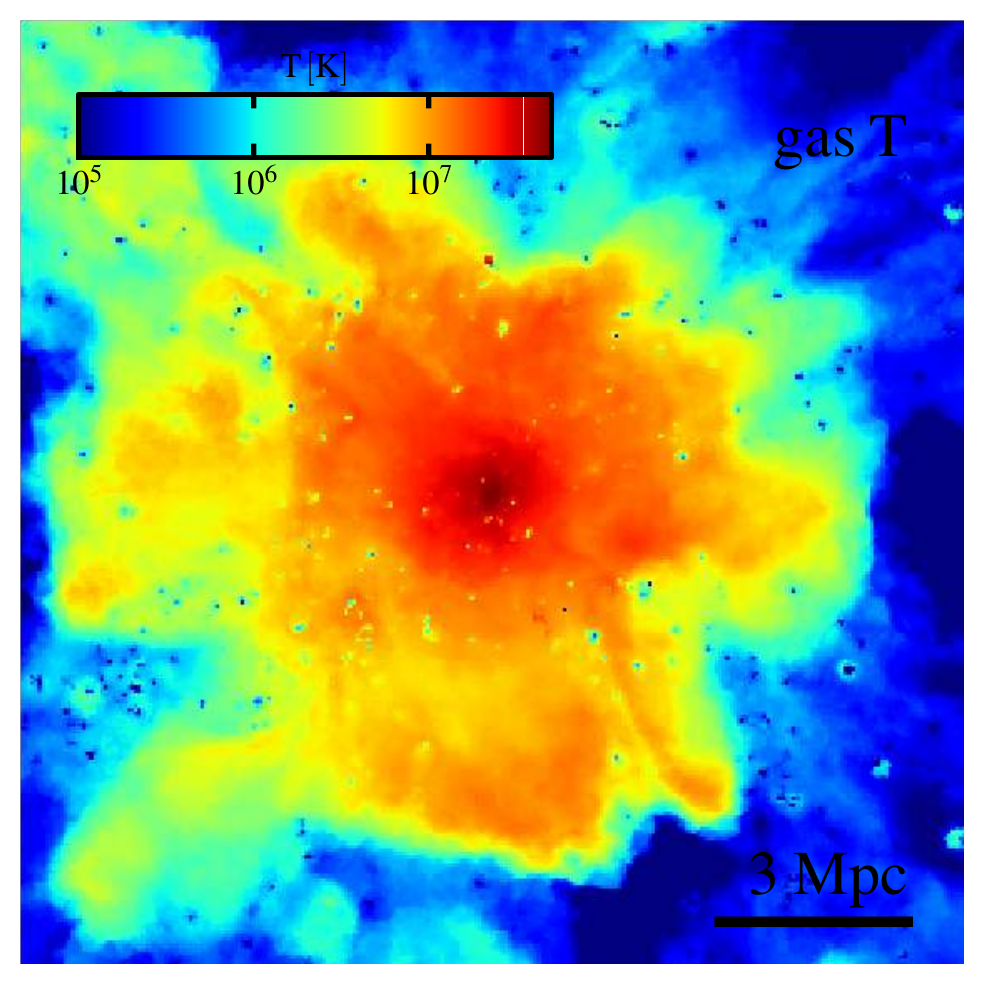}
\includegraphics[width=0.245\textwidth]{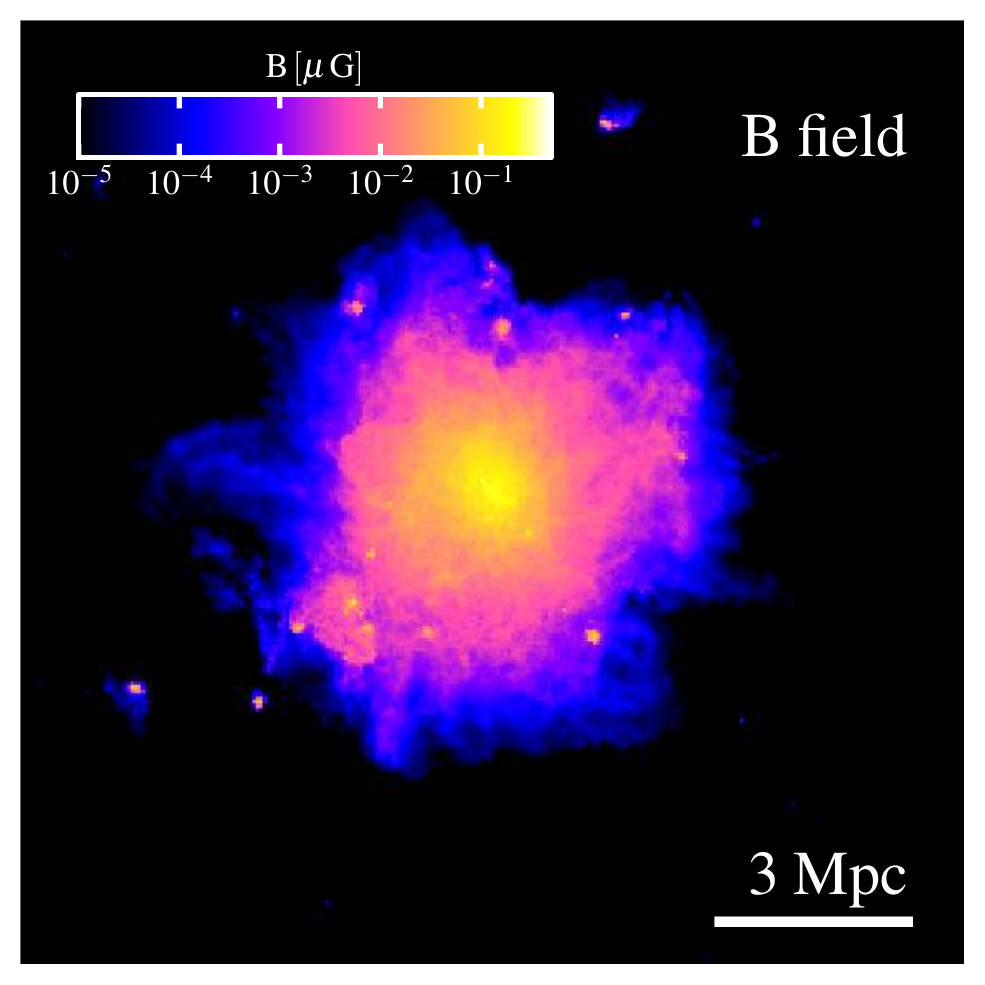}
\includegraphics[width=0.245\textwidth]{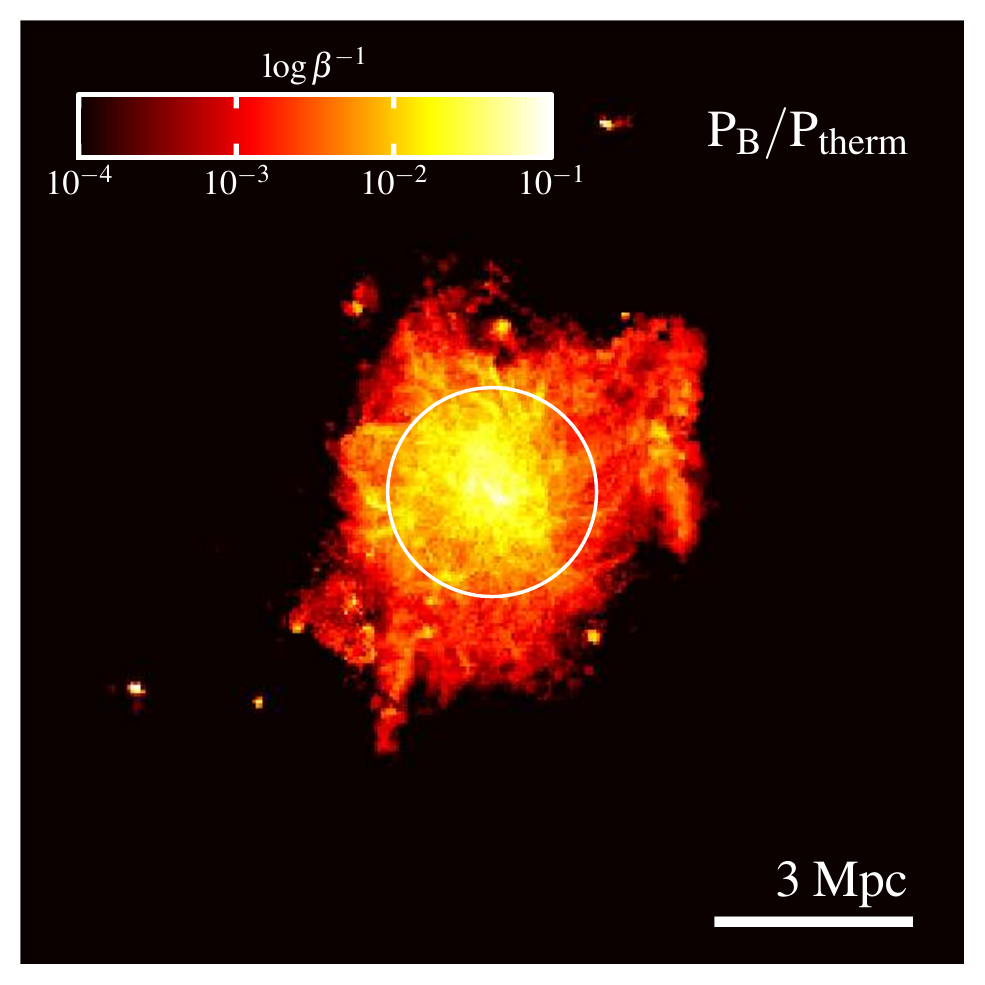}
\caption{Redshift zero projections of gas density, gas temperature, magnetic 
field and the ratio between magnetic and thermal pressure (from left to right) 
for the three most massive haloes in the simulation box-512-fp (ordered by mass from 
top to bottom). The projection region is a cube of $15$ Mpc on a side  (in 
physical units) centred on the halo potential minimum. The white circle in the 
rightmost panels indicates the virial radius ($r_{200}$) of each halo. 
} 
\label{fig:clusterproj}
\end{figure*}

In the adiabatic case (top panel) we recover a tight relationship as in
\cite{Donnert2009} between the halo central temperature and its mean magnetic
field. The weighting procedure adopted for the computation of the {\it B} field does
not lead to a significant difference in the final derived values, although
mass-weighting yields consistently larger values (about a factor of 2) than
volume-weighting. The slope of the relationship is approximately $B \propto
T^3$ with a hint of flattening at the low-temperature end ($\lsim 1\,{\rm keV}$
). The most remarkable difference with respect to \cite{Donnert2009} is that
the values of the central magnetic field reached in our simulation are lower by
about two orders of magnitude.  This is due to a combination of causes. For
instance, we have chosen a larger radius within which to average the field and
this naturally brings down the estimated average value. The most important
factor, however, is the intensity of the seed field.  We have already shown
that in the adiabatic runs this is the crucial quantity that controls the final
intensity of the magnetic field. In our fiducial run we chose a seed field of
$10^{-14}\, \G$, to be compared to values of $\simeq
10^{-12}\, \G$ in \cite{Donnert2009} simulations. Therefore, according to what discussed
in Sect.~\ref{sec:seed field}, our results must be rescaled by approximately this factor, which will
make this discrepancy less severe. Finally, also numerical resolution can play
a role in defining the final {\it B} field strength.

\begin{figure*}
\centering
\includegraphics[width=0.32\textwidth]{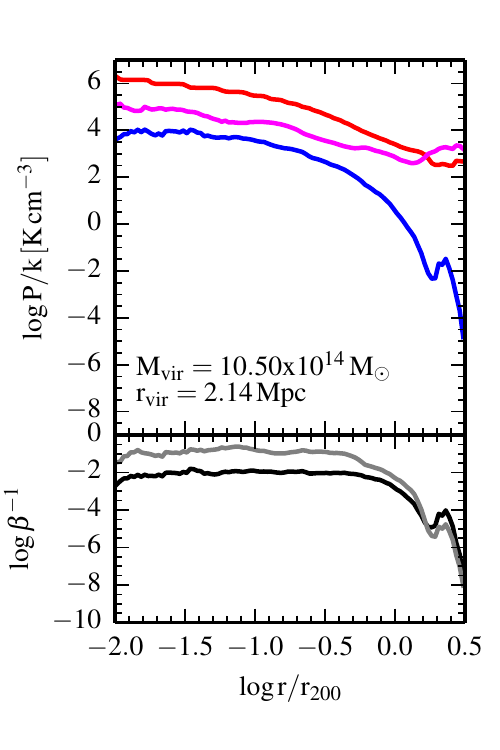}
\includegraphics[width=0.32\textwidth]{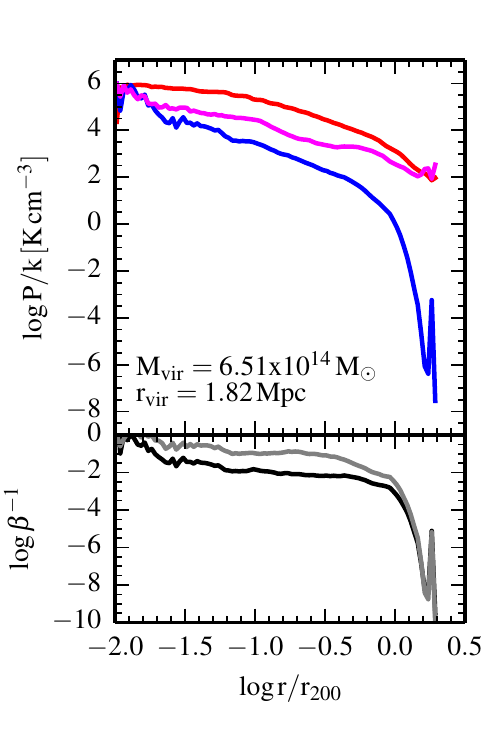}
\includegraphics[width=0.32\textwidth]{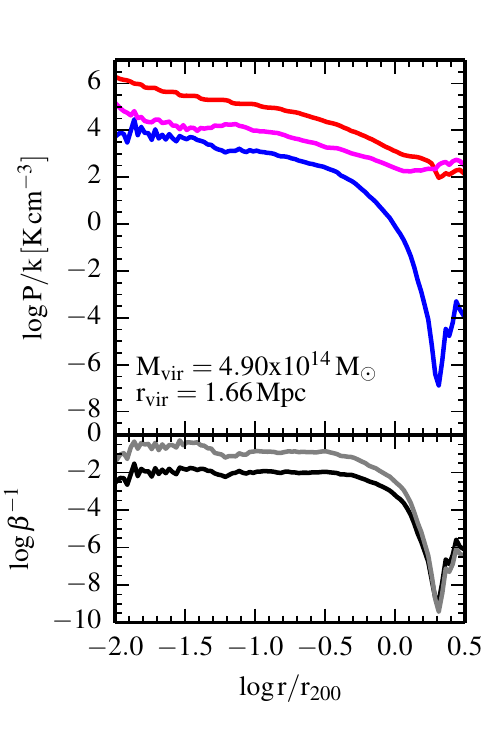}
\caption{
\rev{Volume-weighted thermal pressure, magnetic pressure, and kinetic
energy density median profiles} for the three 
most massive haloes of simulation box-512-fp. The layout of each panel is the same 
as in Fig.~\ref{fig:Bstackedpressure}.}
\label{fig:clusterpprofile}
\end{figure*}

In the full physics run (bottom panel), the monotonically increasing trend of
magnetic field intensity as a function of the temperature is disrupted.  This
shows the dramatic impact that baryon physics, and especially of outflows
originated by galactic winds and AGN feedback, has on the dynamics of the gas
within haloes, which translates in a stark difference in the amplification of
the field with respect to the adiabatic run. There is a much larger difference
between magnetic field values derived by volume-weighted averages versus
mass-weighted ones, with the latter being on average about one order of
magnitude larger. This is the consequence of the cell refinement scheme adopted
in our calculations that keeps the mass per cell within a factor of 2 from a
predefined target. The approximately constant mass per cell implies that cells
are given the same weight to compute the average field regardless of their
distance. This does not hold anymore when magnetic field intensities are
volume weighted. More distant cells, with smaller {\it B} fields, have on average
larger volume and are given more weight in the averaging procedure, thus
reducing the resulting {\it B} field strength. The volume-weighted magnetic field
strength does not show a large variation with temperature remaining on the
level of a few $\muG$. The mass-weighted field declines with temperature 
reaching $\sim 1 \muG$ for the more massive systems. The values of central
magnetic fields for the largest systems (up to a few tens of $\muG$) are
consistent with observational determinations in galaxy clusters
\citep[e.g.][]{Feretti1995, Feretti1999, Murgia2004, Guidetti2008, Govoni2006,
Bonafede2009, Bonafede2011}. 

\subsubsection{B field in individual haloes} \label{sec:haloexamples}
To better understand the trends discussed above and to make a closer connection
with the observations (see also next section), we now analyse some of the
haloes individually by taking the three most massive haloes in the simulation
box-512-fp.  Fig.~\ref{fig:clusterproj} presents different projected quantities
(gas density and temperature, magnetic field strength and the ratio between
magnetic and thermal pressure) at redshift zero. The haloes are ordered
according to their virial mass, which covers the range from $4.9\times
10^{14}\, \mo$ to $1.05\times 10^{15}\, \mo$. 

These three haloes are representative of large groups or clusters of galaxies.
The gas density distributions show that they are almost spherically-symmetric
objects. In the outskirts there are still signs of substructure accretion and
at large scales filaments of material connect the central virialized object to
the cosmic web. The temperature distribution in most cases extends to larger
scales than the gas density. A hotter core can be identified closer to the
central regions, but high temperatures can also be found beyond the halo virial
radius (indicated by the white circles in the rightmost panel). This is a
direct consequence of the radio mode AGN feedback model that we have adopted,
which is particularly effective in ejecting (hot) gas at large distances from
the centre \citep{Genel2014}. The magnetic field closely follows the gas
density distribution. It reaches the largest values in the centres of the
haloes and then rapidly declines with decreasing gas density. Local increases
of the magnetic field can be seen outside the central parts of the haloes in
correspondence of density enhancements due to infalling sub-structures.
Notwithstanding this close connection between magnetic field and gas
properties, and consistently with the results of
Fig.~\ref{fig:Bstackedpressure}, magnetic pressure declines more rapidly than
gas thermal pressure as it is shown in the rightmost panels. Only inside the
halo virial radius the ratios between the two pressures can reach values of a
few percent, to rapidly drop to $10^{-4}$ and below outside a few $r_{200}$.

\begin{figure}
\centering
\includegraphics[width=0.23\textwidth]{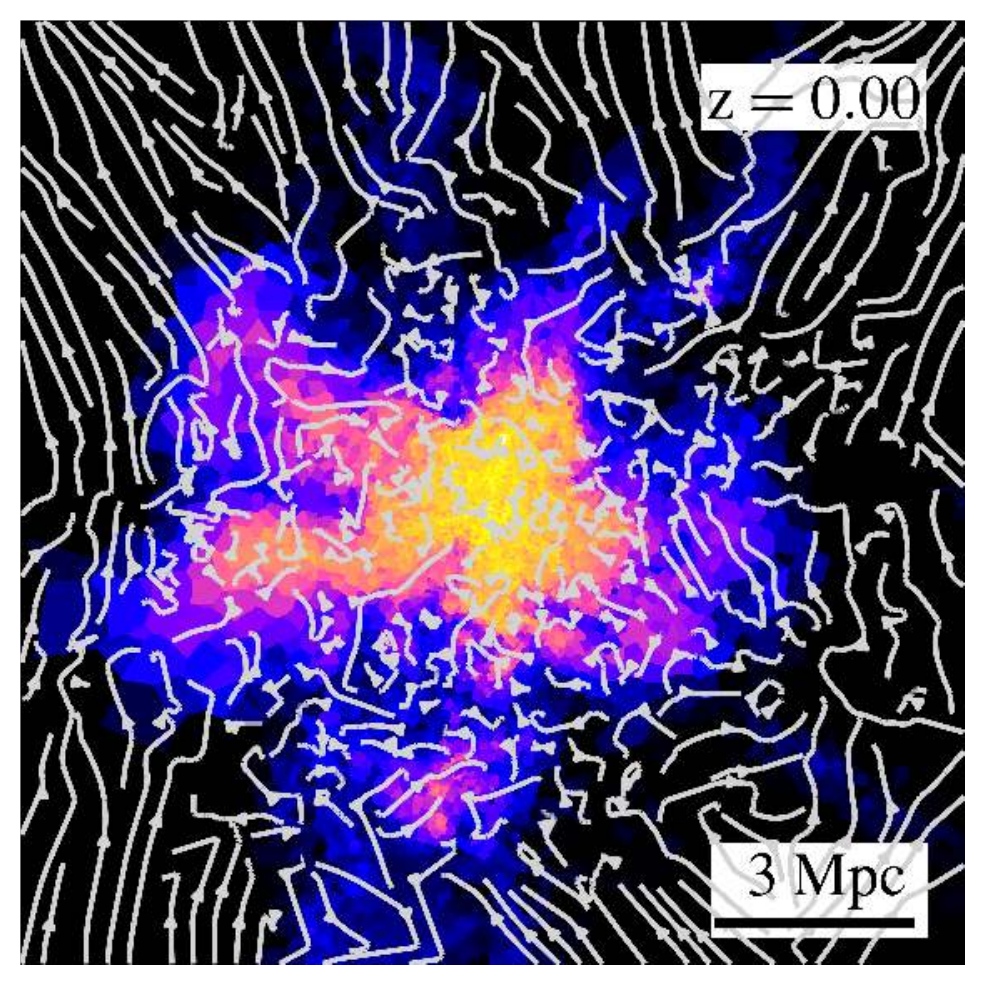}
\includegraphics[width=0.23\textwidth]{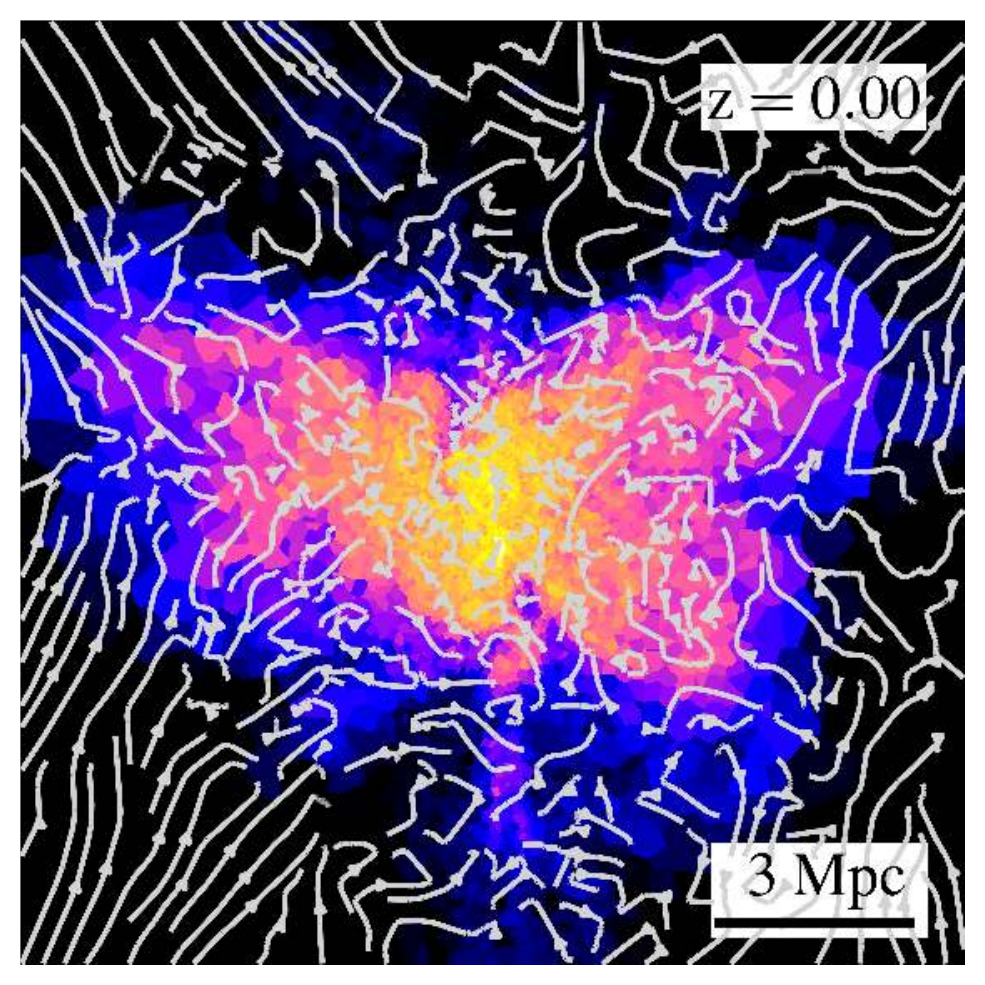}
\includegraphics[width=0.23\textwidth]{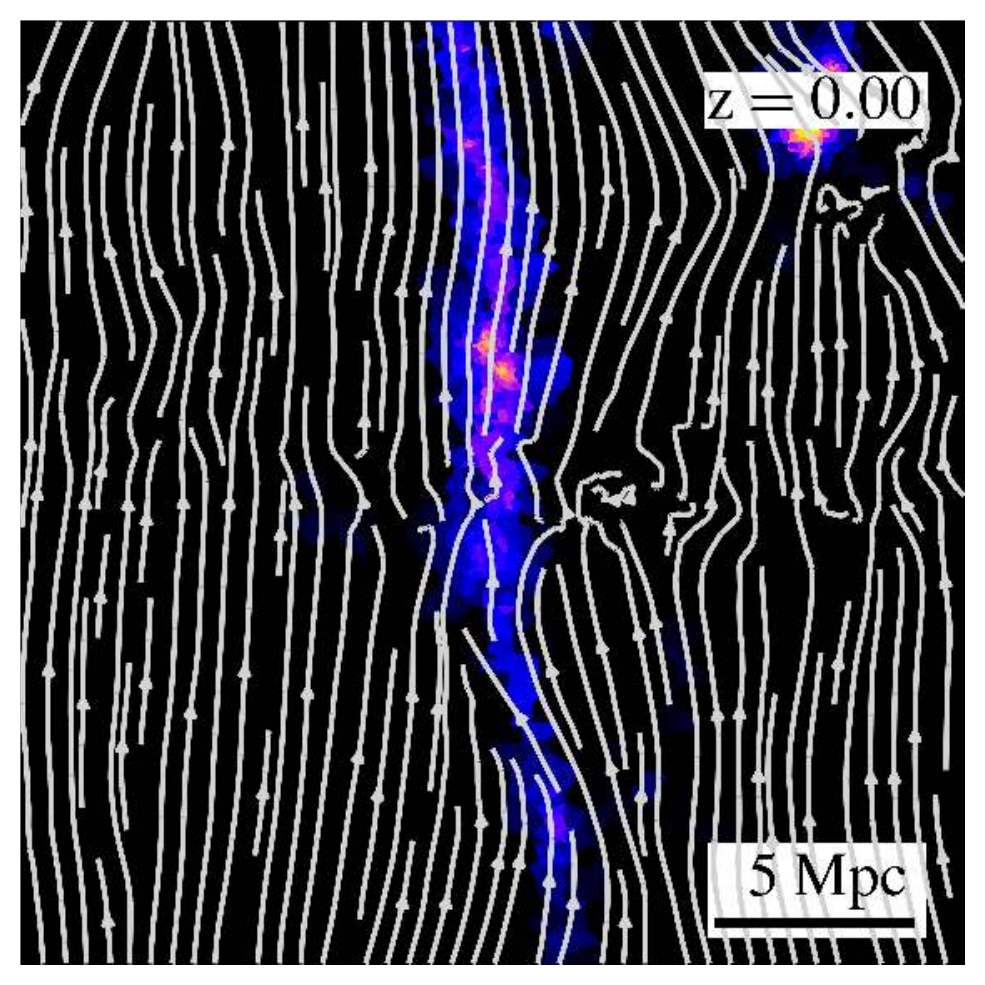}
\includegraphics[width=0.23\textwidth]{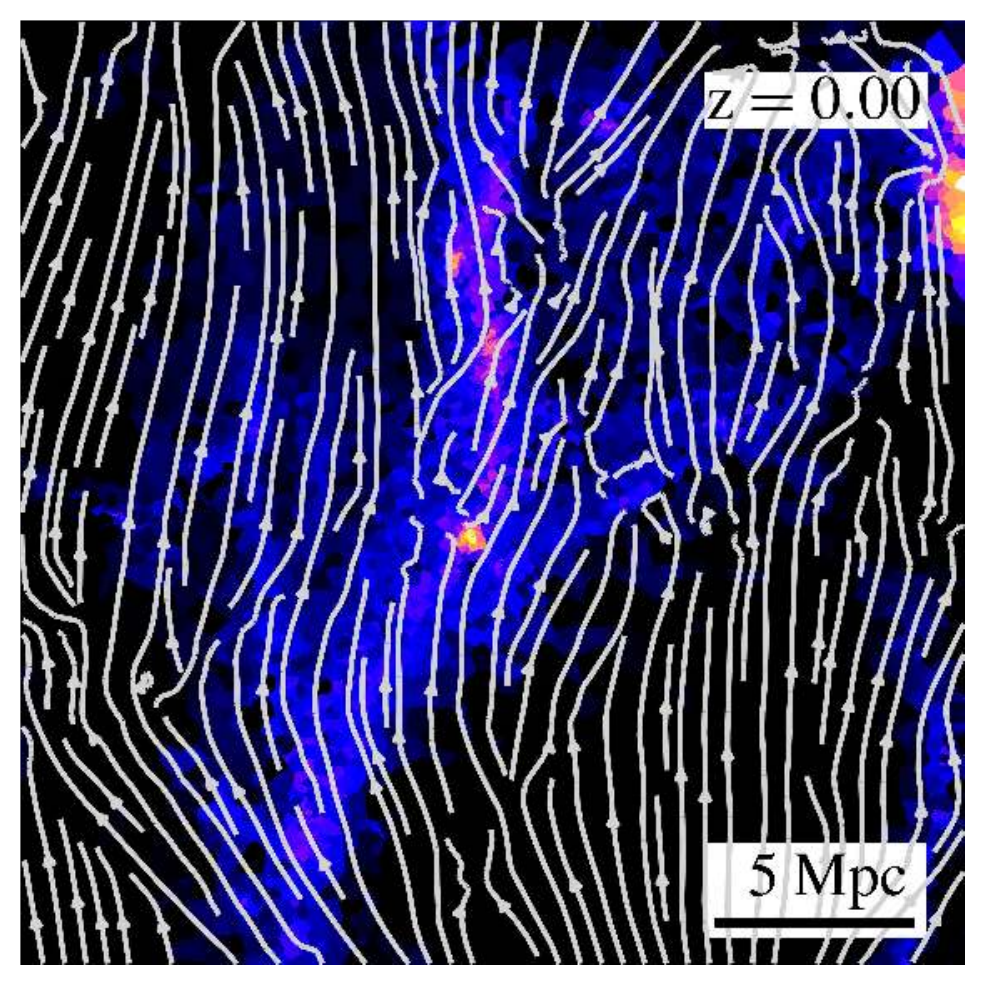}
\caption{Magnetic field slices at redshift zero for a massive halo (top row) 
and a filament (bottom row) selected from the simulation box-256-fp. On the left column
the slice is displayed on the $yz$ plane while on the right-hand column the $xz$ plane
is presented. Colour shades maps the intensity of the magnetic field (the colour scale
is the same for all the panels). Field lines, indicating the direction of the
magnetic field, are also shown.}
\label{fig:clusterslices}
\end{figure}

This trend is confirmed by Fig.~\ref{fig:clusterpprofile}, in which we present 
\rev{thermal pressure, magnetic pressure and kinetic energy density profiles} 
for the three haloes. This figure is analogous to 
Fig.~\ref{fig:Bstackedpressure}, the only difference being that the profiles are 
not resulting from a stacking but computed individually for each halo. Again we 
see that the behaviour of the \rev{three profiles is very similar. 
Interestingly, and in contrast to what we have found for smaller haloes (see 
Fig.~\ref{fig:Bstackedpressure}), the kinetic energy density is in general below 
the gas thermal pressure, except at distances larger than the virial radius. The 
ratios between magnetic and thermal pressure (black line) and between magnetic 
pressure and kinetic energy (grey line) are roughly flat up to the virial 
radius. The ratio between magnetic and thermal pressure} is approximately equal 
to $10^{-2}$, in agreement with galaxy cluster observations \citep{Feretti1995, 
Feretti1999, Murgia2004, Guidetti2008, Bonafede2009}, intracluster medium 
heating models \citep{Kunz2011}, and earlier numerical results 
\citep{Dubois2008,Bonafede2011,Ruszkowski2011,Vazza2014}. \rev{The ratio between 
magnetic pressure and kinetic energy density is larger and in general in the 
range $\sim 0.1-0.4$. We note this range agrees well with numerical work on 
small-scale dynamos \citep[see e.g.][and references therein]{Sur2012}.} Past the 
virial radius the magnetic pressure drops abruptly and becomes negligible with 
respect to the gas thermal pressure \rev{and kinetic energy density}. Consistent 
with the results of the previous section, magnetic fields do not play a major 
role in the dynamics of the gas within these systems.

To give a better idea of the magnetic field configuration within haloes, we show 
in Fig.~\ref{fig:clusterslices} two-dimensional slices of the redshift zero 
magnetic field strength (colour shades) to which magnetic field lines have been 
superimposed. We selected the third most massive halo of the simulation 
box-256-fp (top row) and to highlight the differences with lower overdensity 
environments we contrast its magnetic field properties with those of a 
filament taken from the same simulation (bottom row). Slices are repeated twice 
for each object in the $yz$ (left-hand column) and $xz$ (right-hand column) planes.

It is evident that the magnetic field properties in the two environments are
very different. First of all, the magnetic field reaches a greater level of
amplification within the halo, while the largest value of the field in the
filament is comparable to that found in the halo outskirts (the colour scale is
the same for all the panels). Even more remarkable is the difference in the
morphology of the {\it B} field in the two cases. We start by noting that all the
slices show the $z$-direction, i.e. the initial direction of the seed field. It
can be seen that outside of the cluster the magnetic field retains this
original direction, while it tends to align along the paths where matter is
accreted on to the halo. Inside the halo the orientation of the magnetic field
is more chaotic, and its coherence scale is considerably smaller than the
typical halo size. This is consistent with the picture that magnetic field
amplification is driven by turbulent gas motions -- originated by gravitational
dynamics and the stirring of gas by galactic and AGN outflows -- within
collapsed structures. For the filament this turbulent reordering of the
magnetic field lines is not present and the magnetic field retains its initial
direction, which is also the main direction along which the filament develops.
In the regions where the filament deviates from this direction also a deviation
of the magnetic field is visible.  However, the largest component of the
magnetic field is always oriented along the $z$-axis and for scales that can also
reach $\sim$ tens of Mpc. In conclusion, even though the initial orientation
(and strength) of the seed field is unimportant for
the properties of magnetic field within haloes, this orientation is retained on
larger scales and in low overdensity regions. We will discuss further this
aspect in Sect.~\ref{sec:discussion}. 

\subsection{Faraday RM} \label{sec:faradayrot}

\begin{figure}
\centering
\includegraphics[width=0.48\textwidth]{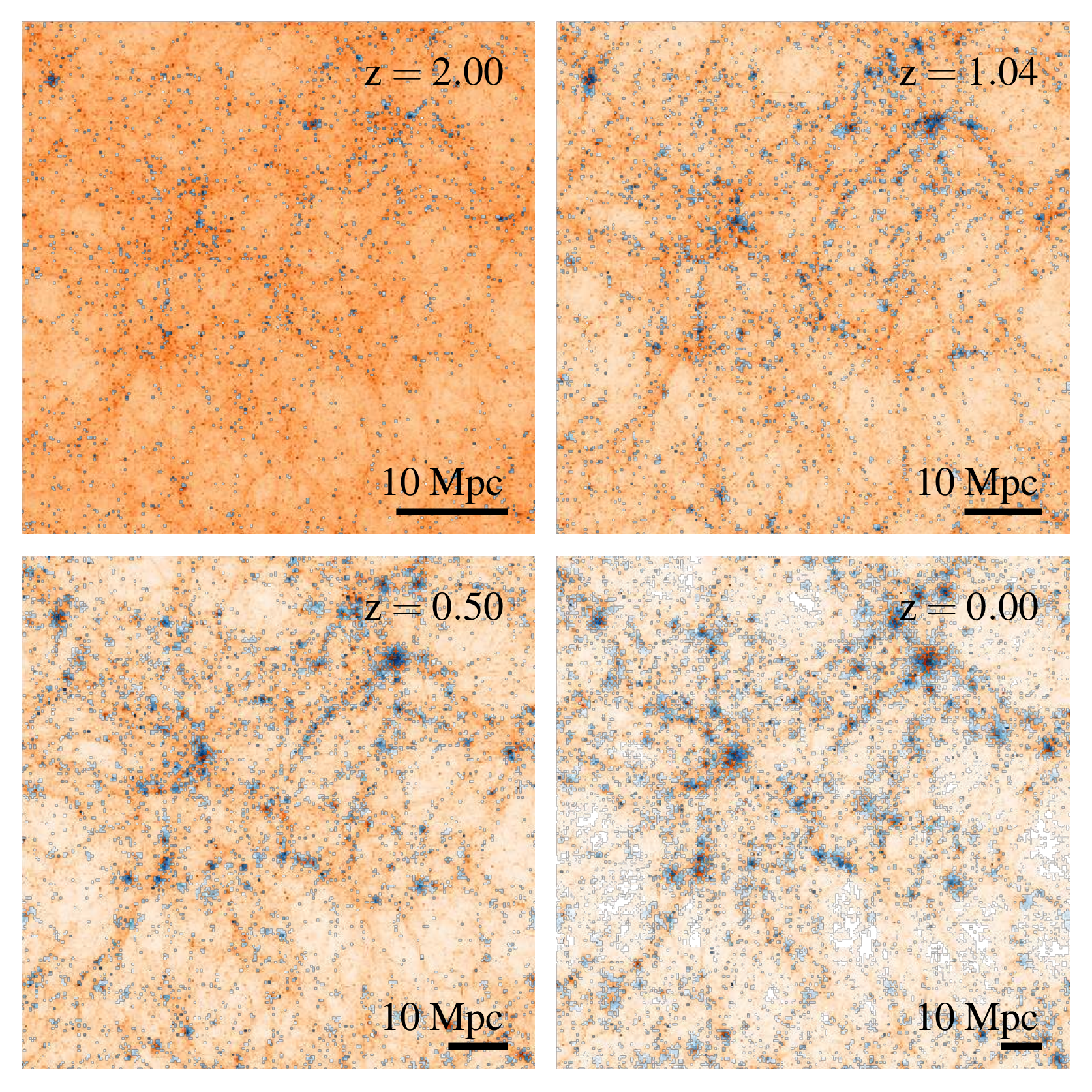}
\caption{RM map at different redshifts, as indicated in the top 
right corner of each panel, for the simulation box-512-fp-high. Each panel is 
$100\,\,h^{-1}{\rm Mpc}$ on a side (in comoving units), the full extent of the 
simulated box. The centre of the projection region corresponds to that of the 
simulated domain. The plots have been obtained by considering all the gas cells 
along the $z$-axis (the initial direction of the seed field) within $25\,\,h^{-1}{\rm 
Mpc}$ (in comoving units) from the centre, for a total thickness of 
$50\,\,h^{-1}{\rm Mpc}$. The physical scale at the corresponding redshift is 
indicated on the bottom right corner of each panel. The colour scheme is the 
same for all the panels and maps logarithmically the absolute value of the 
RM in the interval $[10^{-6}, 10^2]\, {\rm rad\,m^{-2}}$ in orange 
shades for (originally) positive values and in blue shades in the opposite case.} 
\label{fig:faradayrotationfullphys}
\end{figure}

\begin{figure*}
\centering
\includegraphics[width=0.32\textwidth]{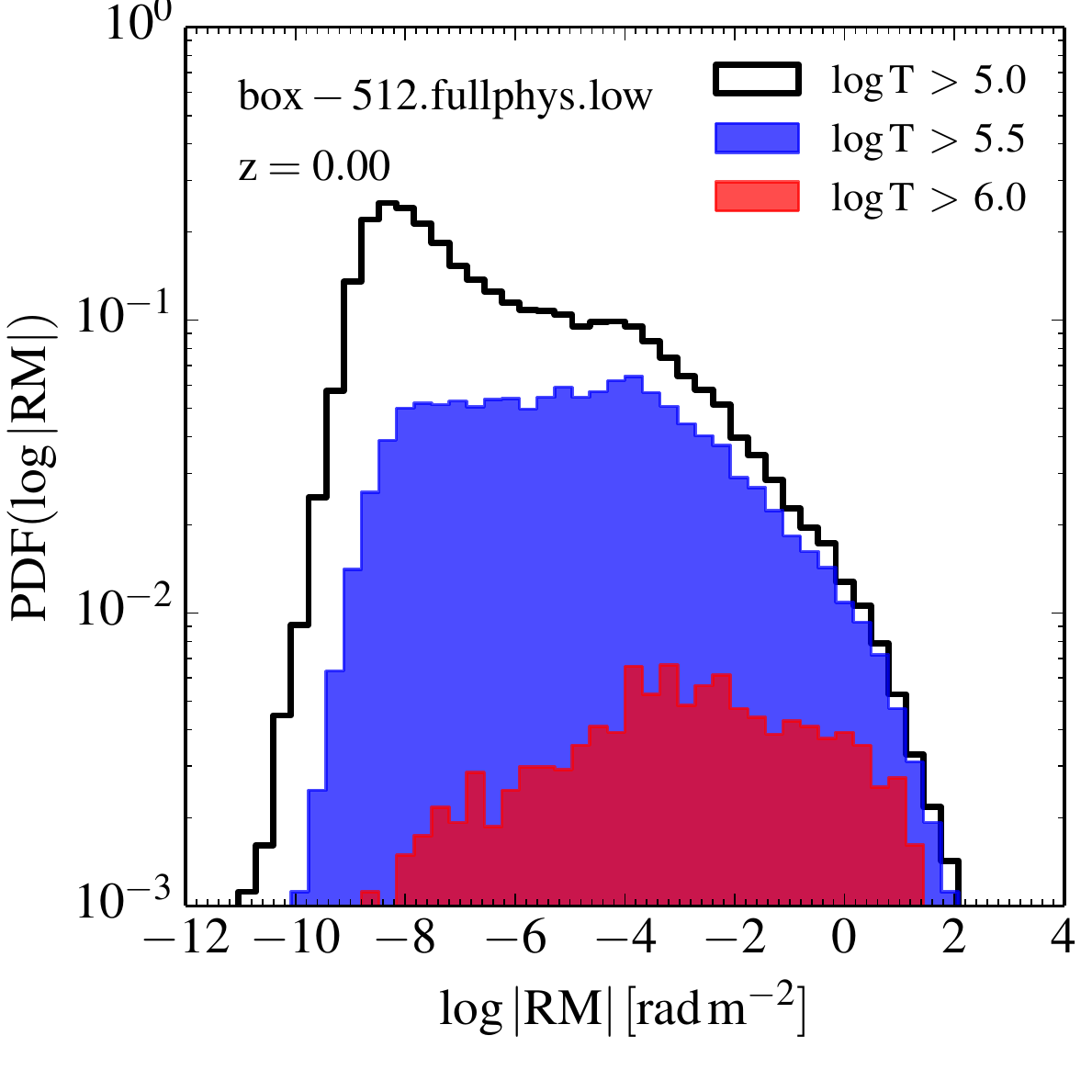}
\includegraphics[width=0.32\textwidth]{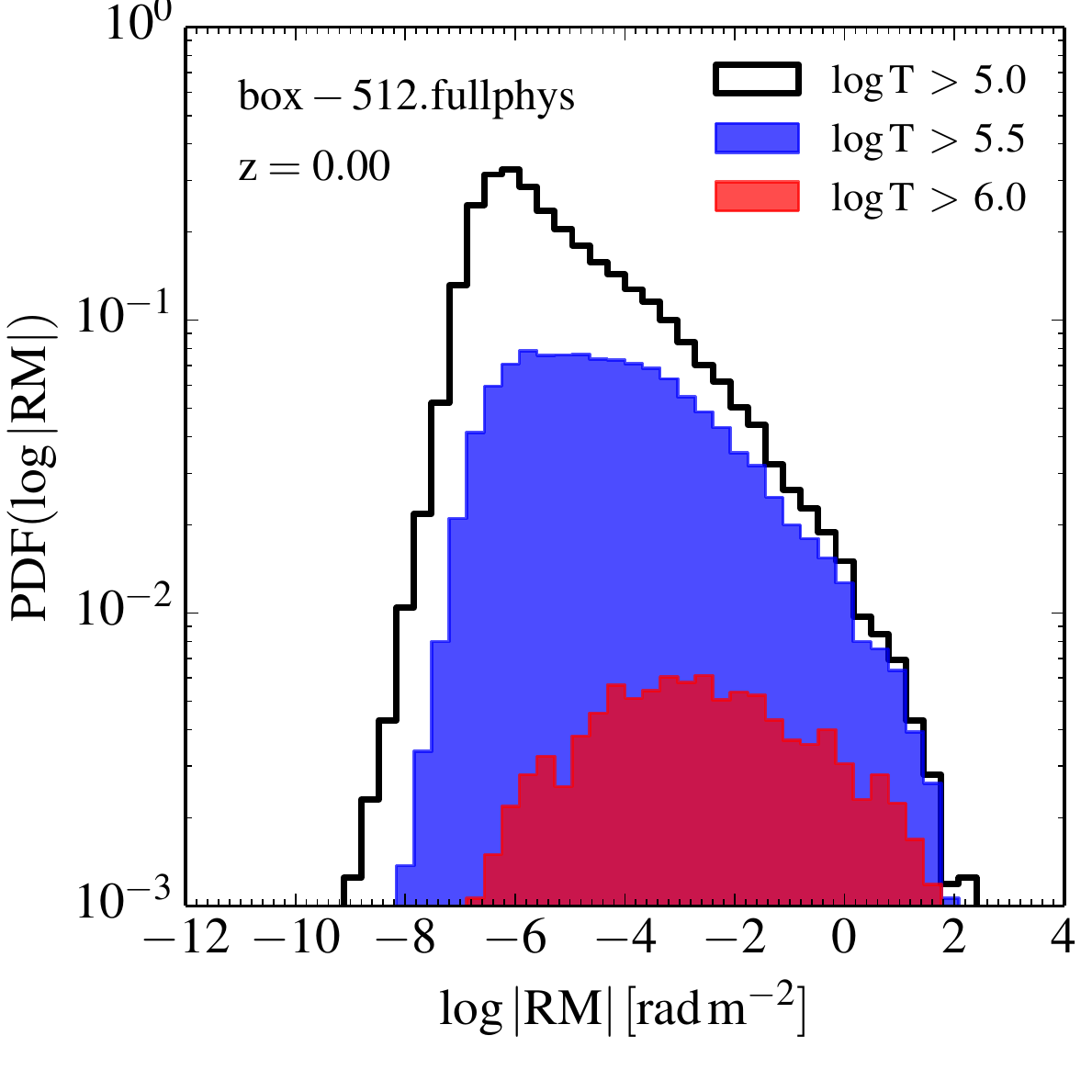}
\includegraphics[width=0.32\textwidth]{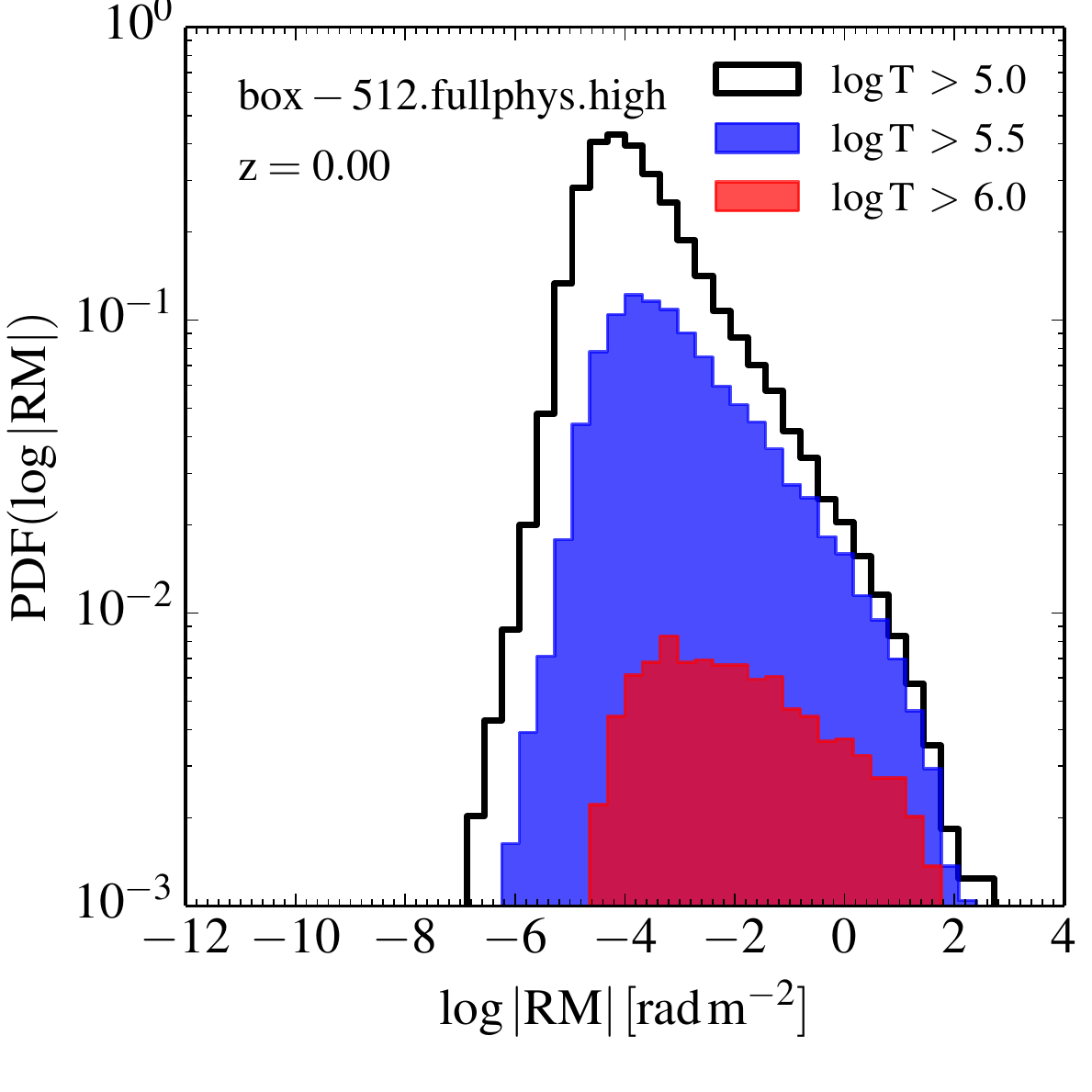}
\caption{RM PDF at redshift zero for the simulations
box-512-fp-low, box-512-fp, and box-512-fp-high (from left to right).
The PDFs are calculated for the whole box projecting along the z axis,
the initial direction of the {\it B} seed field. 
Colour shades show the contribution to the total PDF (black line), computed for
temperatures above $10^5\,\K$, of gas in the temperature ranges indicated in the
legend of each panel. Similarly to Fig.~\ref{fig:BfieldPDFseed}, increasing the
seed field strength shifts the peak of the PDF at low RM values,
while leaving the tail at high RM almost unaffected.
} 
\label{fig:farrotPDF}
\end{figure*}

A powerful tool to indirectly derive magnetic field strengths comes from the
observation of radio sources embedded in or behind massive structures
like galaxy clusters. The magnetized plasma within these structures
is in fact a birefringent medium. When polarized radiation, such as the
linearly polarized synchrotron radiation emitted by a radio source,
passes through it, its polarization vector rotates by an amount
proportional (through $\lambda^2$) to the so-called Faraday rotation 
measure (RM), defined as
\begin{equation}
 \mathrm{RM} = \frac{e^3}{2\pi m_{\rm e}^2 c^4}\int_0^L n_{\rm e}(s)B_{\parallel}(s)\,{\rm d}s,
 \label{eq:rotmeasure}
\end{equation}
where $e$, $m_{\rm e}$ and $c$ are the electron charge, electron mass, and
speed of light, respectively; $n_{\rm e}$ is the electron density and $B_{\parallel}$
is the component of the {\it B} field along the line of sight ${\rm d}s$. By measuring the
shift of the polarization vector at different wavelengths (at least 3 to avoid 
degeneracies), it is then possible to determine the value of the RM and from
it to estimate the magnetic field strength along the line of sight, assuming a
distribution for the electron density.

This method is rather indirect and a number of assumptions enter into it in
order to obtain the {\it B} field strength -- which as such can only be considered an
average along the line of sight.  Nevertheless, comparing the RM
predicted by the simulations with the observed values is extremely useful since
it can constrain several aspects of our calculations. In particular, we can get
further indications about the thermal state of gas in massive structures and
the intensity of the magnetic field there, since both the electron density and
the magnetic field are present in the definition of the RM (see
equation \ref{eq:rotmeasure}). Moreover, RM can also inform on the coherence scale
of the magnetic field. Since the RM is an integral quantity, coherent {\it B} fields
give rise to a strong RM signal, while a more chaotic arrangement of the field
weakens the signal because contributions to the path integral tend to cancel
out. It is worth noting that some degeneracy between magnetic field coherence
and intensity is unavoidable in the final value of the RM: larger but more
chaotic {\it B} fields can give rise to the same signal of weaker but more ordered
field configurations.

In Fig.~\ref{fig:faradayrotationfullphys}, we present RM measure maps for the
simulation box-512-fp-high at different redshifts. The maps have been obtained
by projecting along the $z$-axis (the initial direction of the seed field) the
same region as in Fig.~\ref{fig:Bprojection}.
The colour scheme is the same in all panels, and maps logarithmically the
absolute value of the RM. Since the RM can change sign depending on the
predominant direction of the {\it B} field we use two different colour shades (orange
for positive values and blue for negative ones) to encode this information.   

\begin{figure*}
\centering
\includegraphics[width=0.32\textwidth]{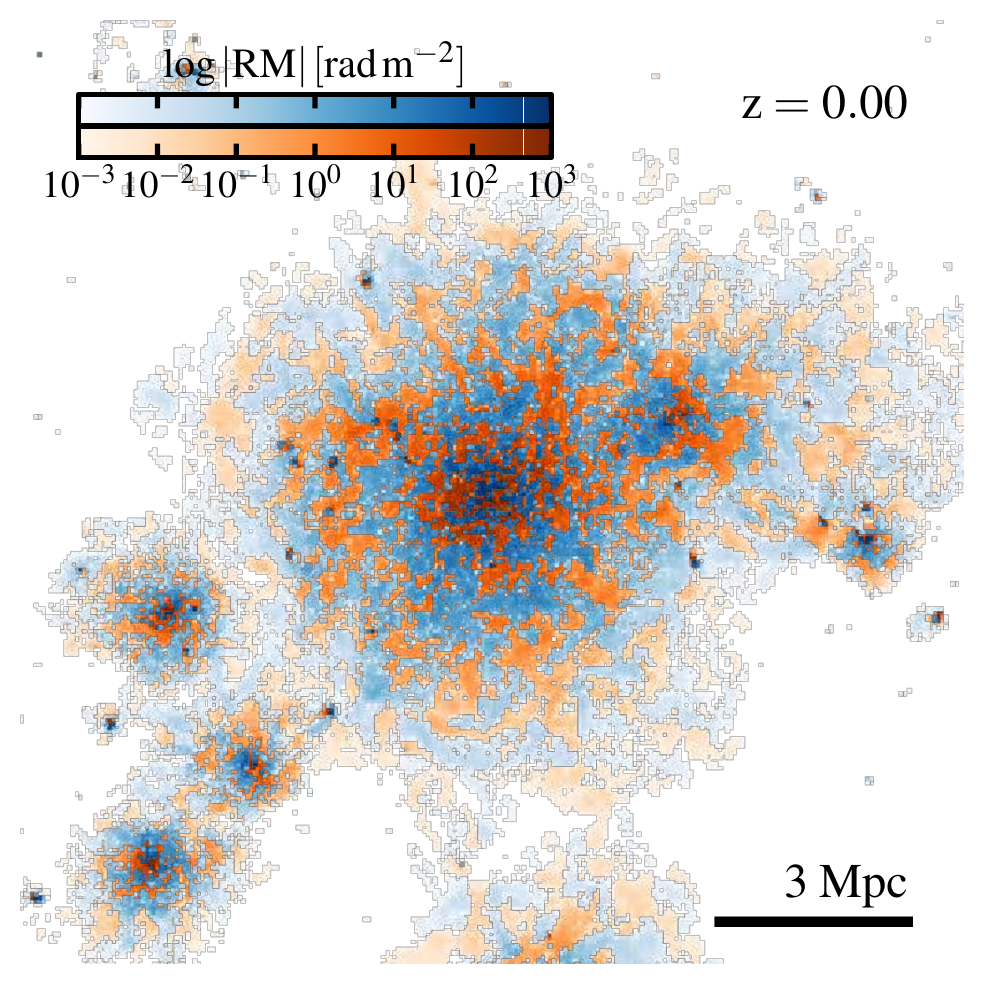}
\includegraphics[width=0.32\textwidth]{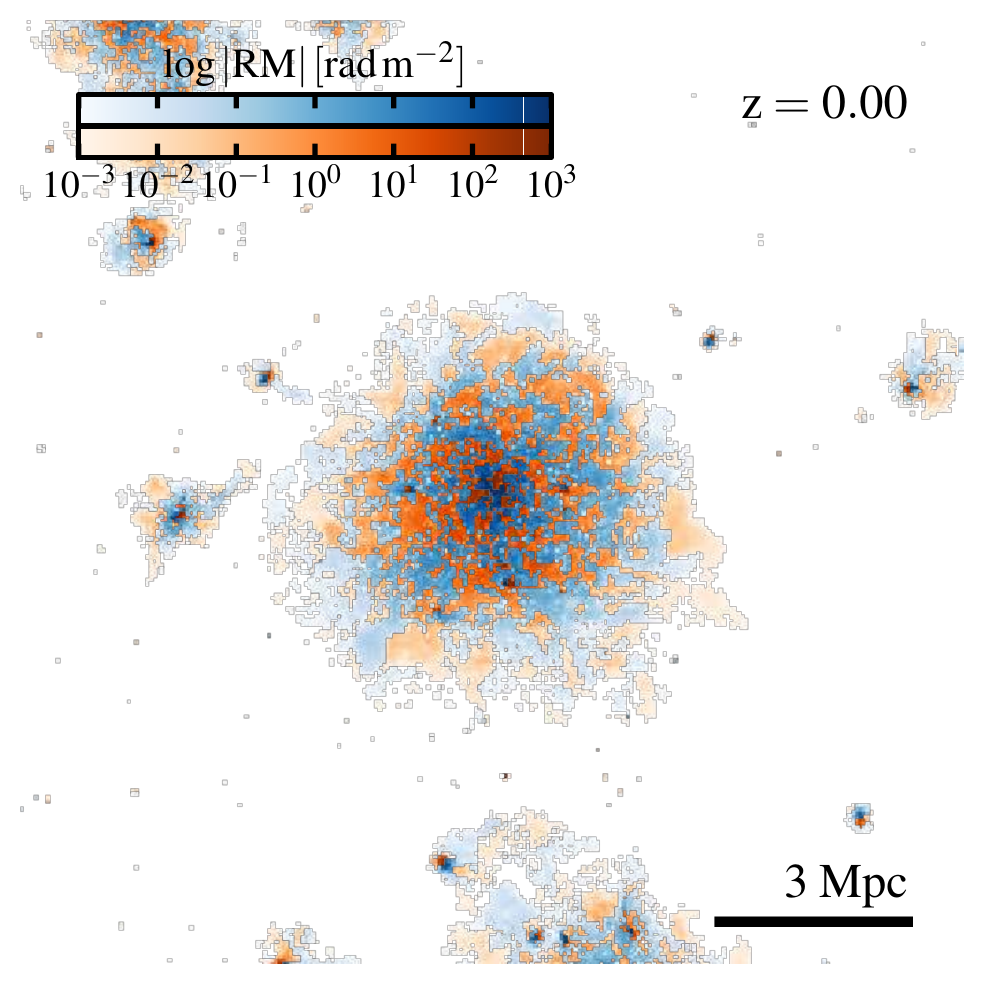}
\includegraphics[width=0.32\textwidth]{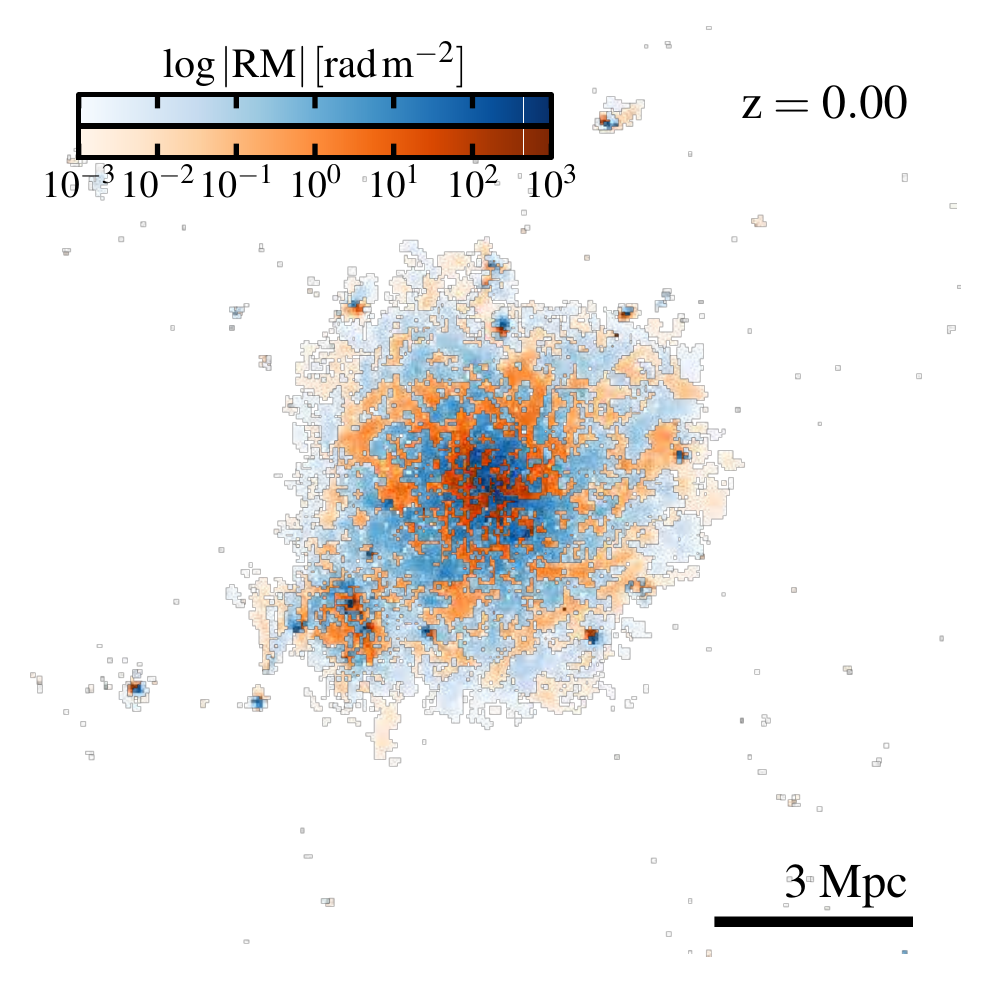}
\caption{RM maps for the three most massive haloes of 
simulation box-512-fp. The maps have been calculated for a cubic box of 15 Mpc 
on a side, centred in the potential minimum of the three haloes. The projection 
direction corresponds to the $z$-axis, the initial direction of the magnetic 
field. Similarly to Fig~\ref{fig:faradayrotationfullphys}, the colour scales 
adopt a logarithmic mapping of the absolute value of the RM with 
orange shades representing positive and blue shades negative values. It is clearly
visible the tangled structure of the magnetic field within the haloes due to
its turbulent amplification.} 
\label{fig:faradayrotationcluster}
\end{figure*}

\begin{figure*}
\centering
\includegraphics[width=0.32\textwidth]{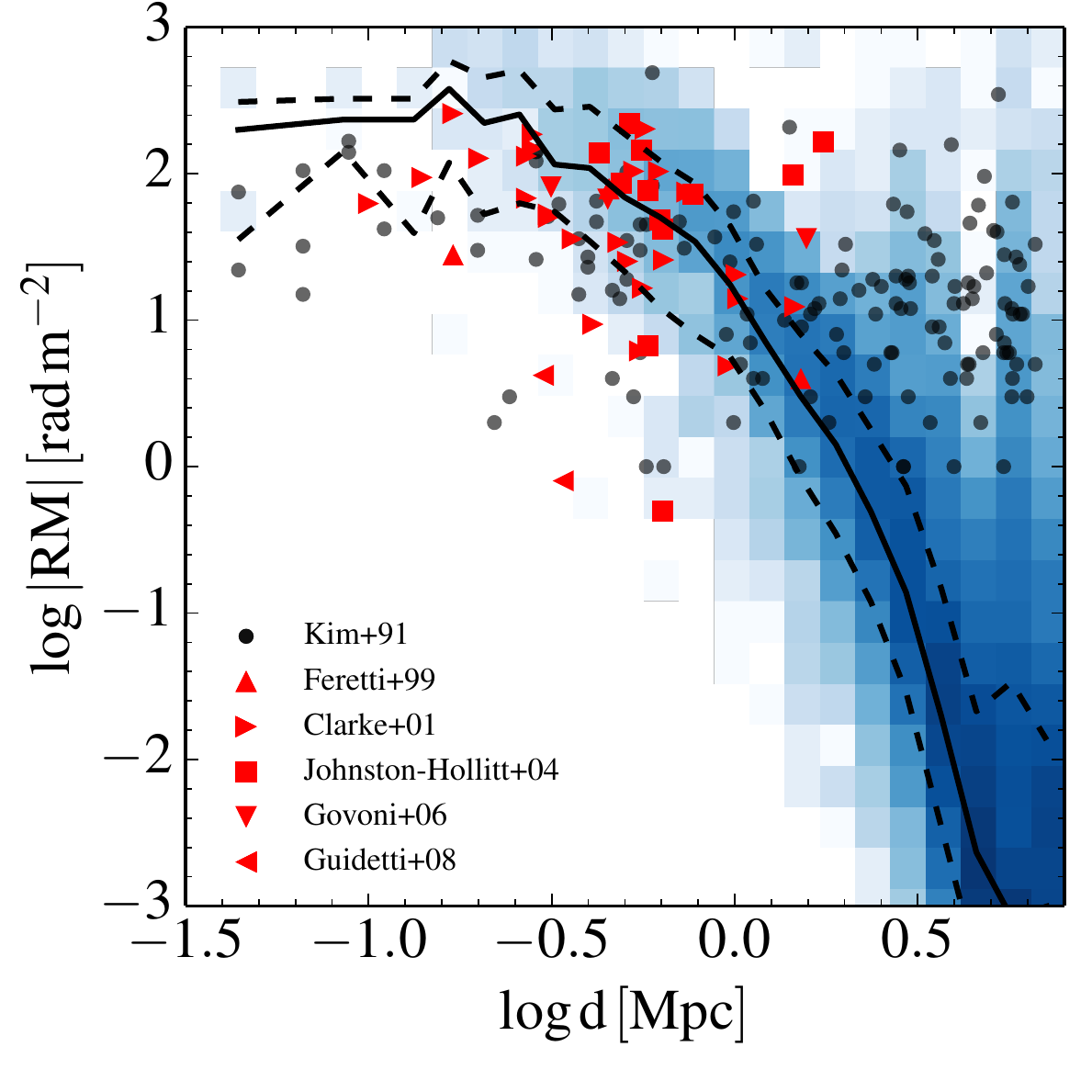}
\includegraphics[width=0.32\textwidth]{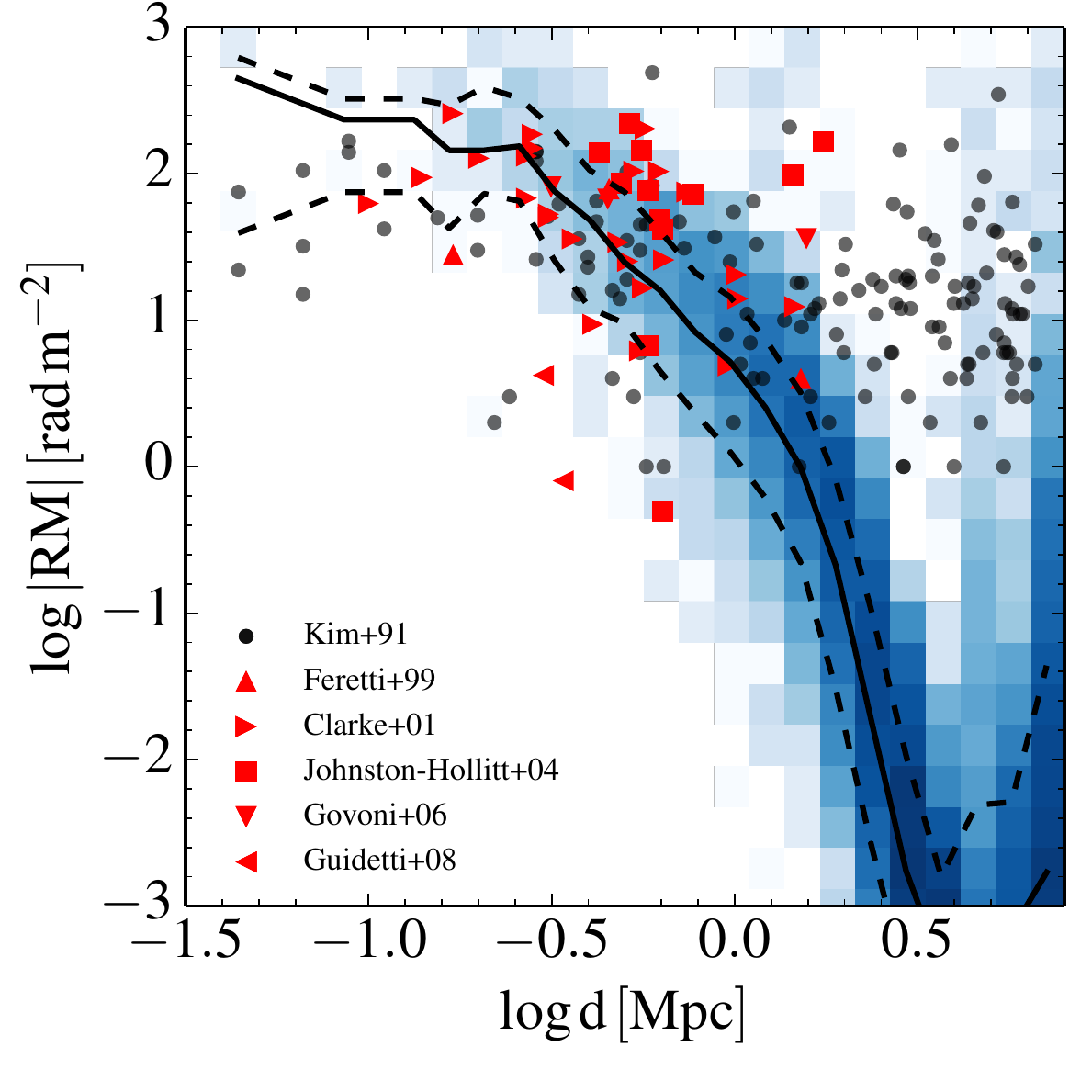}
\includegraphics[width=0.32\textwidth]{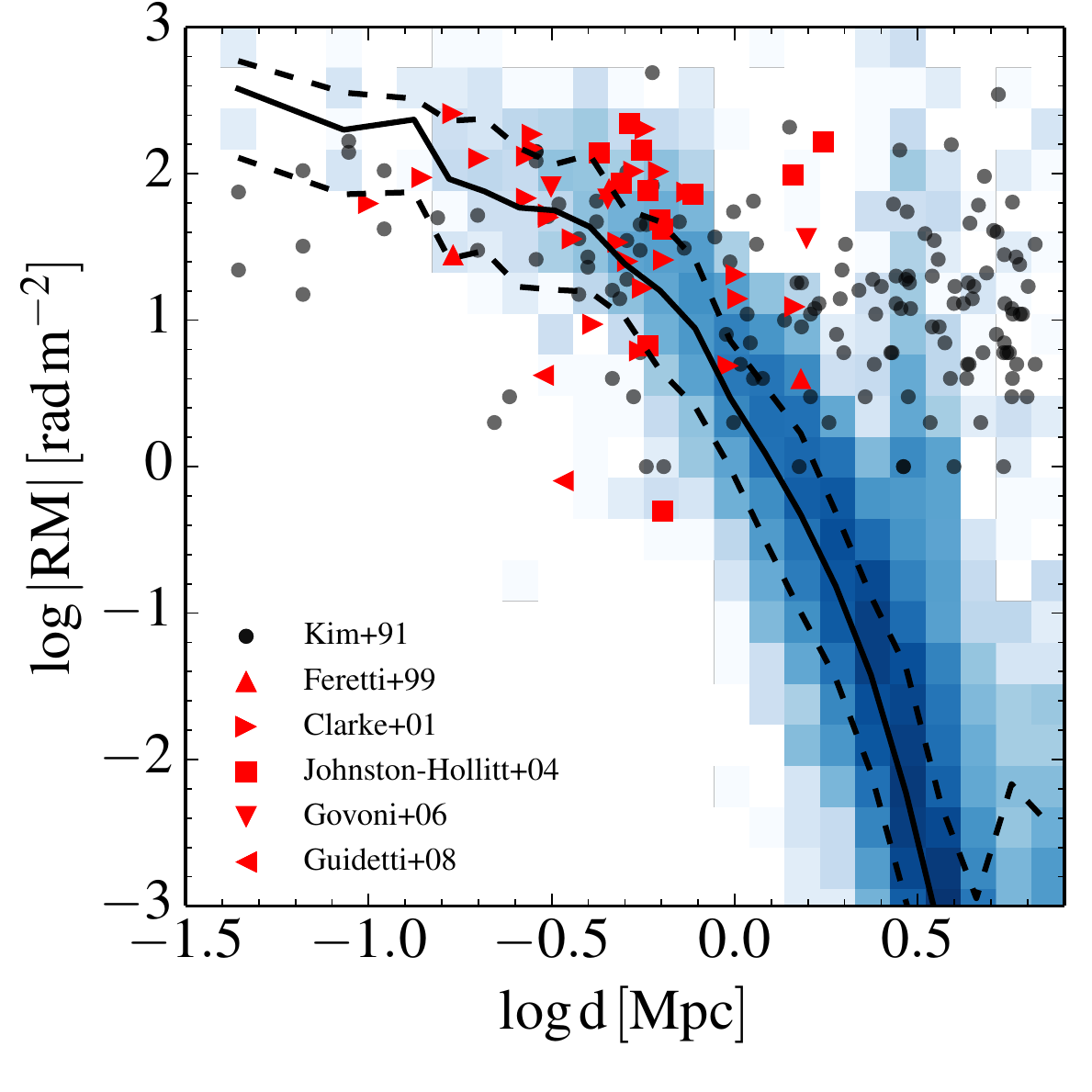}
\caption{Radial profiles, in terms of the projected distance 
from the halo potential minimum, of the RM maps 
in Fig.~\ref{fig:faradayrotationcluster}. Colour shades represent the number of 
pixels falling into each radial and RM bins. Solid black lines indicate the median
trend while dashed black lines show the 16th and 84th percentile of the distribution.
Coloured symbols are actual measurements of RM in 
galaxy clusters (references are indicated in the panel legends). It is 
evident the exponential drop of the RM signal as a function of radius.} 
\label{fig:farrotprofilescluster}
\end{figure*}

From the figure it can be seen that from a rather homogeneous map at high
redshift, the RM signal becomes more and more structured with time. The largest
values of RM tend to follow quite closely the assembly of cosmological
structures, and at redshift zero they tend to concentrate on collapsed
structures and filaments. The large-scale signal and in particular the
contribution from voids declines with time, due to a general decrease of the {\it B}
field caused by the cosmological expansion. Within collapsed structures, and
along the filaments connecting them, positive and negative values of the RM
signal co-exist. Since we chose your projection axis to coincide with the
initial direction of the seed field, negative RM values signal a field
reversals from the original configuration. These reversals are very frequent
inside structures where shear flows and turbulence are more effective in
amplifying magnetic fields, in agreement with the results discussed in
Sect.~\ref{sec:haloexamples}. 

In Fig.~\ref{fig:farrotPDF} we show RM PDFs at redshift zero 
for different full physics simulations (with resolution $2\times 512^3$) 
with increasing seed field strengths (from left to right). To generate the 
PDF the whole computational domain is projected along the $z$-axis, and only
gas with temperature above $10^5\,\K$ is considered to exclude the
contribution of neutral or partially ionized gas. We also show in each 
panel the contribution to the total PDF (black lines) of gas with temperatures above 
$10^{5.5}\,\K$ (blue shades) and $10^{6}\,\K$ (red shades). The PDF is computed
for the absolute value of the RM. We explicitly checked that the choice of the 
projection direction does not alter the properties of the resulting PDFs.

In agreement with the results in Sect.~\ref{sec:seed field}, changing the
intensity of the seed field strongly alters the PDF at lower values of RM. The
peak of the PDF is shifted by approximately two orders of magnitude in each
panel -- the amount by which the seed field is varied. Low values of the RM are
found in low-density regions and this indicates that the RM signal in these
regions is very sensitive to the initial {\it B} field strength. In low-density
regions low or little (turbulent) {\it B} field amplification, which also set the
coherence scale of the field, is expected. Hence, this sensitivity of the PDF
to the value of the seed field is ultimately related to the intensity that the
magnetic field can attain. We showed earlier that at low baryon overdensities
full physics and adiabatic runs have a similar behaviour, and that the
intensity of the seed field basically sets the final {\it B} field strength.

At the opposite end (i.e. for large RM values) the dependence of PDF on the
seed field disappears and the distributions look rather similar. This is again
consistent with our earlier findings on the saturation of the amplification of
the {\it B} field at high baryon overdensities. The maximum value of the RM shows a
little residual dependence on the seed field strength, perhaps due to a more
coherent field configuration for larger seed field value that slightly
increases the RM signal. Temperature seems to play a secondary role for the
properties of the RM PDF, but this can be just a reflection of
the fact that neutral and partially ionized gas has been excluded from the
calculation of the PDFs. Not surprisingly, gas at $T > 10^{5.5}\,\K$ has a
broader distribution than gas selected with a larger temperature cut ($T >
10^{6}\,\K$). The latter also peaks approximately at $\log|{\rm RM}| \sim -3\,
{\rm rad\,m^{-2}}$ independently of the seed field value, while the peak in the
distribution of colder gas (which can be also located in lower density regions)
shifts towards higher RM values with increasing seed field strength.

In Figs~\ref{fig:faradayrotationcluster} and \ref{fig:farrotprofilescluster} we 
analyse the RM signal of the three most massive haloes of the simulation 
box-512-fp (see also Sect.~\ref{sec:haloexamples}). We start, in 
Fig.~\ref{fig:faradayrotationcluster}, with the Faraday RM maps of 
the three haloes at redshift zero. The maps have been obtained by integrating 
along the $z$-axis equation (\ref{eq:rotmeasure}) in a cubic box of 15 Mpc on a side,
centred on the haloes potential minimum and aligned with the simulation box. 
Again, the particular choice of the projection 
direction leaves essentially unaltered the main features of the maps (see also 
the top row of Fig.~\ref{fig:slicesdiscussion}). We use the same colour
scheme of Fig.~\ref{fig:faradayrotationfullphys} and the resulting value of the 
RM (in ${\rm rad\,m^{-2}}$) is indicated on the colour bar on the top-left corner 
of each panel.

It is clearly visible that the three haloes, although quite different in mass, 
show rather similar features in their RM measure maps. First of all, RM signal 
extends roughly in the same region where a significant {\it B} field is present 
(compare the maps with the third column of Fig.~\ref{fig:clusterproj}). This is 
not surprising since the dependence on the electron density is secondary here 
(due the high temperature the gas is fully ionized). However, these maps also 
clearly show the tangled structure of the magnetic field within the haloes as a 
results of its turbulent amplification. Regions with coherent sign of the RM can 
be as small as a few pixels ($\sim$ 50 kpc at the current map resolution), which 
is a much lower value than the virial radii of the haloes ($\sim$ few Mpc). Values 
of the RM as high as $10^{3}\,{\rm rad\,m^{-2}}$ are not uncommon in the halo 
centres, but the signal declines considerably with radius and drops very rapidly 
outside the virial radius, closely mimicking the behaviour of the magnetic 
field. Local enhancements of the RM signal due to infalling substructures are 
also present. 

To study more closely the radial variation of the RM, in
Fig.~\ref{fig:farrotprofilescluster} we present radial profiles of the maps
above as a function of the projected distance from the halo centre. Colour
shading shows the number of pixels falling into each radial and RM bins, while
black solid lines show median trends and black dashed lines the 16th and 84th
percentiles. Symbols show observed values of RM detected in
galaxy clusters \citep{Kim1991, Feretti1999, Clarke2001, J-Hollitt2004,
Govoni2006, Guidetti2008}. As inferred from the maps, RM is a rapidly declining
function of distance. From the median relation we can infer that this decline
is almost exponential and little signal is left past the halo virial radius.
Only for the most massive halo RM bins are significantly populated for
$r\gsim\,r_{200}$. A comparison with the observations shows that our
simulations predict the correct RM intensity, especially in the halo central
regions. However, our predictions are marginally inconsistent with one of the
data set \citep{Kim1991}, which features a more extended distribution of RM in
the radial direction. Only the most massive halo of the simulation, thanks to
the RM signal originating from the prominent substructures that surround the
main body of the halo, can partially predict this signal at radial distances
comparable to the virial radius and beyond. Note, however, that in the 
previous analysis we did not include the Galactic and extragalactic foreground 
contributions to the RM signal. Outside the Galactic plane these contributions 
can account for $\sim 6-8\,\,{\rm rad\,m^2}$ \citep{Schnitzeler2010}, thus 
rendering the discrepancy between the observed and predicted RM radial profiles 
less severe.

\section{Discussion} \label{sec:discussion}

An essentially arbitrary aspect in cosmological ideal \mhd\ simulations
is represented by the freedom in the choice of the initial seed field. The
chosen initial field configuration must be divergence free, but except for this
requirement only scant and somewhat uncertain observational constraints on the
magnetic field strength are currently available in very low density
environments or at high redshift \citepalias{Neronov2010, Planck2015}. Also,
theoretical models of cosmic magnetogenesis come in a variety of flavours with
different predictions for both the intensity and the configuration of
primordial fields \citep[see][and references therein]{Widrow2012}.  Numerical
simulations can potentially exploit this freedom to constrain theoretical
models by testing their predictions for the resulting magnetic field properties
at large baryon overdensities, where observational limits are more stringent
and abundant. 

In our simulations we have chosen the simplest divergence-free field
configuration: a uniform magnetic field oriented along the $z$-axis of the
simulation box. We have extensively discussed how the strength of the magnetic
seed field influences the results of non-radiative and full physics runs. We
concluded that, while in non-radiative runs the seed field intensity provides
an overall normalization factor to the values of the final magnetic field
strength, in full physics runs magnetic field within structures always reached
\rev{roughly the same maximum} value \textit{regardless} of the initial seed field, whose
strength is consistent with the observational constraints cited above.  The
magnetic field loses memory of its initial configuration even for large seed
field variations \citep[see also][]{Pakmor2014}.  Although this is desirable
feature from a numerical modelling point of view, because results are
independent of the particular choice of the seed field, it makes simulations a
less stringent test for cosmic magnetogenesis models.  Different authors, using
different numerical techniques, baryon physics prescriptions and seeding
strategies find similar results \citep[e.g][]{Dolag1999, Donnert2009, Xu2010}.
The situation is more promising in less dense regions such as filaments and
voids, where the dependence on seed field is stronger, and allows us to
discriminate among different scenarios \citep[][]{Ryu1998,Donnert2009,Dubois2010,
Akahori2011,Akahori2014}. However, we stress again that numerical
simulations are confronted here with the scarcity of data although this will
improve with the current and forthcoming generation of radio instruments, such
as LOFAR and SKA, which are potentially able to detect magnetic
fields of the order of $1\,\nG$ or below \citep{Beck2007, Pritchard2015,
Sethi2009, Vazza2015b}.

\begin{figure}
\centering
\includegraphics[width=0.23\textwidth]{fig19a.pdf}
\includegraphics[width=0.23\textwidth]{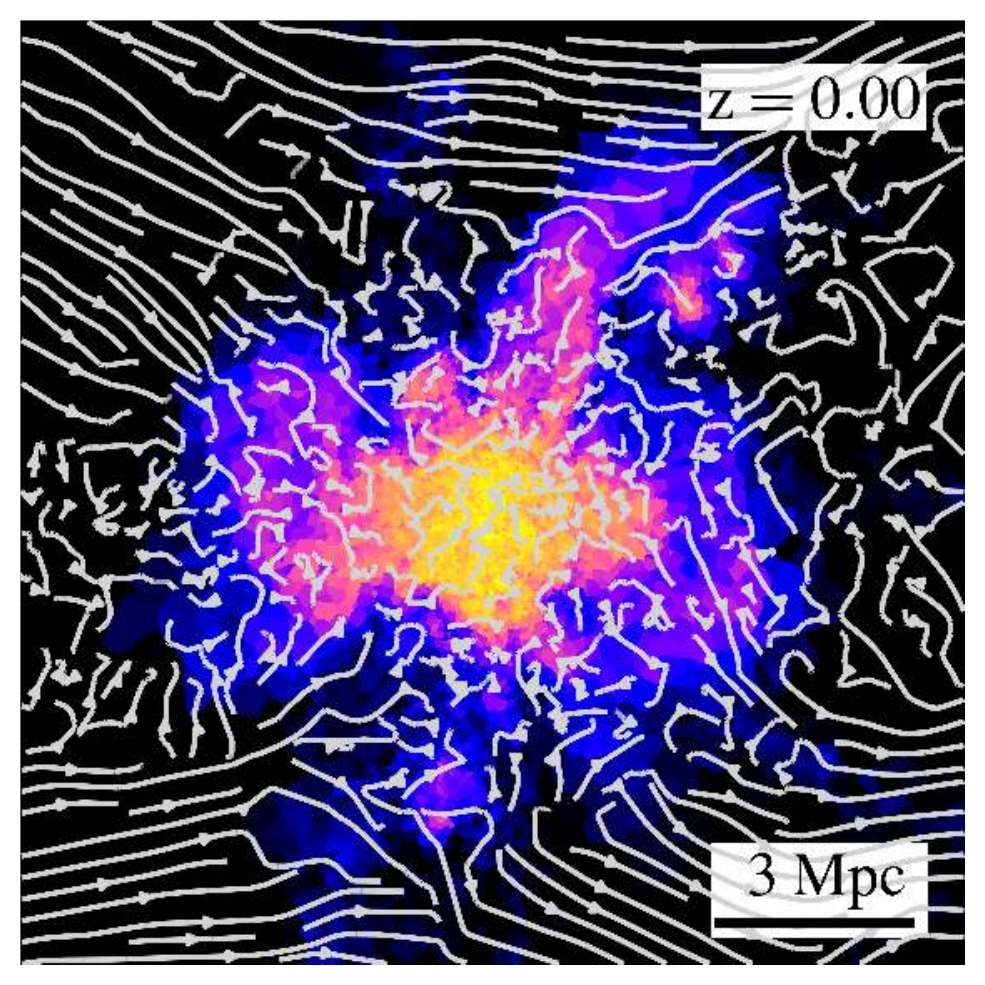}
\includegraphics[width=0.23\textwidth]{fig19c.pdf}
\includegraphics[width=0.23\textwidth]{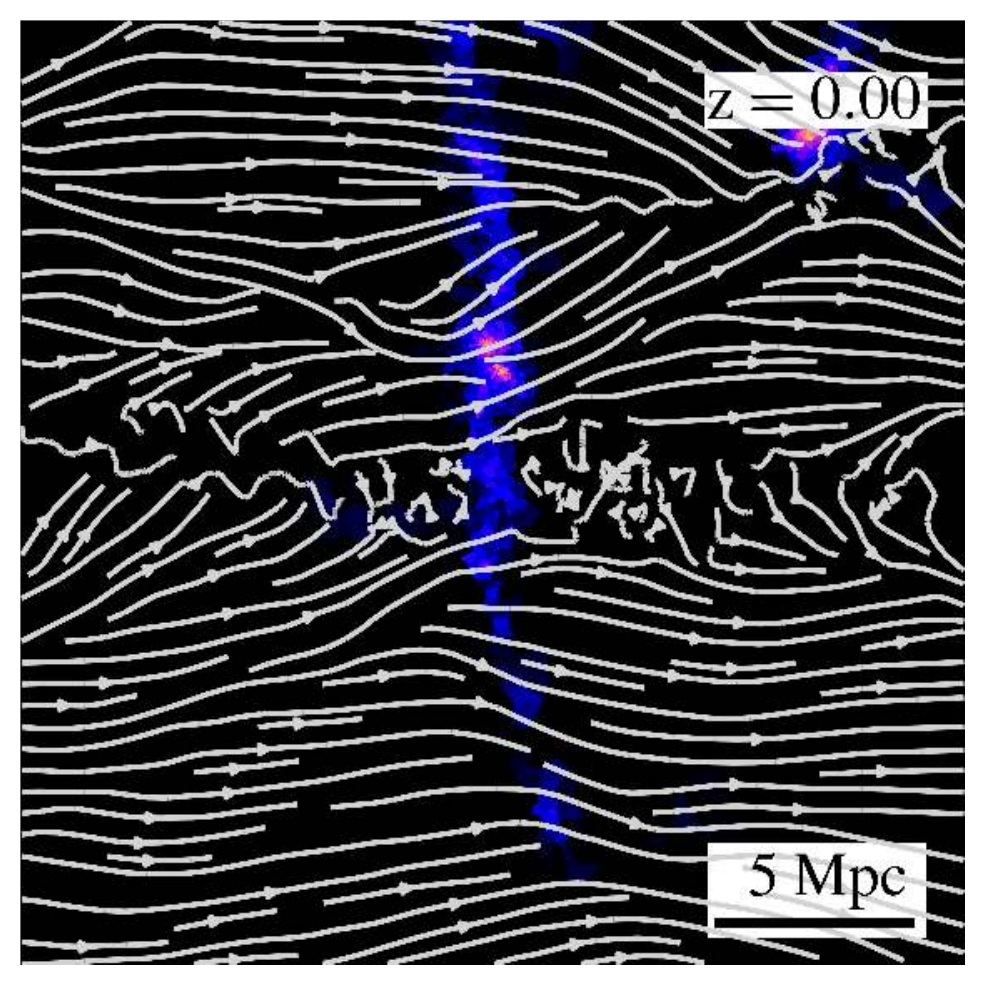}
\caption{The same as Fig.~\ref{fig:clusterslices}, but showing the simulation
box-256-fp with seed field oriented along the $z$-axis (left-hand column), 
and the simulation box-256-fp-ydir with exactly the same set up but 
seed field aligned with the $y$-axis (right-hand column). For all the slices only 
the $yz$ plane is displayed.}
\label{fig:slicesdiscussion}
\end{figure}

As an example of how low density environments retain a stronger imprint of 
the initial magnetic seed field, we repeat in Fig.~\ref{fig:slicesdiscussion} 
the analysis carried out in Fig.~\ref{fig:clusterslices}. We present
two-dimensional slices of the redshift zero magnetic field and its orientation
on the $yz$ plane, but for a seed field with initial direction along
the $z$-axis (left-hand column) and the $y$-axis (right-hand column), respectively. 
The slices corresponding to the same object are centred at exactly the same
point for both field directions. Hence, as all the other parameters are not
varied between simulations, the differences in magnetic field strength and 
orientation are due only to the different choice of the initial direction 
of the seed field.

The agreement between the magnetic field intensities in the two runs is very 
good. In the halo, the maximum strength reached by both simulations is 
essentially the same and also the characteristic tangled orientation of
the field is present. Minor differences are also visible, as for instance a more
developed magnetic filament in the upper-right part of the halo in the case
of the $y$-axis oriented seed field or a larger magnetized substructure in the
lower region for the simulation with the default magnetic field orientation. 
Also, the magnetic field in the halo appears to be more extended in the $y$-direction 
in the latter case. The most striking difference, however, is in the orientation
of the field just outside the halo, which is strongly aligned with 
the seed field initial direction in regions where matter accretes on to
the virialized structure as well.

The effect is even stronger for the filament. Here,  the prevailing direction
of the {\it B} field is that of the initial seed, as field lines flip by $90^{\circ}$
between panels. The magnetized structure in the upper-right corner looks very
similar in both cases, while the filament, although conserving the overall
shape, shows a larger magnetization for the default {\it B} field orientation. This
is because the magnetic field amplification is the result of gas compression in
the filament, which mostly occurs perpendicularly to the filament itself. In
default magnetic field orientation field lines are perpendicular to the motion
of gas and are thus dragged along with it. This dragging results in a stronger
amplification effect than in the other simulation where one of the gas velocity
component is aligned with the magnetic field lines which, as a result, are
compressed to a lesser degree. The final maximum {\it B} field strength in the
filament is $\sim 1\,\nG$ for the default orientation of the seed field, and
about a factor of 3 less for the $y$-axis orientation. These results are in
agreement with \cite{Brueggen2005} and \cite{Vazza2014} 
findings\footnote{\cite{Vazza2014} used a seed field strength $10^{4}$ larger that our
default choice, but also the more diffusive {\sc HLL} \citep{Harten1983} Riemann 
solver that could have in principle reduced the {\it B} field amplification.}, 
and about a factor of 10
smaller than the values quoted by \cite{Akahori2010}, \cite{Akahori2011}, and
\cite{Ryu2010}, who however did not follow the field evolution
self-consistently in their cosmological simulations but used a prescription to
estimate the strength of the magnetic field based on \mhd\ turbulence
simulations. Note that at filament overdensities, our simulations are still
sensitive to the value of the initial seed field. Thus, the discrepancy with
simulations predicting higher {\it B} field intensities can be reconciled by choosing
a larger initial seed field.

We conclude that our general results are robust with respect to both the
orientation and the strength of the adopted seed field.  However, a more
detailed level of analysis reveals some subtle differences in the evolution of
the properties of the magnetic field, especially in low-density environments.
Also, the choice of a uniform seed field implies that the field structure in
underdense regions of the Universe is coherent on large spatial scales and it
is unclear whether this is an accurate description of primordial cosmological
magnetic fields
\citep{Beck2013b,Cho2014,Kronberg1999,Kulsrud1997,Schlickeiser2012,Tsagas2014}. 

Another important aspect of our simulations for magnetic field amplification is
their resolution. We have stressed in many places that, in addition to the gas
compression provided by gravitational collapse, shear and turbulent motions
of the gas are equally if not more important for the final amplification level
reached by the magnetic field. Even at the highest level of resolution, our
simulations contain a much lower number of resolution elements than
cosmological simulation presented in \cite{Vazza2014}. These authors find that
in order to develop a small-scale dynamo in cosmological simulations a minimum
spatial resolution, depending on the size of the simulated halo, must be
achieved (for a $10^{14}\,\mo$ halo the minimum spatial resolution
is $\sim 10$ kpc). While their results may have been affected by the choice of
a relative diffusive Riemann solver, this resolution requirement to capture
(small-scale) \mhd\ turbulence remains an important point. We note that we
indeed observe a variation in the final strength of the magnetic field as a
function of the resolution for halo masses below $10^{13}\, \mo$,
especially in the central regions. The discrepancy is less severe for larger
haloes. Also, non-radiative simulations show that the degree of amplification
of the magnetic field at large baryon overdensity is affected by resolution as
well, while the effect is less prominent in the full physics runs. We believe
that this is exactly the effect that \cite{Vazza2014} inferred from their
simulations. 

However, we would like to point out that it is difficult to do a
straightforward comparison between the maximum resolution achieved in our
calculations and those in a fixed-grid setup as in \cite{Vazza2014}. The
quasi-Lagrangian nature of \arepo, and its default refinement scheme that
keeps the mass per cell roughly constant, naturally leads to a configuration in
which the bulk of the resolution elements is put in the most dense regions
(i.e. the halo centres), which thus are better resolved than the lower density
ones. This automatic adaptive resolution allows us to follow more faithfully the
clustering of matter into cosmological structures, but it might not be the
optimal choice to capture all the thermal and dynamical processes occurring in
gaseous haloes \citep{Nelson2015}. This degradation of the (spatial) resolution
of our simulation at large distances from the halo centres, might be the reason
for the steep decline of the RM signal in the haloes examined in
Fig.~\ref{fig:farrotprofilescluster}. Simulations with better resolution
throughout the halo are needed to fully prove this point, but the fact that in
lower resolution runs the decline of the RM profiles is slightly faster can be
taken as an indication that properly capturing the small-scale \mhd\
turbulence is a crucial factor to get the correct level of amplification of the
{\it B} field \rev{\citep[see also][]{Federrath2011}}.

Finally, we would like to mention that in our numerical setup AGN feedback, and 
in particular the radio mode channel \citep[see][]{Vogelsberger2013}, does not 
provide a source term of magnetic field for the gas in the haloes of massive 
objects. Therefore, our simulations are currently neglecting a potential channel 
to seed a large fraction of the IGM (on the Mpc scale) and filaments with 
magnetic fields that are stronger than the primordial values during the quasar 
era at $z\sim2-3$ \citep{Gopal-Krishna2001, Mocz2011}, which could alleviate
the tension between the observed and the simulated values of the RM at large 
radii in massive haloes.

\section{Summary and conclusions} \label{sec:conclusions}
We have analysed the properties of magnetic 
fields in uniformly sampled cosmological box simulations performed with the 
moving-mesh code \arepo. Our analysis is based on a set of simulations that 
include magnetic fields in the ideal \mhd\ approximation. The magnetic field is 
introduced at the beginning of the simulations as a uniform seed field that is 
subsequently amplified by the formation of cosmological structures. We contrast 
the final magnetic field properties in two main types of simulations: adiabatic 
runs including only non-radiative (magneto)hydrodynamics, and full physics runs 
featuring the fiducial model of baryon physics used in the recent {\sc 
illustris} simulation suite. We repeat both adiabatic and full physics 
simulations at different resolution levels and by varying the strength and  
direction of the initial seed field in order to assess their importance for the 
final {\it B} field properties. Our main results are as follows.
\begin{enumerate}
 \item The intensity of the magnetic field in both adiabatic and full physics
 runs traces very well the underlying distribution of matter, indicating a close
 connection between the amplification mechanism of the field and the formation
 of structures in the Universe.
 \item The amplification of magnetic field intensity in non-radiative runs is 
 well described by magnetic flux conservation ($B \propto \rho^{2/3}$) at low baryon 
 overdensities. Within collapsed structures magnetic flux conservation under-predicts
 the magnetic field strength indicating that additional amplification takes place 
 thanks to gas turbulent motions and shear flows. 
 \item In the full physics simulations, {\it B} field amplification reaches saturation in 
 collapsed structure. Magnetic field intensities are much larger (up to a factor 
 $10^3$) with respect to non-radiative runs. Such a large degree of amplification 
 is attained due to radiative cooling leading to high baryon overdensities and the 
 increased level of turbulent and shear gas motions triggered by galactic 
 outflows and AGN feedback. 
 \item Varying the strength of the initial seed field (up to four orders of magnitude)
 does not affect the {\it B} field saturation value inside haloes in full physics runs. 
 On the other hand, in adiabatic runs the initial seed feed strength provides an 
 overall normalization factor for magnetic field intensities at all overdensities.
 The same holds for full physics runs at low overdensities.
 \item The initial direction of the seed field does not play any role for the 
 average {\it B} field properties and for its orientation inside haloes, since 
 turbulent gas motions rapidly delete any memory of the initial direction. 
 However, the original field orientation is retained in low overdensity 
 environments such as filaments and voids. 
 \item The average {\it B} field intensity correlates well with intrinsic halo
 properties such as viral masses. The introduction of baryon physics 
 (cooling and feedback processes) may disrupt trends that are present
 in non-radiative runs as, for instance, the {\it B} field temperature relation in
 massive haloes.
 \item Even in the full physics runs, magnetic fields are sub-dominant as far as 
 the global gas dynamics is concerned. The ratio between magnetic field and 
 gas thermal pressures is in the vast majority of cases well below the percent 
 level. Only in one of the examined halo mass bins for the (reference) full physics 
 simulation  the value of the magnetic pressure is larger than its thermal counterpart.
 \rev{The ratio between magnetic and gas kinetic energy 
 agrees well with what found in small-scale dynamo studies.}
 \item {\it B} field values predicted in full physics simulations in the centres of 
 massive haloes span a range from a few up to tens of ${\rm \muG}$, in agreement with  
 observations of galaxy cluster. 
 \item The predicted Faraday RM signal for the three most massive 
 haloes of the fiducial full physics simulation matches quite well galaxy cluster
 constraints. However, at radii comparable to the virial radius the inferred
 RM profiles decline more rapidly than the observational findings.
\end{enumerate}

For the first time, our simulations include magnetic fields within a
comprehensive model of galaxy formation physics in large-scale cosmological
simulations. This allows us to study the co-evolution of magnetic fields and
galaxies on different scales, a possibility that we will further exploit in 
future work.

\section*{Acknowledgements}
We thank the referee Robi Banerjee for a constructive report and Dylan Nelson, 
Annalisa Pillepich, Lars Hernquist, Volker Springel, and Rainer Beck
for their valuable suggestions. We further thank Volker Springel for giving us 
access to the \arepo\ code. RP acknowledges support by the European Research 
Council under ERC-StG EXAGAL-308037, by the DFG Research Centre SFB-881 `The 
Milky Way System', and by the Klaus Tschira Foundation. The simulations were 
performed on the joint MIT-Harvard computing cluster supported by MKI and FAS, 
the Magny cluster at the Heidelberg Institute for Theoretical Studies, and the 
Stampede supercomputer at the Texas Advanced Computing Center as part of XSEDE 
projects TG-AST14007S8 and TG-AST140082. All the figures in this work were 
produced by using the {\sc matplotlib} graphics environment \citep{Matplotlib}.

\bibliographystyle{mnras}
\bibliography{paper}
    
\label{lastpage}

\end{document}